\documentclass{aa}  

\usepackage{graphicx}
\usepackage{txfonts}
\newcommand{\kms}{km\,s$^{-1}$}
\newcommand{\vr}{V$_{\rm r}$}
\newcommand{\macc}{$\dot{\rm M}_{\rm acc}$}
\newcommand{\prot}{P$_{\rm rot}$}
\newcommand{\msunyr}{M$_\odot$.yr$^{-1}$}

\begin{document}

   \title{A multi-kiloGauss magnetic field driving the magnetospheric accretion process in EX Lupi}

   \author{K. Pouilly\inst{1}
          \and
          M. Audard\inst{1}
          \and
          Á. Kóspál\inst{2}\fnmsep\inst{3}
          \and
          A. Lavail\inst{4}
          }

   \institute{Department of Astronomy, University of Geneva, Chemin Pegasi 51, CH-1290 Versoix, Switzerland\\
              \email{Kim.Pouilly@unige.ch}
        \and
        Konkoly Observatory, HUN-REN Research Centre for Astronomy and Earth Sciences, MTA Centre of Excellence, Konkoly-Thege Mikl\'os \'ut 15-17, 1121 Budapest, Hungary
        \and
        Institute of Physics and Astronomy, ELTE E\"otv\"os Lor\'and University, P\'azm\'any P\'eter s\'et\'any 1/A, 1117 Budapest, Hungary
        \and
        Institut de Recherche en Astrophysique et Planétologie, Université de Toulouse, CNRS, IRAP/UMR 5277, 14 avenue Edouard Belin, 31400 Toulouse, France
             }

   \date{Received 16 July 2024; accepted 04 September 2024}

 
  \abstract
   {EX~Lupi is the prototype of EX Lup-type stars, meaning classical T Tauri stars (cTTSs) showing luminosity bursts and outbursts of 1 to 5 magnitudes lasting for a few months to a few years. These events are ascribed to an episodic accretion that can occur repeatedly but whose physical mechanism is still debated.}
   {In this work, we aim to investigate the magnetically-driven accretion of EX~Lup in quiescence, including for the first time a study of the small and large-scale magnetic field. This allows us to provide a complete characterisation of the magnetospheric accretion process of the system.}
   {We use spectropolarimetric times series acquired in 2016 and 2019 with the Echelle SpectroPolarimetric Device for the Observation of Stars and in 2019 with the SpectroPolarimètre InfraRouge at the Canada-France-Hawaii telescope, during a quiescence phase of EX~Lup. We were thus able to perform a variability analysis of the radial velocity, the emission lines and surface averaged longitudinal magnetic field along different epochs and wavelength domains. We also provide a small-scale magnetic field analysis using Zeeman intensification of photospheric lines and large-scale magnetic topology reconstruction using Zeeman-Doppler Imaging.}
   {Our study reveals a typical magnetospheric accretion ongoing on EX~Lup, with a main accretion funnel flow connecting the inner disc to the star in a stable fashion and producing an accretion shock on the stellar surface close to the pole of the magnetic dipole component. We also measure one of the strongest fields ever observed on cTTSs. Such a strong field indicates that the disc is truncated by the magnetic field close but beyond the corotation radius, where the angular velocity of the disc equals the angular velocity of the star. Such a configuration is suitable for a magnetically-induced disc instability yielding episodic accretion onto the star.}
   {}

   \keywords{Stars: variables: T Tauri --
                Stars: individual: EX Lup --
                Stars: magnetic field --
                Accretion, accretion disks --
                Techniques: spectroscopic --
                Techniques: polarimetric  
               }

   \maketitle
%

\section{Introduction}
   
    EX Lup-type objects (EXors) are classical T Tauri stars (cTTSs), meaning low-mass pre-main sequence stars surrounded by an accretion disc, that show burst and outburst events ascribed to episodic accretion. 
    During these phases, they can increase their optical luminosity from 1 to 5 magnitudes, lasting typically for a few months to a few years, and can occur repeatedly \citep[for a review, see, e.g.,][]{Fischer23}. 
    These events are therefore more moderate, both in duration and luminosity increase, than the FU Orionis-type stars (FUors).
    
    While the magnetospheric accretion of cTTSs, where the strong stellar magnetic field truncates the disc, forcing the accreted material to follow the magnetic field lines \citep[see the review by][]{Hartmann16}, seems to be ongoing on EXors as well, the origin of this episodic accretion is still highly debated. 
    The different hypotheses can be gathered in three groups \citep[see review by][]{Audard14}: (i) the magnetospheric accretion itself, being inherently episodic when a strong magnetic field truncates the disc close to the corotation radius \citep{DAngelo10}. (ii) The disc, showing viscous-thermal \citep{Bell94}, or gravitational and magneto-rotational \citep{Armitage01} instabilities, or accretion clumps in a gravitationally unstable environment \citep{Vorobyov05, Vorobyov06}. (iii) The presence of a companion, perturbing the accretion trough tidal effect \citep{Bonnell92},  or thermal instabilities \citep{Lodato04}.

    This work aims to investigate the magnetospheric accretion process of the prototypical EXors, \object{EX Lup}, using high-resolution spectropolarimetric time series, as it was done for cTTSs not displaying episodic accretion \citep[e.g.,][]{Pouilly20, Pouilly21}.
    This object is a young M0.5-type star \citep{GrasVelazquez05}, known to have both moderate and short-timescale variability and rare extreme episodic outbursts.
    It is located at 154.7$\pm$0.4 pc \citep[\textit{Gaia} DR3 parallax 6.463$\pm$0.015 mas,][]{Gaia23} and has a rotation period of 7.417 days, from its radial velocity modulation that was first ascribed to a low-mass companion \citep{Kospal14}, before being ascribed to stellar activity \citep{SiciliaAguilar15}.
    The system has a moderate-to-low inclination of its rotation axis (between 20$^\circ$ and 45$^\circ$) according to modelling of the spectral energy distribution \citep{Sipos09} and emission lines analysis \citep{Goto11,SiciliaAguilar15}, and a projected rotational velocity $v\sin i$=4.4$\pm$2.0 \kms\ \citep{Sipos09}.
    
    This object is one of the most studied EXors using spectroscopy, both in quiescence \citep[i.e.,][]{Kospal14,Sipos09,SiciliaAguilar12, SiciliaAguilar15, SiciliaAguilar23, Campbell21, Wang23} and in outburst \citep[][for the last two outbursts in 2008 and 2022]{SiciliaAguilar12, Cruz23, Wang23, Singh24}, but the present study is the first one including spectropolarimetry, giving access to information on the magnetic field together with the accretion diagnostics.
    Its spectrum contains the typical accretion-related emission lines observed in cTTSs, in addition to numerous neutral metallic emission lines superimposed to photospheric absorption in quiescence, and overwhelming any absorption feature in outburst \citep{SiciliaAguilar12}.
    The present dataset was acquired during quiescence, meaning that we will characterise the "stable" accretion of the system, even if \cite{SiciliaAguilar12, SiciliaAguilar23} have shown that the accretion pattern seems stable in quiescence and outburst, only the amount of accreted material is affected.
    This accretion pattern is consistent with the typical cTTS magnetospheric accretion, through accretion funnel flows connecting the disc to the stellar surface, except that "clumps" of material are also accreted through these funnel flows, detected thanks to the day-to-day variation of the emission lines' broad component (BC) \citep[see Fig. 17 of][]{SiciliaAguilar12}.

    This article is organised as follows: we describe the observations in Sect.~\ref{sec:obs}, the analysis and results are presented in Sect.~\ref{sec:results} and discussed in Sect.~\ref{sec:discussion}.
    We conclude this work in Sect.~\ref{sec:conclusion}.

\section{Observations}
\label{sec:obs}

    The spectropolarimetric time series used in this work were acquired at the Canada-France-Hawaii Telescope at two different epochs (2016 and 2019, proposals 16AF03 and 19AF50, respectively).
    The second data set is composed of two subsets using two different instruments: the Echelle SpectroPolarimetric Device for the Observation of Stars \citep[ESPaDOnS,][]{Donati03} in the optical and the SpectroPolarimètre InfraRouge \citep[SPIRou, ][]{Donati20b} in the near-infrared, both used in polarimetric mode, while the first one only used ESPaDOnS.
    This means that each observation is composed of four sub-exposures taken in different polarimeter configurations, which are then combined to obtain the intensity (Stokes \textit{I}), the circularly polarised (Stokes \textit{V}), and the null polarisation spectra.
    A complete journal of observation is provided in Table~\ref{tab:logObs}.

    \subsection{ESPaDOnS}
    The ESPaDOnS observations, which cover the 370 to 1050 nm wavelength range and reach a resolving power of 68\,000, consist of 11 nights between 2016 June 09 and 2016 June 24, with an approximately nightly cadence, and 6 nights between 2019 May 31 and 2019 June 12, the 5 latter respecting a 1-day sampling.
    The signal-to-noise ratio (S/N) of the 2016 (2019) observations ranges between 69 and 142 (111 and 140) for the Stokes \textit{I} at 731 nm. 
    Each observation was reduced using the \texttt{Libre-ESpRIT} package \citep{Donati97}.

    \subsection{SPIRou}
    The SPIRou observations are covering the 960 to 2350 nm wavelength range with R$\sim$75\,000.
    They were acquired during 8 consecutive nights, between 2019 June 14 and 2019 June 21, and the S/N of the unpolarised spectra in the H-band range between 107 and 175.
    The observations were reduced using the \texttt{APERO} pipeline \citep{Cook22}.

    \begin{table}
        \centering
            \caption{Log of EX~Lup observations.}
        \begin{tabular}{l l r l l l  }
            \hline
            \hline
            Date & HJD & S/N$_{\rm I}$ & S/N$_{\rm LSD}$ & Inst.\\ 
             & ($-$2\,450\,000 d) & & &   \\
            \hline
            09 Jun & 7548.91042 & 133 &  7035 & E \\
            10 Jun & 7549.89640 & 139 &  7873 & E \\
            11 Jun & 7550.90738 & 131 &  7771 & E \\
            12 Jun & 7551.87135 & 69  &  3794 & E \\
            16 Jun & 7555.85625 & 122 &  6485 & E \\
            19 Jun & 7558.86916 & 128 &  7441 & E \\
            20 Jun & 7559.88712 & 142 &  8422 & E \\
            21 Jun & 7560.90209 & 126 &  7114 & E \\
            22 Jun & 7561.90205 & 120 &  6732 & E \\
            23 Jun & 7562.86001 & 125 &  6721 & E \\ 
            24 Jun & 7563.86197 & 129 &  6858 & E \\
            \hline
            31 May & 8634.94852 & 126 &  8872 & E \\
            08 Jun & 8642.94545 & 140 &  9815 & E \\
            09 Jun & 8643.94843 & 121 &  7931 & E \\ 
            10 Jun & 8644.94641 & 119 &  7624 & E \\
            11 Jun & 8645.87939 & 111 &  6384 & E \\ 
            12 Jun & 8646.87437 & 115 &  6482 & E \\
            14 Jun & 8648.83143 & 175 &  3643 & S\\
            15 Jun & 8649.83910 & 173 &  3999 & S \\
            16 Jun & 8650.85516 & 172 &  3785 & S \\
            17 Jun & 8651.86939 & 107 &  1247 & S \\
            18 Jun & 8652.82038 & 152 &  3551 & S \\
            19 Jun & 8653.82673 & 140 &  2417 & S \\
            20 Jun & 8654.80603 & 166 &  4125 & S \\
            21 Jun & 8655.87664 & 160 &  2966 & S \\
   
        \hline
        \end{tabular}
        \tablefoot{The S/N$_{\rm I}$ corresponds to the peak S/N by spectral pixel at 731~nm (in H-band) for ESPaDOnS (SPIRou) observations. The S/N$_{\rm LSD}$ is the effective S/N of the Stokes \textit{V} LSD profiles (see Sect.~\ref{subsubsec:largescale}). The last column indicates which instrument was used for each observation (E-ESPaDOnS, S-SPIRou). The horizontal line separates the 2016 from the 2019 observations.}
        \label{tab:logObs}
        \end{table}

\section{Results}
\label{sec:results}
    
    In this section, we present the results obtained from the analysis of the observations described in Sect.~\ref{sec:obs}. They consist of the analysis of the radial velocity, emission lines, and stellar magnetic field.
    
    \subsection{Radial Velocity}
    \label{subsec:rv}
\begin{figure}
            \centering
            \includegraphics[width=.45\textwidth]{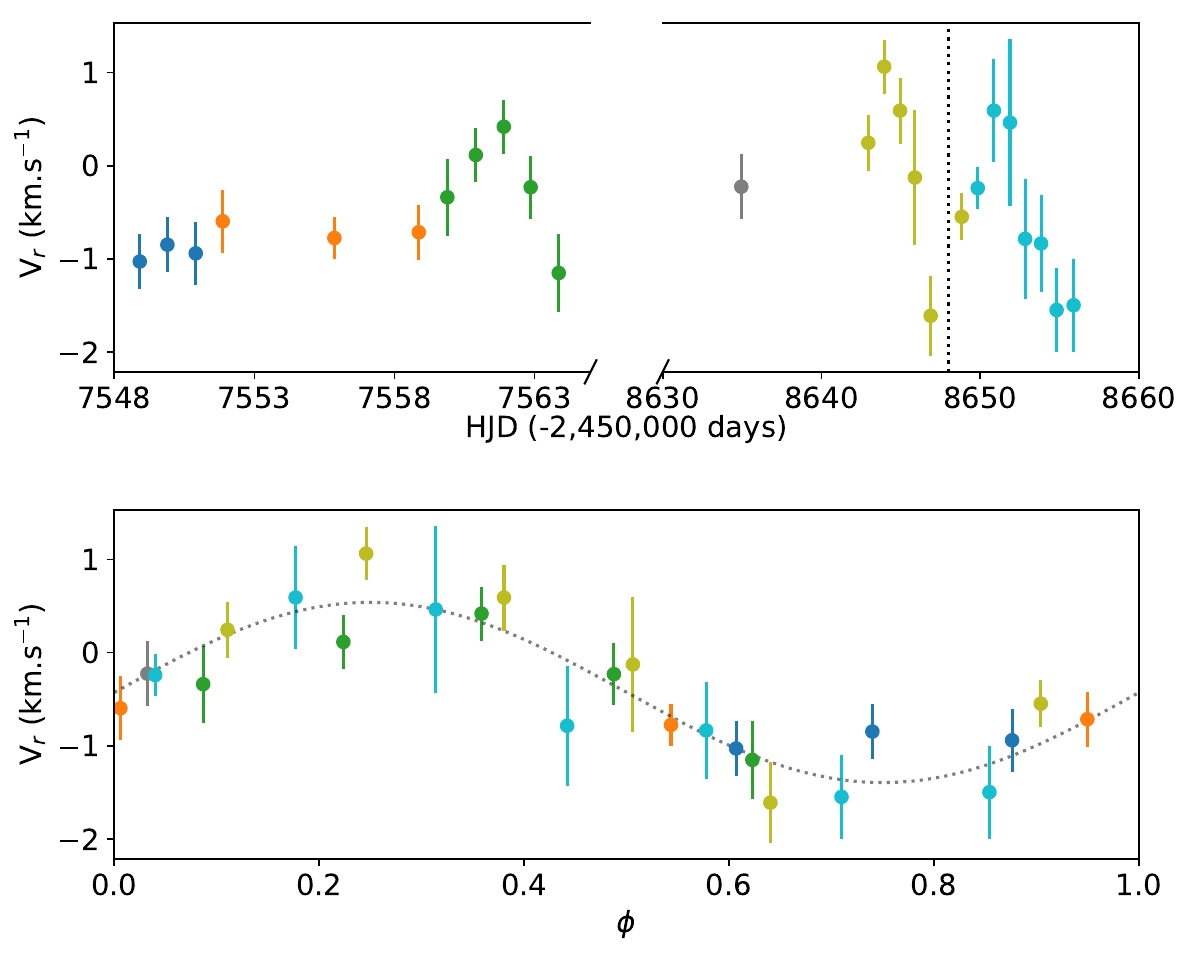}
            \caption{Radial velocities curves determined for the observation summarised in Table~\ref{tab:logObs}. \textit{Top:} radial velocity vs. HJD, the vertical dotted line marks the switch from ESPaDOnS to SPIRou, and the different colours represent different rotation cycles. \textit{Bottom:} same as above but folded in phase with P=7.417 d and T0=2\,457\,544.40981. The dotted curve shows the sinus fit.}
            \label{fig:RVtot}
        \end{figure}

        To determine the radial velocity of EX~Lup, we cross-correlated each spectrum with a synthetic spectrum computed using the \texttt{ZEEMAN} code \citep{Landstreet88, Wade01, Folsom12}, with \texttt{MARCS} atmospheric models \citep{Gustafsson08} and \texttt{VALD} \citep{Ryabchikova15} line lists adapted to EX~Lup stellar parameters for ESPaDOnS and SPIRou wavelengths.
        We computed the cross-correlation function (CCF) over 27 (14) wavelength windows of about 10 nm ranging from 441 to 890 nm (1150 to 2290 nm) for ESPaDOnS (SPIRou) observations.
        Then we performed a sigma-clipping across all the CCFs for each observation and used the mean and standard deviation of the remaining values as measurement of the radial velocity and its uncertainty.
        The results are plotted in Fig.~\ref{fig:RVtot}.
        A quick sinusoidal fit of the values for each data set allowed us roughly measure the periodicity and mean value of the radial velocity and yielded P=7.55$\pm$0.24\,d and $\langle$\vr$\rangle$=$-$0.48$\pm$0.07\,\kms\ (ESPaDOnS 2016), P=7.63$\pm$0.18\,d and $\langle$\vr$\rangle$=$-$0.59$\pm$0.14\,\kms\ (ESPaDOnS 2019), and P=8.36$\pm$0.97\,d and $\langle$\vr$\rangle$=$-$0.58$\pm$0.14\,\kms\ (SPIRou).
        These measures are consistent within the uncertainties with previous results obtained by \cite{Kospal14}, \prot=7.417$\pm$0.001\,d and $\langle$\vr$\rangle$=$-$0.52$\pm$0.07\,\kms, we will thus adopt the latter values for our analysis.
        Finally, we folded all the measurements in phase using \prot=7.417~d and an arbitrary T0, and we fitted this curve with a sinus to estimate the T0 required to set $\phi$ = 0.5 at the mean velocity between the maximum and minimum of the modulation \citep[when the spot modulating the curve is facing the observer, see Fig. 7 of][]{SiciliaAguilar15}.
        The resulting T0 is HJD 2\,457\,544.40981.
        We will thus use the following ephemeris for the rest of this work:
        \begin{equation}
            \rm{HJD (d)} = 2\,457\,544.40981 + 7.417\,\rm{E},
            \label{eq:ephemeris}
        \end{equation}
        where E is the rotation cycle.
        All our radial velocity measurements are in phase (Fig.~\ref{fig:RVtot}), but the 2019 measurements show a larger amplitude of its modulation.
        This indicates an evolving feature modulating the radial velocity but located at the same longitude in 2016 and 2019.
        The values are summarised in Table~\ref{tab:RV}.

        \begin{table}
        \centering
            \caption{Radial velocities measured for each observation and their uncertainties.}
        \begin{tabular}{l l l l l l l  }
            \hline
            \hline
            HJD & \vr & $\delta$\vr & $\phi$ & Inst.\\ 
            ($-$2~450~000 d) & (\kms) & (\kms) & &  \\
            \hline
            7548.91042 & $-$1.03 & 0.30 & 0.607 & E \\
            7549.89640 & $-$0.85 & 0.29 & 0.740 & E \\
            7550.90738 & $-$0.94 & 0.34 & 0.876 & E \\
            7551.87135 & $-$0.60 & 0.34 & 0.006 & E \\
            7555.85625 & $-$0.78 & 0.22 & 0.543 & E \\
            7558.86916 & $-$0.71 & 0.30 & 0.949 & E \\
            7559.88712 & $-$0.34 & 0.41 & 0.087 & E \\
            7560.90209 &  0.12 & 0.29 & 0.224 & E \\
            7561.90205 &  0.42 & 0.29 & 0.358 & E \\
            7562.86001 & $-$0.23 & 0.34 & 0.488 & E \\ 
            7563.86197 & $-$1.15 & 0.42 & 0.623 & E \\
            \hline
            8634.94852 & $-$0.23 & 0.35 & 0.032 & E \\
            8642.94545 &  0.25 & 0.30 & 0.111 & E \\
            8643.94843 &  1.06 & 0.29 & 0.246 & E \\ 
            8644.94641 &  0.59 & 0.35 & 0.380 & E \\
            8645.87939 & $-$0.13 & 0.73 & 0.506 & E \\ 
            8646.87437 & $-$1.61 & 0.43 & 0.640 & E \\
            8648.83143 & $-$0.55 & 0.25 & 0.904 & S \\
            8649.83910 & $-$0.24 & 0.22 & 0.040 & S \\
            8650.85516 &  0.59 & 0.55 & 0.177 & S \\
            8651.86939 &  0.46 & 0.90 & 0.314 & S \\
            8652.82038 & $-$0.79 & 0.64 & 0.442 & S \\
            8653.82673 & $-$0.83 & 0.52 & 0.578 & S \\
            8654.80603 & $-$1.55 & 0.45 & 0.710 & S \\
            8655.87664 & $-$1.50 & 0.50 & 0.854 & S \\
   
        \hline
        \end{tabular}
        \tablefoot{The phases $\phi$ are computed using the ephemeris provided in Eq.~\ref{eq:ephemeris}.  The horizontal line separates the 2016 from the 2019 observations.}
        \label{tab:RV}
        \end{table}

    \subsection{Emission line variability}
    \label{subsec:emLines}
        In this section, we present the analysis of emission lines tracing either the accretion funnel flow, here the Balmer lines \citep{Muzerolle01}, or the accretion shock such as the \ion{Ca}{II} infrared triplet (IRT), the \ion{He}{I} D3 \citep{Beristain01}, or the \ion{He}{I} at 1083 nm lines.
        For each line, we analysed the profile variability, their periodicity (except for the 2019 ESPaDOnS data set which does not cover a sufficient time span), and the correlations of these variabilities.
        Most of the analyses of this section were performed using \texttt{PySTEL(L)A}\footnote{\url{https://github.com/pouillyk/PySTELLA}}, a Python tool for SpecTral Emission Lines (variabiLity) Analysis.
        
        \subsubsection{Balmer lines}
\begin{figure*}
                \centering
                \includegraphics[width=.3\textwidth]{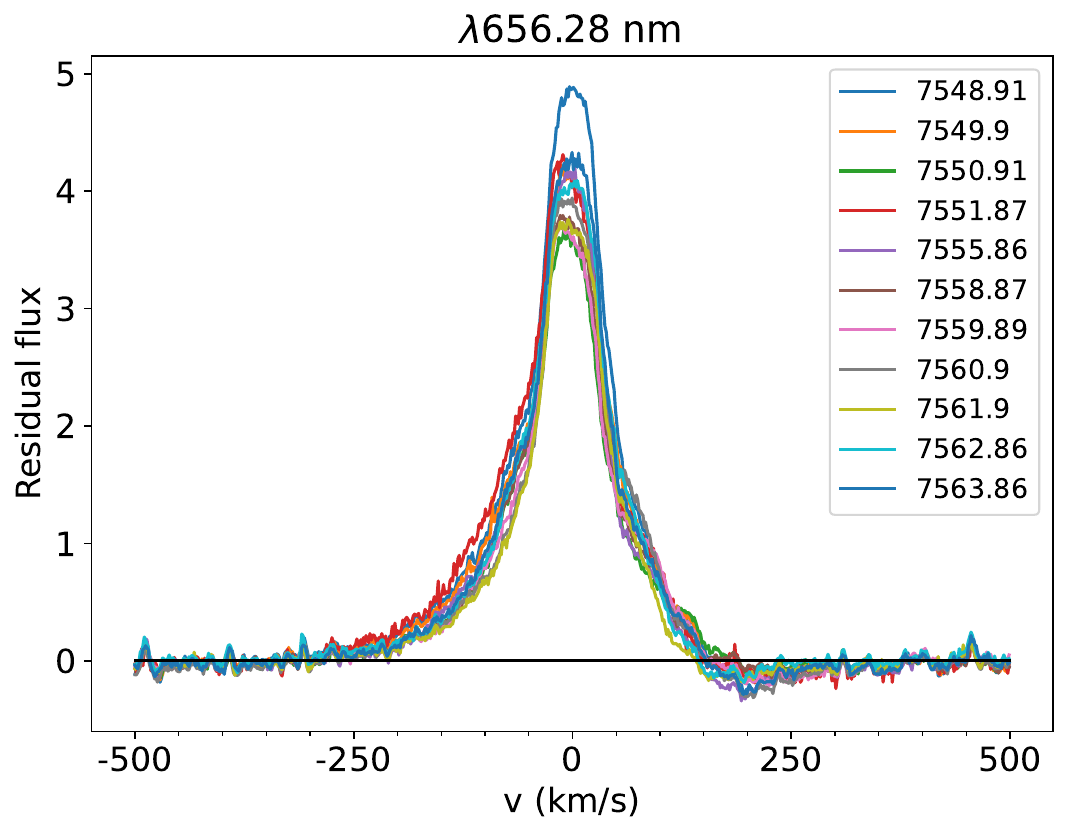}
                \includegraphics[width=.3\textwidth]{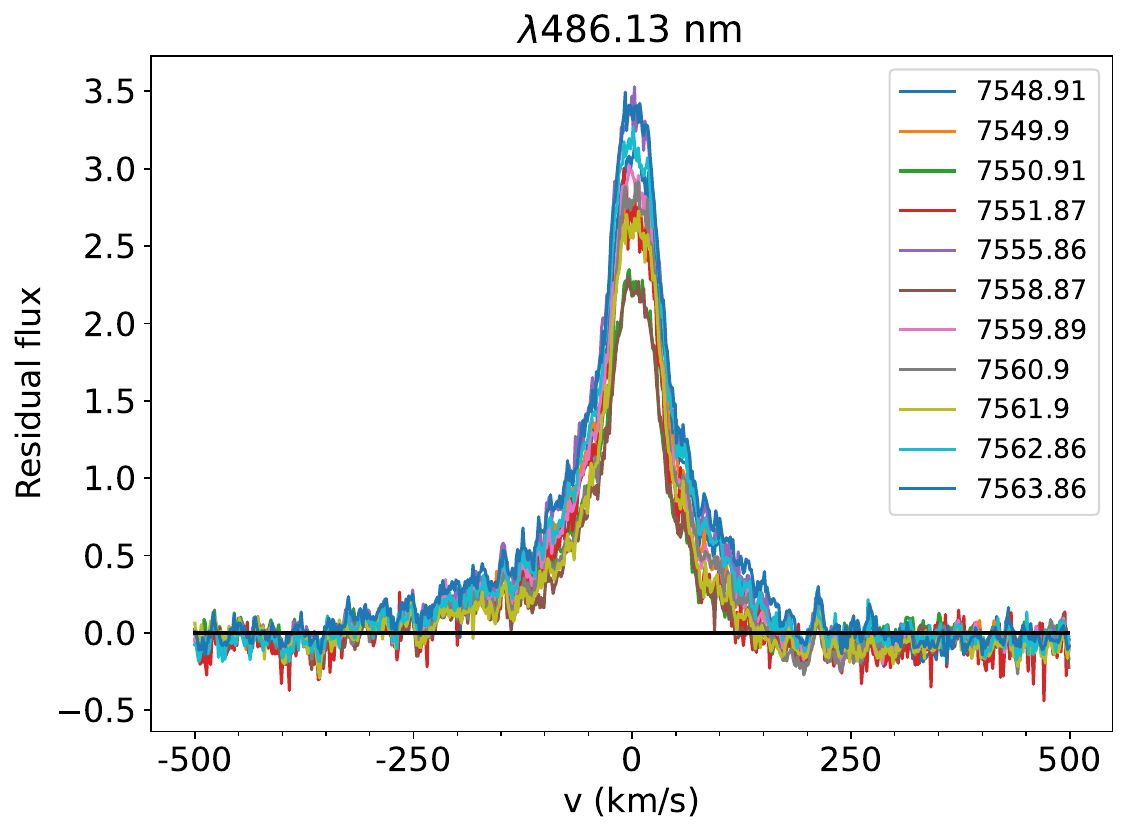}
                \includegraphics[width=.3\textwidth]{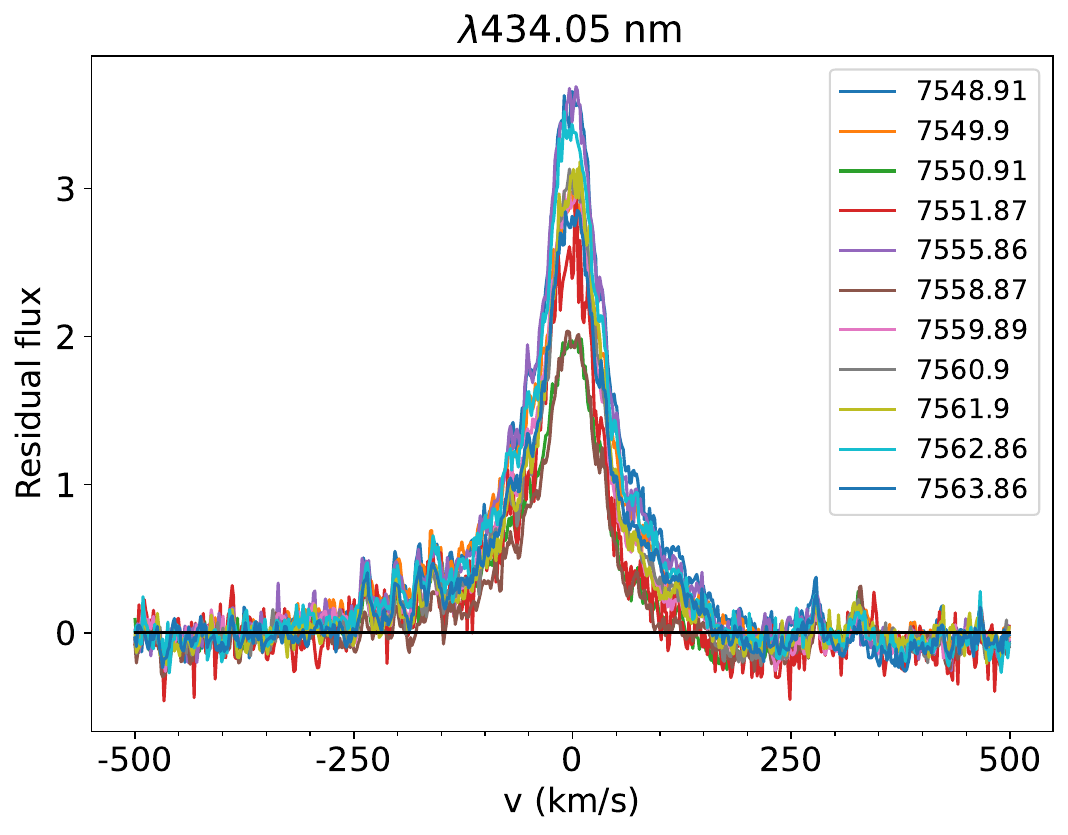}
                \includegraphics[width=.3\textwidth]{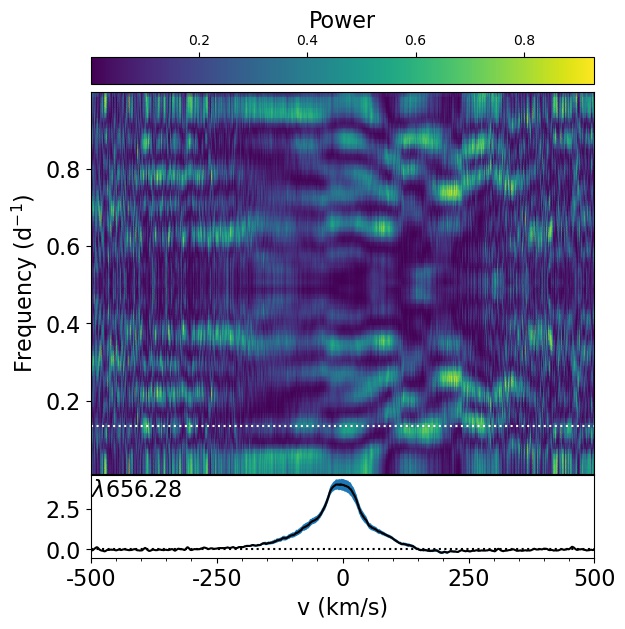}
                \includegraphics[width=.3\textwidth]{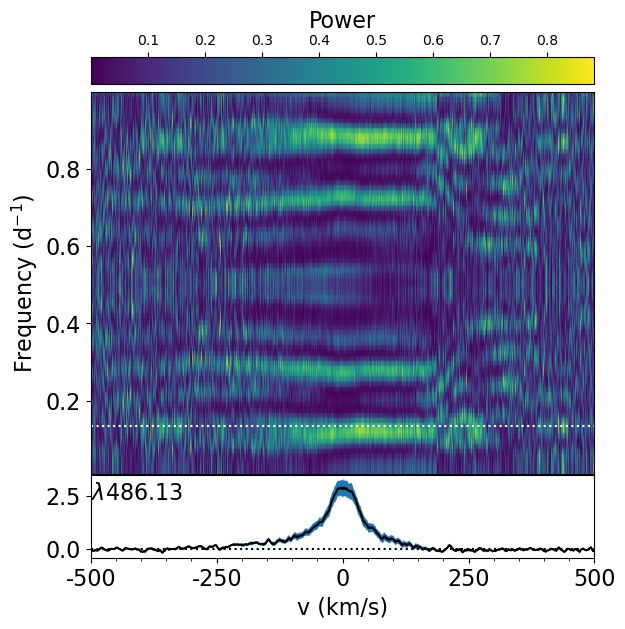}
                \includegraphics[width=.3\textwidth]{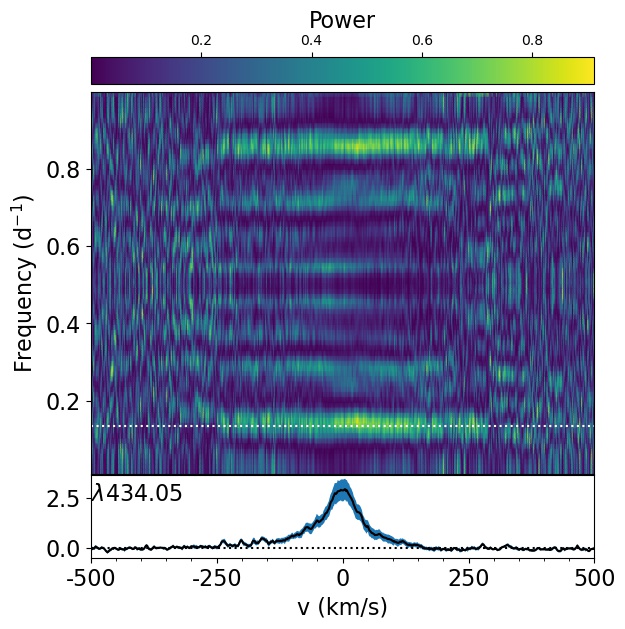}
                \caption{Variability of 2016 Balmer lines of EX~Lup. \textit{Top row:} H$\alpha$ \textit{(left)}, H$\beta$ \textit{(middle)}, and H$\gamma$ \textit{(right)} residual lines profiles. Each colour represents a different observation. \textit{Bottom row:} 2D-periodograms of H$\alpha$ \textit{(left)}, H$\beta$ \textit{(middle)}, and H$\gamma$ \textit{(right)} residual lines. The white dotted line marks the rotation period of 7.417 d.}
                \label{fig:Balmer16}
            \end{figure*}

\begin{figure*}
                \centering
                \includegraphics[width=.3\textwidth]{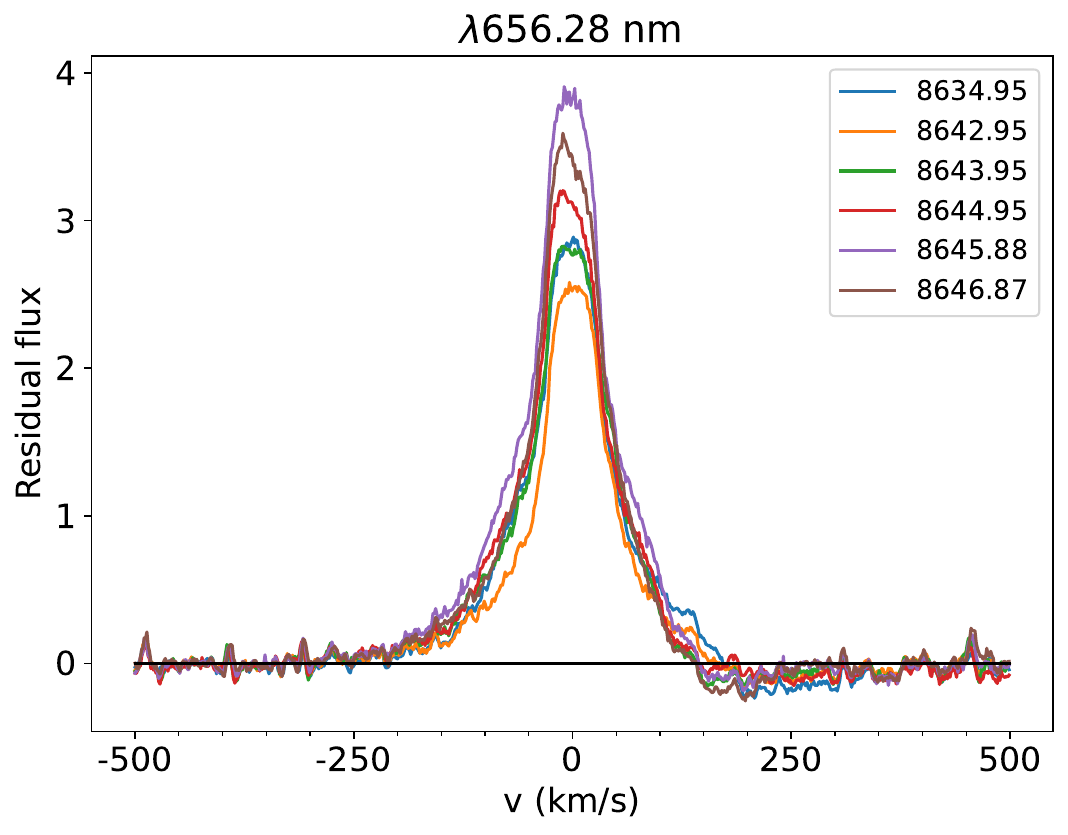}
                \includegraphics[width=.3\textwidth]{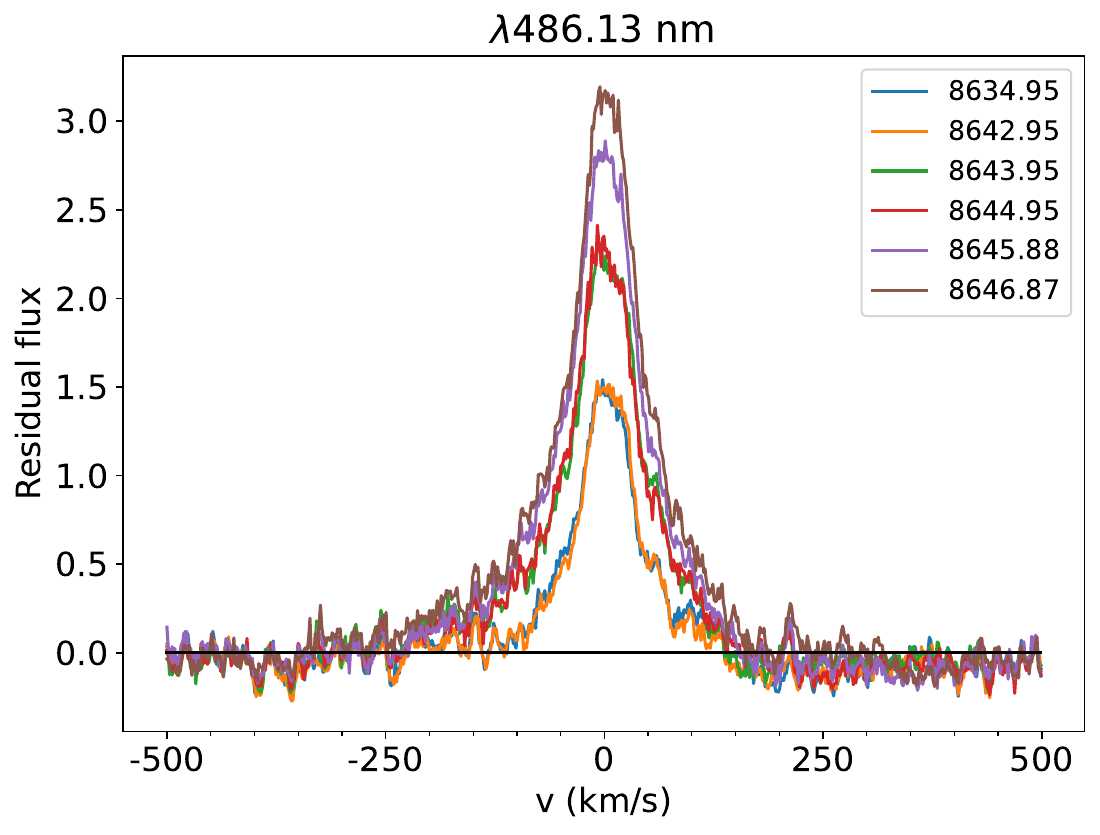}
                \includegraphics[width=.3\textwidth]{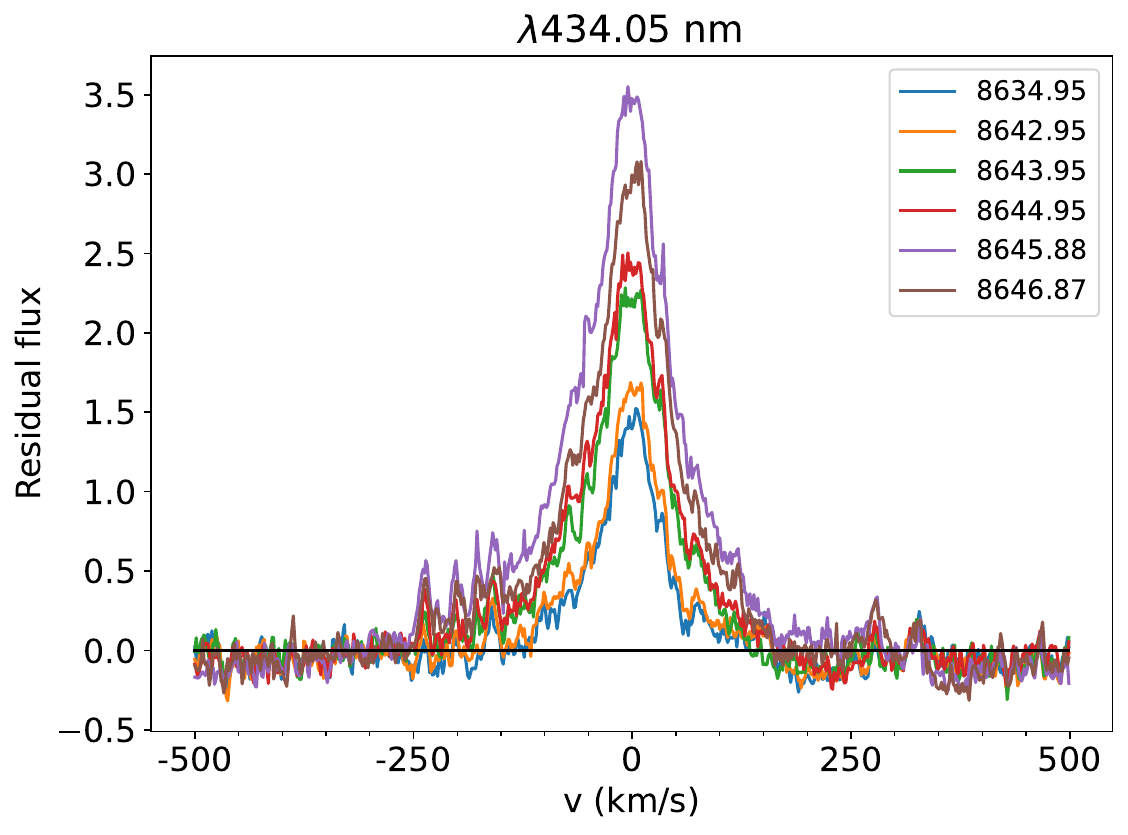}
                \includegraphics[width=.3\textwidth]{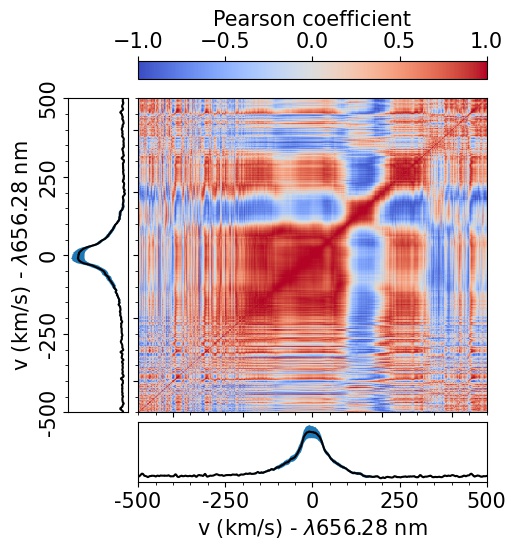}
                \includegraphics[width=.3\textwidth]{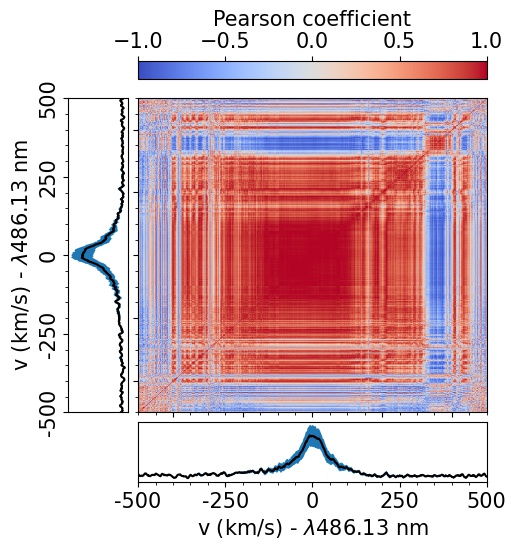}
                \includegraphics[width=.3\textwidth]{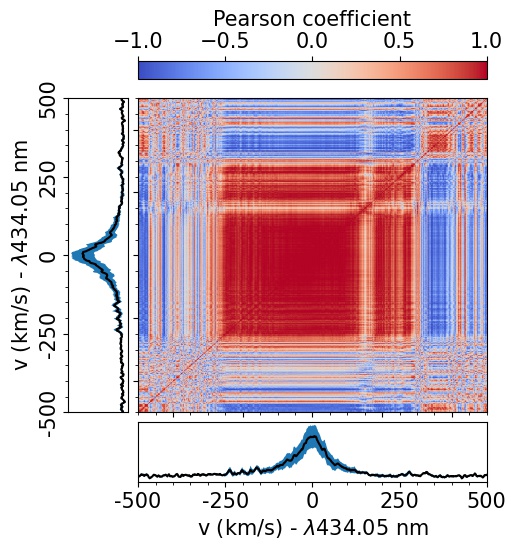}
                \caption{Variability of 2019 Balmer lines of EX~Lup. \textit{Top row:} H$\alpha$ \textit{(left)}, H$\beta$ \textit{(middle)}, and H$\gamma$ \textit{(right)} residual lines profiles. Each colour represents a different observation. \textit{Bottom row:} H$\alpha$ \textit{(left)}, H$\beta$ \textit{(middle)}, and H$\gamma$ \textit{(right)} residual lines auto-correlation matrices. The colour code is scaling the correlation coefficient. Light yellow represents a strong correlation and dark purple a strong anti-correlation. Please note that the strong anti-correlation around $+$350\,\kms\ on H$\beta$ and H$\gamma$ matrices is probably an artefact as located in a poorly varying part of the continuum.}
                \label{fig:Balmer19}
            \end{figure*}

            The Balmer lines are partly formed in the accretion funnel flow and are thus tracing the magnetospheric accretion process.
            In this work, we focussed on H$\alpha$, H$\beta$, and H$\gamma$, which we corrected from the photospheric contribution using the moderately active M-dwarf HD~42581 as template \citep[T$_{\rm eff}$ = 3822 K, $v\sin i$ = 2.6 \kms,][]{Manara21}, broadened to the $v\sin i$ of EX~Lup.
            The 2016 and 2019 profiles are shown in Fig.~\ref{fig:Balmer16} and Fig.~\ref{fig:Balmer19}, respectively.
            On both data sets, the profiles are composed of a broad and a narrow component, both highly variable. 
            Furthermore we can notice a flux depletion, going below the continuum, around $+$200~\kms, and extending up to $+$300~\kms.
            This behaviour is characteristic of the so-called inverse P Cygni (IPC) profiles, the red-shifted absorption forming due to infalling material.
            
            The 2D-periodograms, consisting of a Lomb-Scargle periodogram computed in each velocity channel, of the 2016 lines are presented in Fig.~\ref{fig:Balmer16}, and show a periodic signal along the whole H$\beta$ and H$\gamma$ lines consistent with the rotation period of the star, with a false alarm probability \citep[FAP, computed from][prescriptions]{Baluev08} of 0.04 and 0.01, respectively. 
            The H$\alpha$ line is showing this signal in a less continuous way with a much higher FAP (0.25).
            The symmetric signal around 0.9 d$^{\rm -1}$ is the 1-day alias, a spectral leakage of the Fourrier transform reconstructing (with the real period) the observation sampling, meaning approximately one observation per day.
            
            To separate the various parts of the line profile undergoing different variability patterns, we computed the auto-correlation matrices of each line.
            This tool consists of computing a linear correlation coefficient (here a Pearson coefficient) between the velocity channels of the line.
            A correlated region (close to 1) indicates a variability dominated by a given physical process. 
            An anti-correlated region (close to $-$1) indicates a variability dominated by a given physical process or, at least, linked physical processes.
            The auto-correlation matrices of 2019 H$\alpha$, H$\beta$, and H$\gamma$ lines are shown in Fig~\ref{fig:Balmer19}.
            Here again, H$\alpha$ is showing a different behaviour than H$\beta$ and H$\gamma$.
            H$\alpha$ shows three main correlated regions between $-$100~\kms\ and $+$100~\kms, corresponding to the core of the line, around $+$150~\kms, corresponding to a slight emission excess in the IPC profile (occurring around HJD 2\,458\,8634.95 and 2\,458\,641.95), and around $+$250~\kms, corresponding to the IPC profile itself. 
            The two latter regions are anti-correlated between them, but the $\sim$$+$150~\kms\ region is also slightly anti-correlated with the line center.
            This means that this region might be a broadening of the BC invoked by \cite{SiciliaAguilar12}, causing a global decrease of the line (and so a anti-correlation with the whole line profile).
            H$\beta$ and H$\gamma$ are showing the same correlation around the line centre and $+$350~\kms, without significant anti-correlation around $+$150~\kms, and a moderate anti-correlation of the profile with a region around $+$350~\kms\ which seems to be an artefact as it is located in the continuum.
            
        \subsubsection{\ion{He}{I} D3 587.6 nm}
        \label{subsubsec:heid3}
\begin{figure*}
                \centering
                \includegraphics[width=0.4\textwidth]{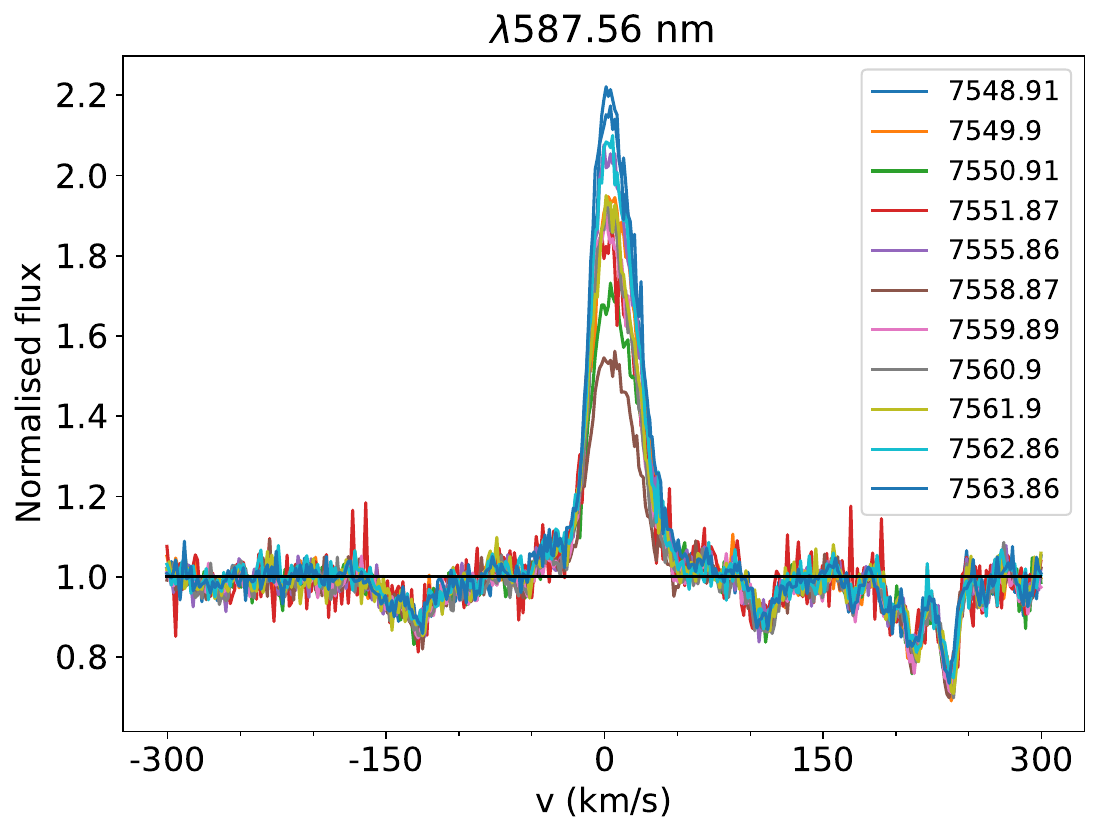}
                \includegraphics[width=0.4\textwidth]{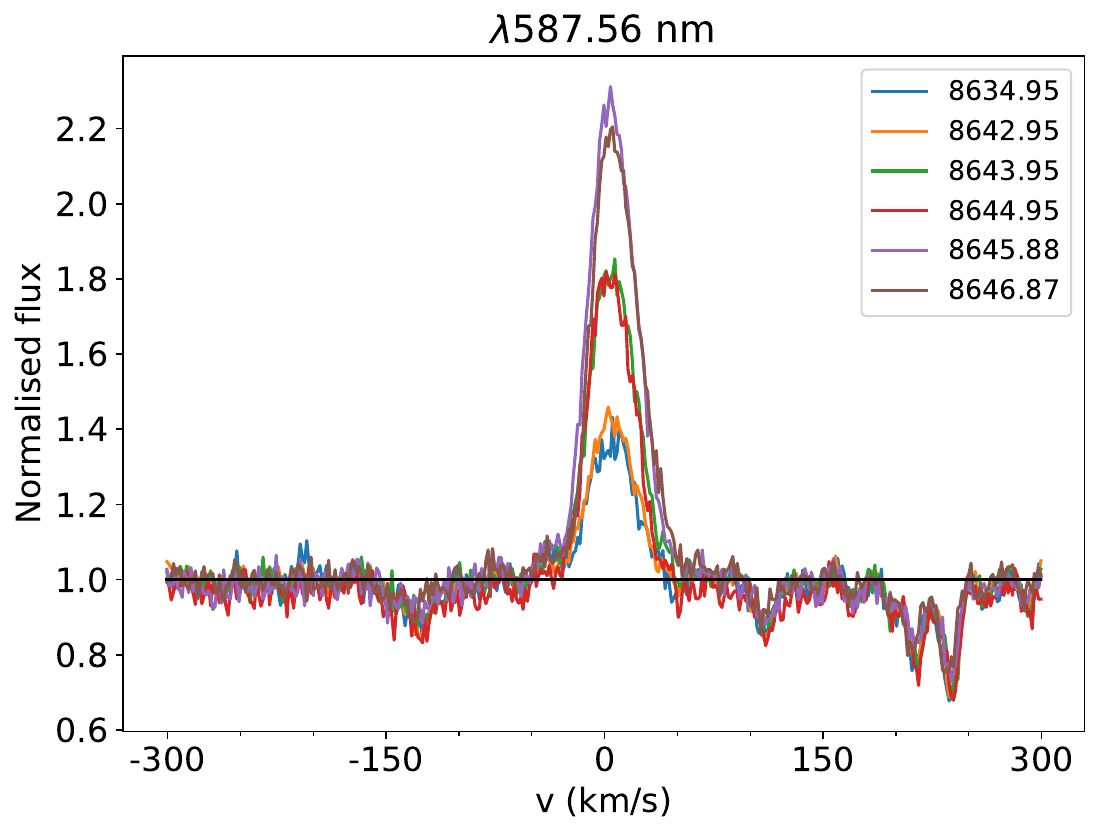}
                \includegraphics[width=0.4\textwidth]{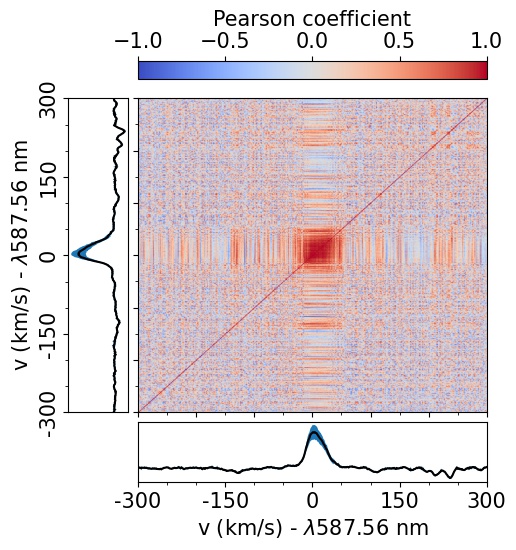}
                \includegraphics[width=0.4\textwidth]{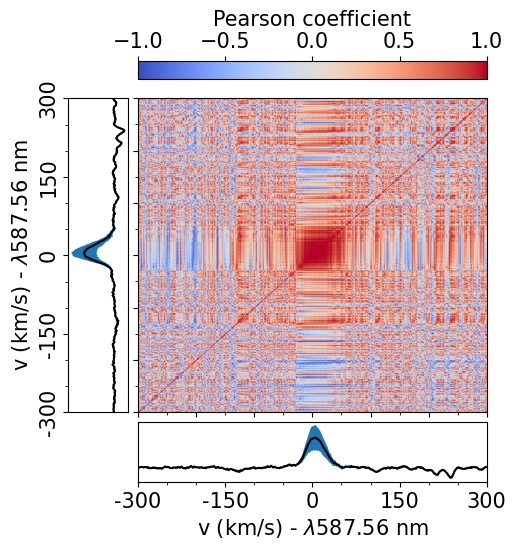}
                \caption{EX Lup \ion{He}{I} D3 line variability. \textit{Top row:} 2016 \textit{(left)} and 2019 \textit{(right)} lines profiles. \textit{Bottom row:} Auto-correlation matrices of 2016 \textit{(left)} and 2019 \textit{(right)} lines.}
                \label{fig:HeID3}
            \end{figure*}

            The \ion{He}{I} D3 lines of 2016 and 2019 data sets are presented in Fig.~\ref{fig:HeID3} and are composed of a narrow component (NC) only, extending from approximately $-$35 to  $+$50~\kms, without significant BC.
            The NC is formed in the post-shock region of the accretion spot \citep{Beristain01}, and can thus trace the accretion close to the stellar surface. 
            One can note the strong variability of this NC with a maximum reached around $\phi$=0.6 \footnote{HJD 2\,457\,548.91, 2\,458\,645.88, and 2\,458\,646.87}.
            
\begin{figure}
                \centering
                \includegraphics[width=0.4\textwidth]{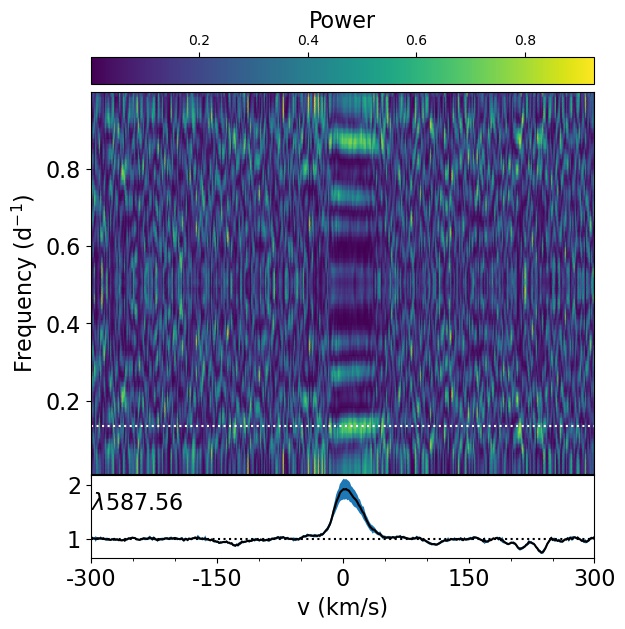}
                \caption{\ion{He}{I} D3 2016 2D-periodogram.}
                \label{fig:2DPHeID3}
            \end{figure}

            The auto-correlation matrices shown in Fig.~\ref{fig:HeID3} confirm that this region is formed by one physical process, and the periodicity of the 2016 NC's variation, consistent with the stellar rotation period (see Fig.~\ref{fig:2DPHeID3}, FAP=0.06), indicates that this component is tracing an accretion shock at the stellar surface.

\begin{figure}
                \centering
                \includegraphics[width=0.4\textwidth]{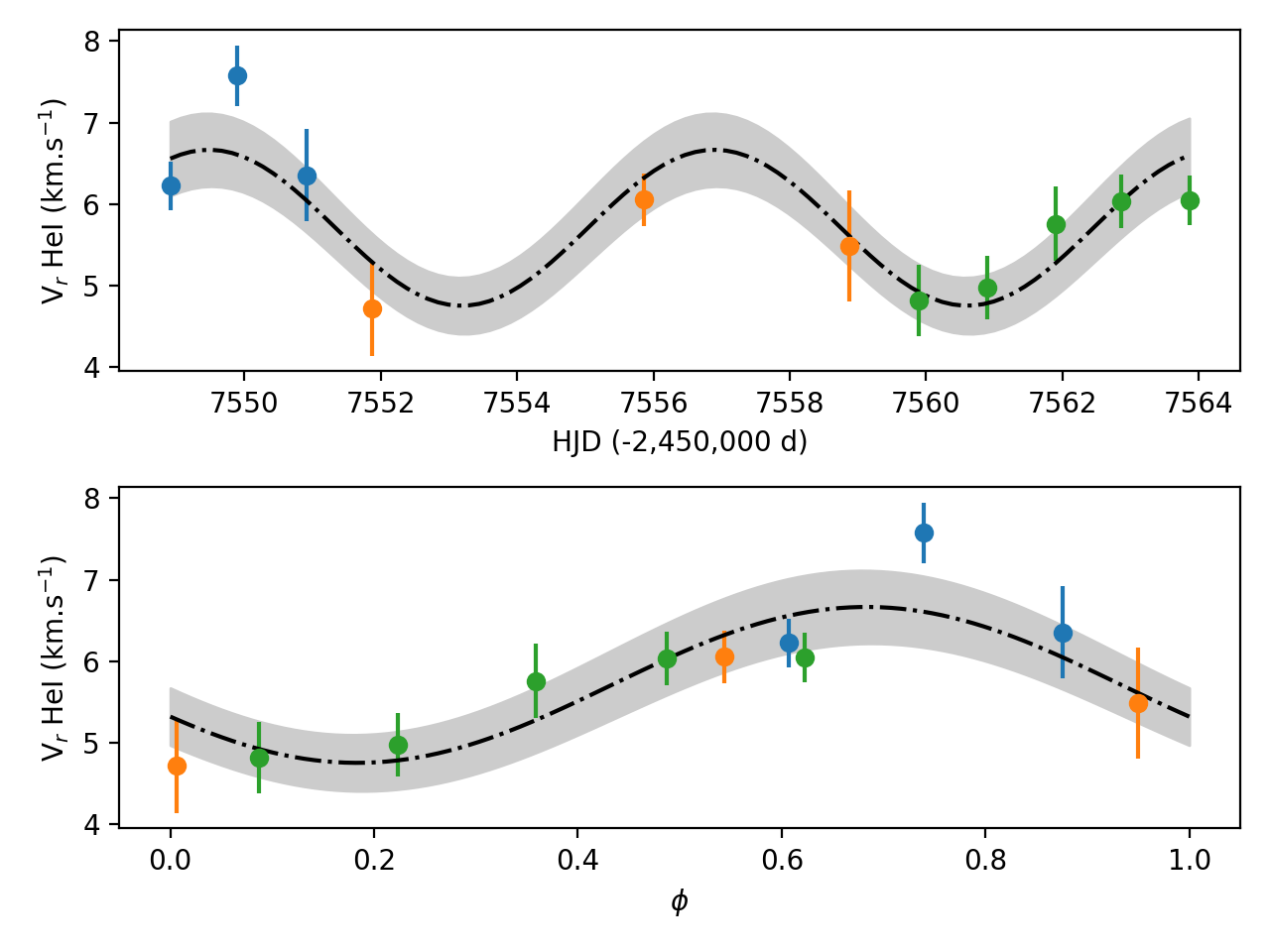}
                \caption{Radial Velocity fit of \ion{He}{I} D3 2016 NC \textit{(top)} and its version folded in phase \textit{(bottom)} following the ephemeris given in Eq.~\ref{eq:ephemeris}}
                \label{fig:vrHeNC}
            \end{figure}

            We thus performed a fit of the \ion{He}{I} D3 NC's radial velocity following the method described in \cite{Pouilly21} to recover the emitting region's location.
            The results are shown in Fig.~\ref{fig:vrHeNC} and summarised below:
            \begin{itemize}
                \item V$_{\rm flow}$ = 7.091$_{-1.03}^{+2.00}$ \kms,
                \item V$_{\rm rot}$ = 3.85 $\pm$ 5.0 \kms,
                \item d$\phi$ = 0.10$_{-0.11}^{+0.15}$,
                \item $\theta$ = 12.05$_{-6.7}^{+48.00}$ $^\circ$,
                \item $\alpha$ = 55.40$_{-16.36}^{+20.00}$ $^\circ$,
            \end{itemize}
            where V$_{\rm flow}$ is the velocity of the material in the post-shock region, V$_{\rm rot}$ the equatorial velocity, 0.5+d$\phi$ the phase where the emitting region is facing the observer, $\theta$ the colatitude of the spot, and $\alpha$ = 90$^\circ$$-$i, i being the inclination of the rotation axis.
            This means that the emitting region is located at $\phi$ = 0.6,  70$^\circ$ latitude.
            These results are consistent within uncertainties with the measurements by \cite{Campbell21}, who give for the \ion{He}{I} emitting region\footnote{Values estimated from their Fig.~7.} a latitude of 60$\pm$25$^\circ$, longitude 40$\pm$5$^\circ$, meaning  $\phi$ = 0.1$\pm$0.01.
            However, the authors are using the first date of observation as T0, translated to our ephemeris (Eq.~\ref{eq:ephemeris}), this yields $\phi$ = 0.7.

        \subsubsection{\ion{Ca}{II} infrared triplet}
\begin{figure*}
                \centering
                \includegraphics[width=.39\textwidth]{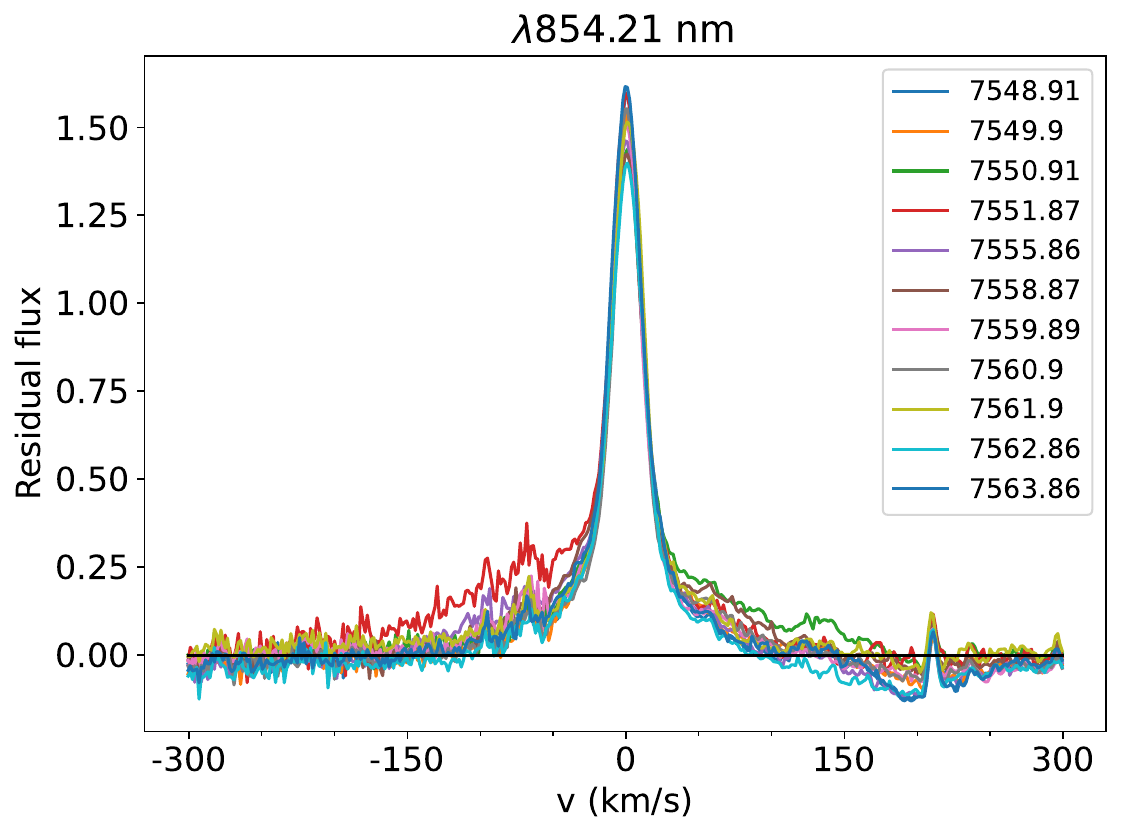}
                \includegraphics[width=.29\textwidth]{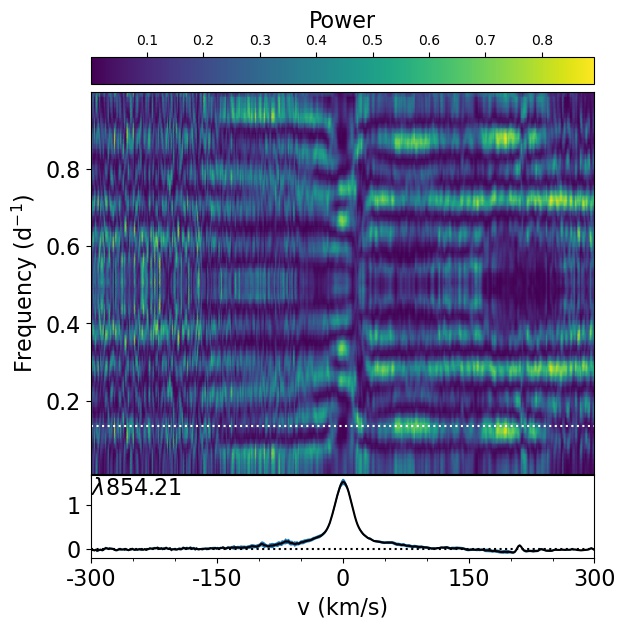}
                \includegraphics[width=.29\textwidth]{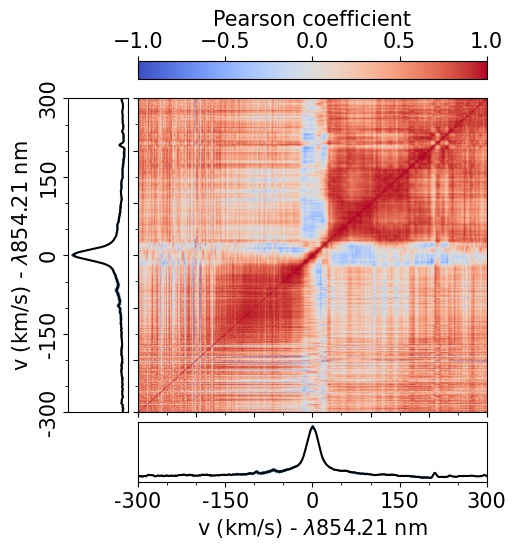}
                \caption{ESPaDOnS 2016 \ion{Ca}{II} IRT (854.2 nm) residual line profiles \textit{(left)}, 2D-periodogram \textit{(middle)}, and auto-correlation matrix \textit{(right)}.}
                \label{fig:CaIRT16}
            \end{figure*}

\begin{figure}
                \centering
                \includegraphics[width=.4\textwidth]{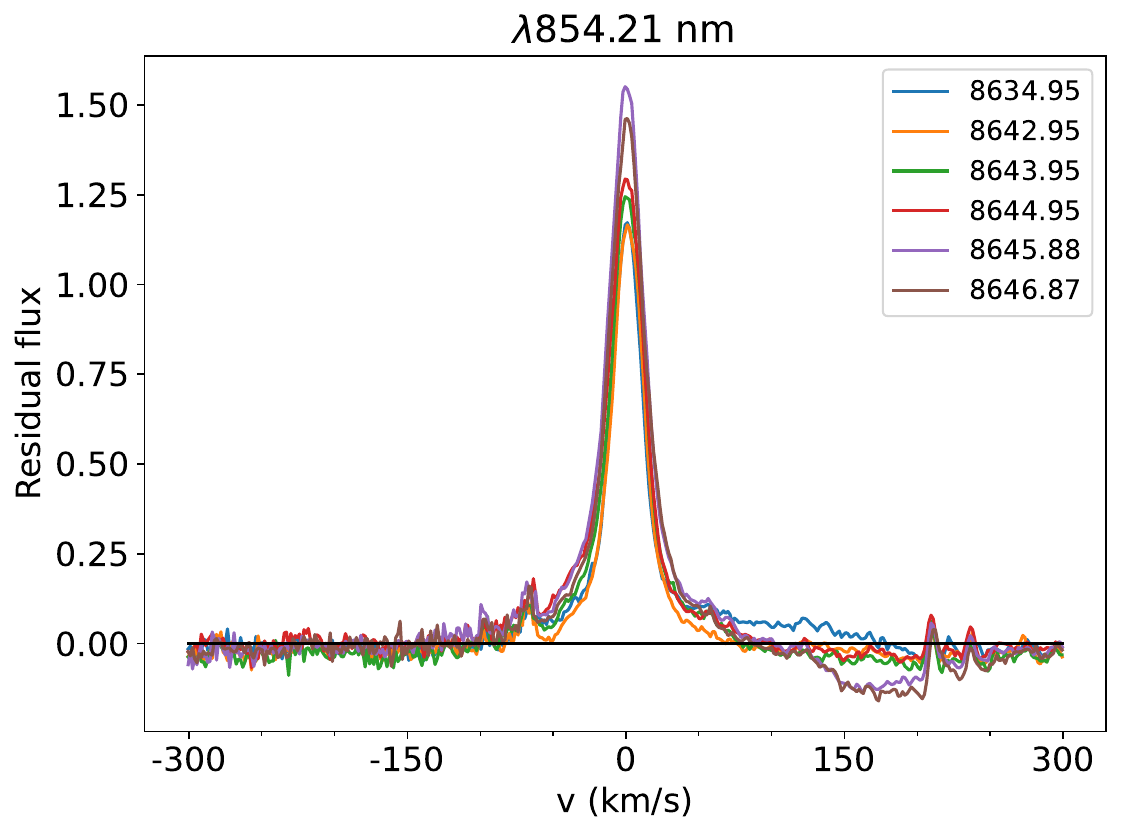}
                \includegraphics[width=.4\textwidth]{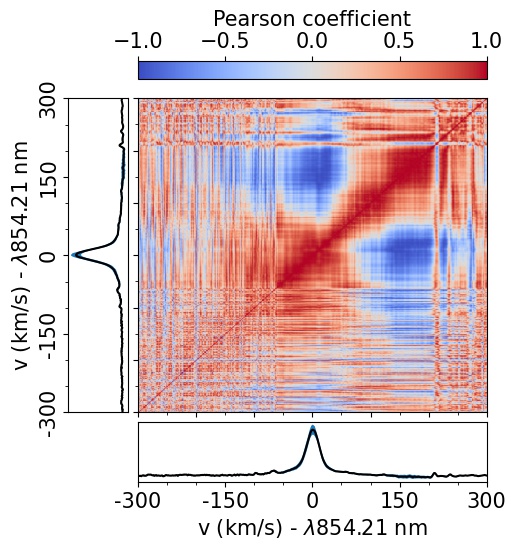}
                \caption{ESPaDOnS 2019 \ion{Ca}{II} IRT (854.2 nm) residual line profiles \textit{(top)} and auto-correlation matrix \textit{(bottom)}.}
                \label{fig:CaIRT19}
            \end{figure}

            As the \ion{He}{I} D3 NC, the \ion{Ca}{II} IRT NC is formed in the post-shock region, we thus studied these lines as well. 
            The three components of this triplet show identical shape and variability, we thus focussed on one of them located at 854.209 nm. 
            The 2016 and 2019 residual line profiles are shown in Fig.~\ref{fig:CaIRT16} and Fig.~\ref{fig:CaIRT19}, respectively.
            The two sets of lines show an IPC profile around 200~\kms, which is, for the 2016 line, periodic with the stellar rotation period (see the 2D-periodogram in Fig.~\ref{fig:CaIRT16}, FAP=0.05).

            The 2016's auto-correlation matrix (Fig.~\ref{fig:CaIRT16}) exhibits several correlated regions: from $-$130 to $-$60, $-$50 to $-$10, $+$50 to $+$100, $+$110 to $+$160, and $+$170 to $+$200 \kms.
            However, these regions are less numerous on the 2019 matrix (Fig.~\ref{fig:CaIRT19}), with only three correlated regions from $-$40 to $+$40, $+$80 to $+$130, and $+$150 to $+$200 \kms, but we retrieve the correlated region around the IPC profile in both matrices, which is anti-correlated with the NC in 2019.
            This goes with the smaller (larger) variability of the NC (BC) observed in 2016 compared to 2019, showing the different origins of the two components and a small change in the accretion pattern between the two epochs.
            
        \subsubsection{\ion{He}{I} 1083 nm}
\begin{figure*}
                \centering
                \includegraphics[width=0.39\textwidth]{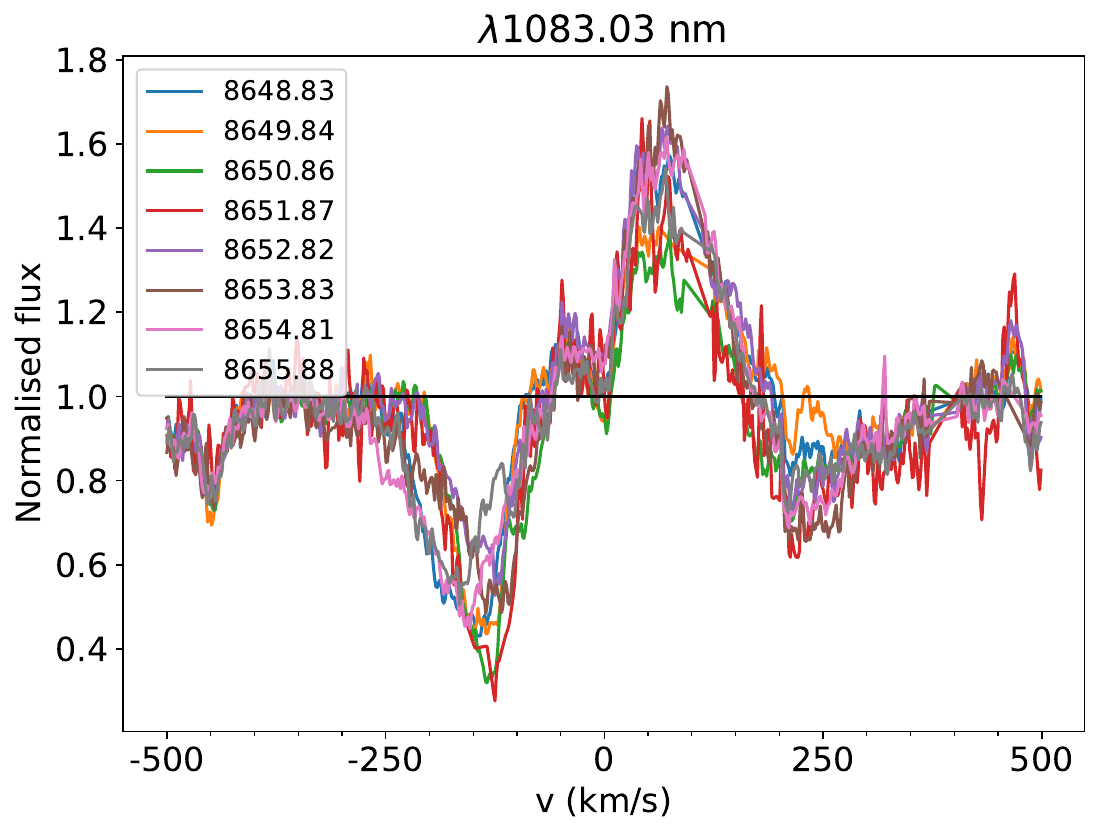}
                \includegraphics[width=0.29\textwidth]{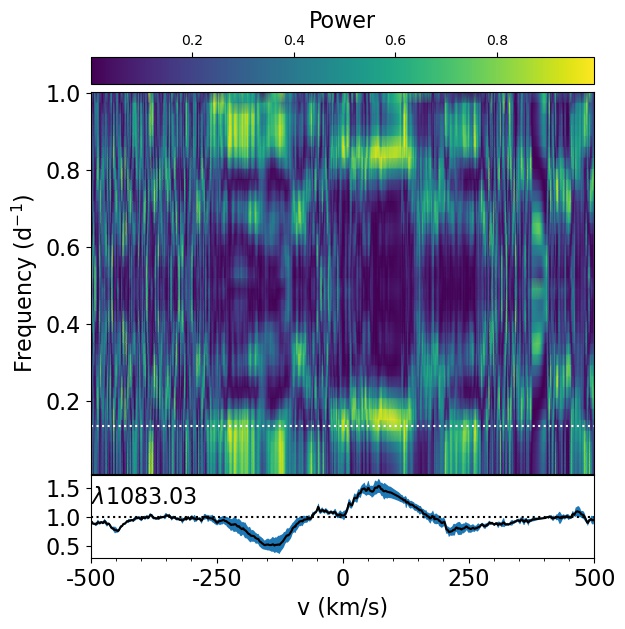}
                \includegraphics[width=0.29\textwidth]{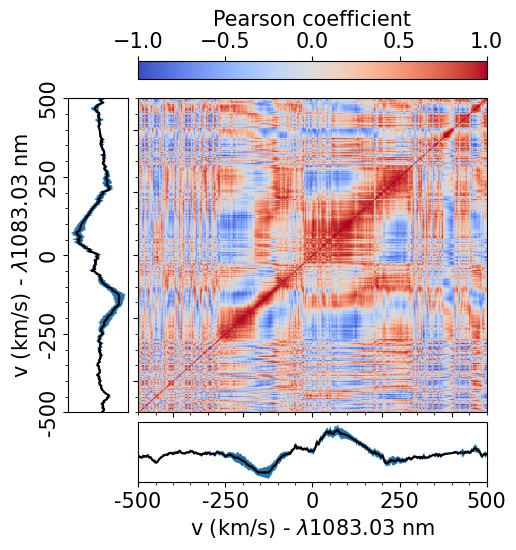}
                \caption{SPIRou \ion{He}{I} (1083 nm) line profiles \textit{(left)}, 2D-periodogram \textit{(middle)}, and auto-correlation matrix \textit{(right)}.}
                \label{fig:HeI1083}
            \end{figure*}

            The only accretion tracer in emission in EX~Lup's SPIRou observation is the \ion{He}{I} line at 1083 nm. 
            The profiles, the 2D-periodogram and the auto-correlation matrix are shown in Fig.~\ref{fig:HeI1083}.
            The profiles seem composed of two peaks blue- and red-shifted around $-$50 and $+$100\kms, and two absorptions, blue-and red-shifted at higher velocity ($-$150 and $+$200\kms).
            
            The red-shifted peak and the two absorptions display significant variability and seem modulated on the stellar rotation period with FAPs reaching 0.001, 0.01, and 0.02 for the red-shifted peak, the blue- and the red-shifted absorption, respectively.
            
            The auto-correlation matrix revealed a more complex decomposition.
            Indeed, if the four substructures seen in the profiles are represented, it seems that the two absorptions can both be separated in two regions, from $-$250 to $-$160 and $-$160 to $-$110 \kms\ for the blue-shifted absorption, and from $+$130 to $+$170 \kms\ and $+$210 to $+$270 \kms\ for the red-shifted absorption. 
            Furthermore, the main peak at $\sim$$+$100~\kms\ is anti-correlated with the most blue- and red-shifted regions only.
            This can be interpreted as follows: the blueshifted absorption, probably a P-Cygni profile traditionally ascribed to a wind, is also associated with a redshifted emission excess, producing the two substructures seen in the redshifted absorption. 
            The IPC profile, as the opposite physical phenomenon, is also associated with a blueshifted emission excess, responsible for the two substructures in the blueshifted absorption.

        \subsubsection{Correlation matrices ESPaDOnS}
        \label{subsubsec:cmESP}
            As the several lines studied are tracing different regions of the accretion, we can compute correlation matrices between two different lines to analyse the link between the different regions identified from the auto-correlation matrices.
            The correlation matrices of the 2016 and 2019 lines are presented in
            Appendix~\ref{ap:CM}.

            In 2016, the H$\alpha$ line centre is correlated with the \ion{He}{I} D3 and the \ion{Ca}{II} IRT NCs, and slightly anti-correlated with the region of \ion{Ca}{II} IRT IPC profile.
            The latter is also strongly anti-correlated with the \ion{He}{I} D3 NC.
            The region of the H$\alpha$'s IPC profile is also anti-correlated with the \ion{He}{I} D3 NC, and correlated with the region of the \ion{Ca}{II} IRT IPC profile.
            In 2019 matrices, we observe the same behaviour between the NCs and IPC regions of the various lines, but both the correlation and anti-correlation coefficients are stronger.
            

    \subsection{Magnetic field}
        In this section, we present the magnetic analysis of EX~Lup.
        This was done at two scales: the large scale using the Zeeman-Doppler Imaging technique \citep[ZDI,][]{Donati11}, and the small scale using the Zeeman intensification of photospheric lines.
        
        \subsubsection{Large-scale}
        \label{subsubsec:largescale}
\begin{figure*}
                \centering
                \includegraphics[width=0.45\textwidth]{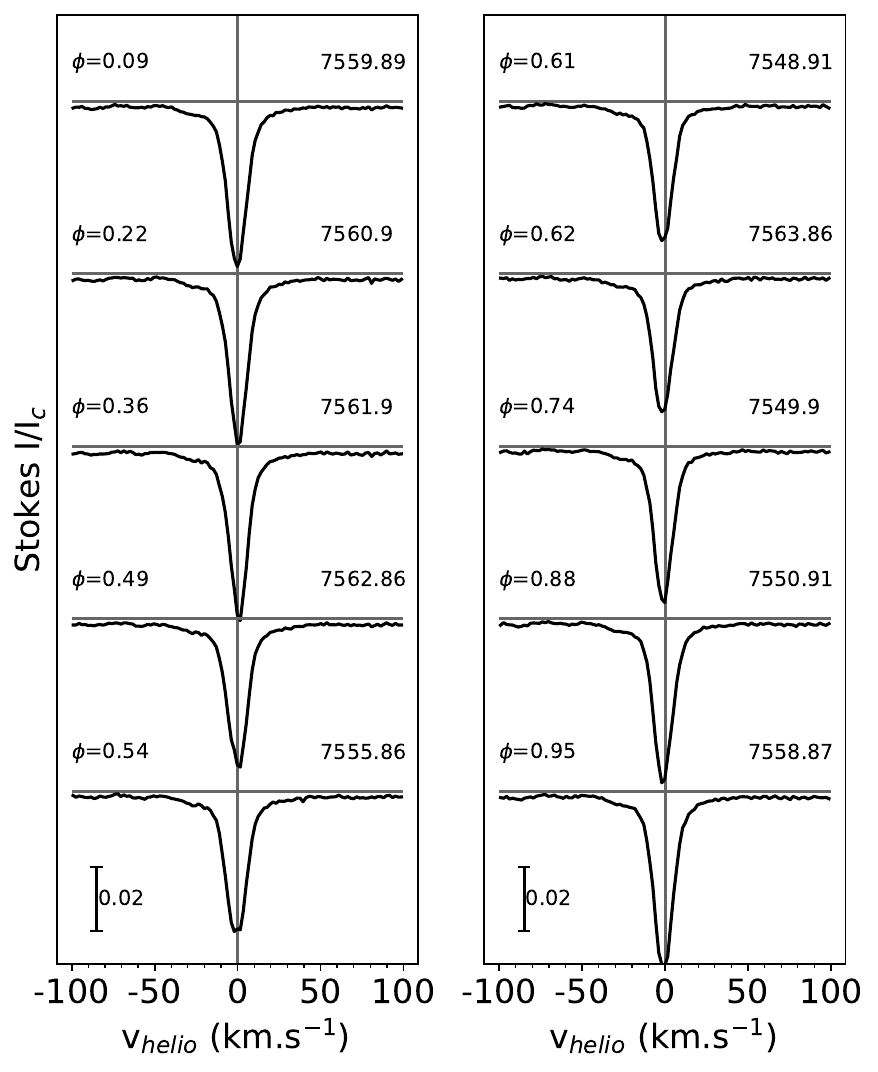}
                \includegraphics[width=0.45\textwidth]{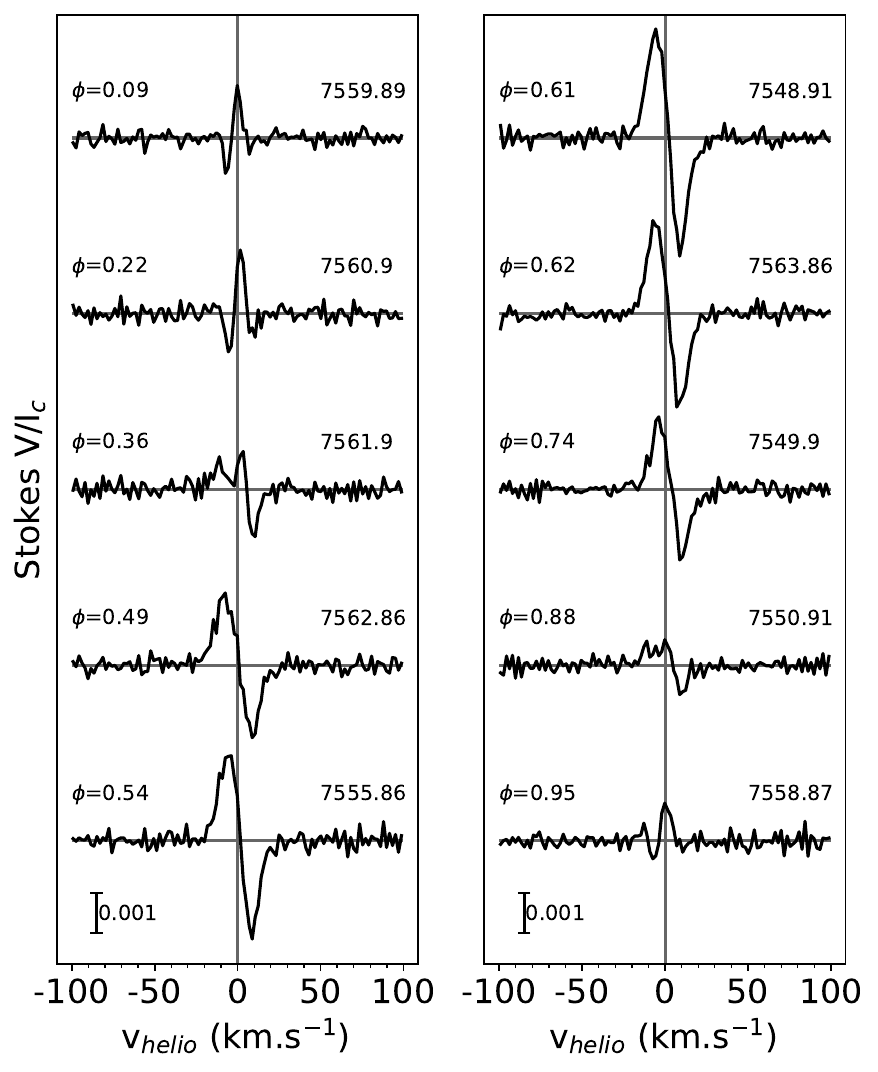}
                \caption{ESPaDOnS 2016 LSD Stokes \textit{I} \textit{(left)} and V \textit{(right)} profiles. The rotation phase and the HJD are indicated at the left and right, respectively, of each profile. The scale is indicated at the bottom left of each plot.}
                \label{fig:LSD16}
            \end{figure*}

\begin{figure}
                \centering
                \includegraphics[width=0.24\textwidth]{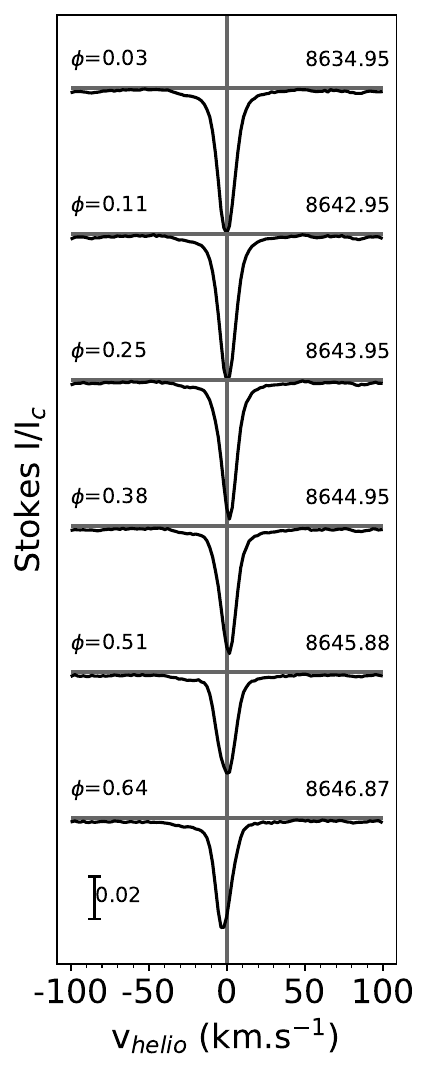}
                \includegraphics[width=0.24\textwidth]{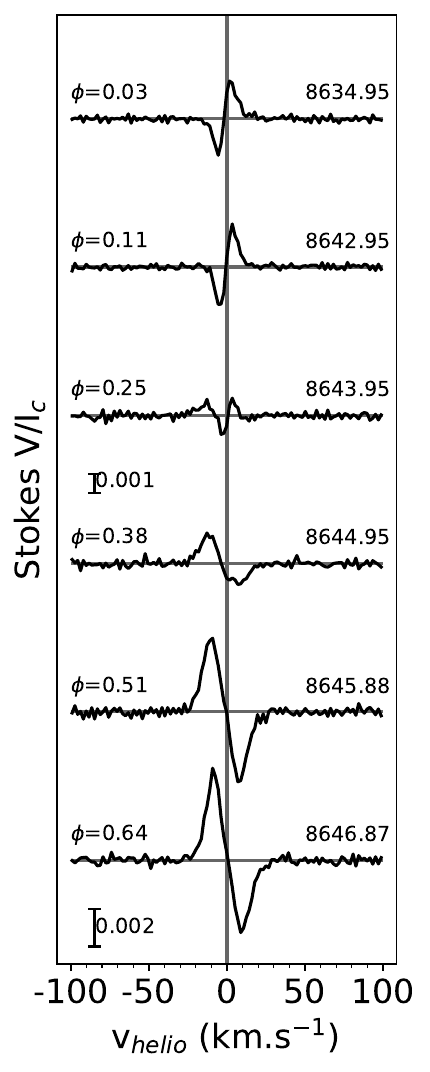}
                \caption{Same as Fig.~\ref{fig:LSD16} for ESPaDOnS 2019 observations.}
                \label{fig:LSD19}
            \end{figure}

\begin{figure*}
                \centering
                \includegraphics[width=0.45\textwidth]{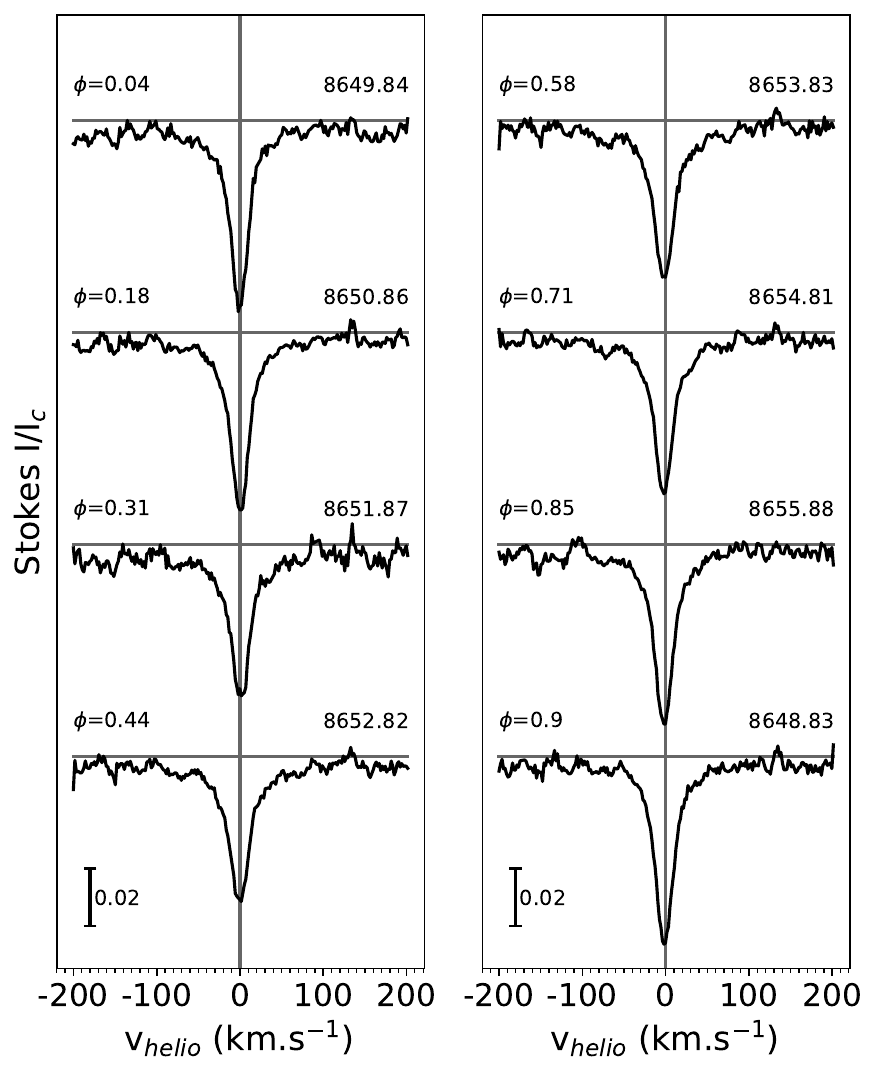}
                \includegraphics[width=0.45\textwidth]{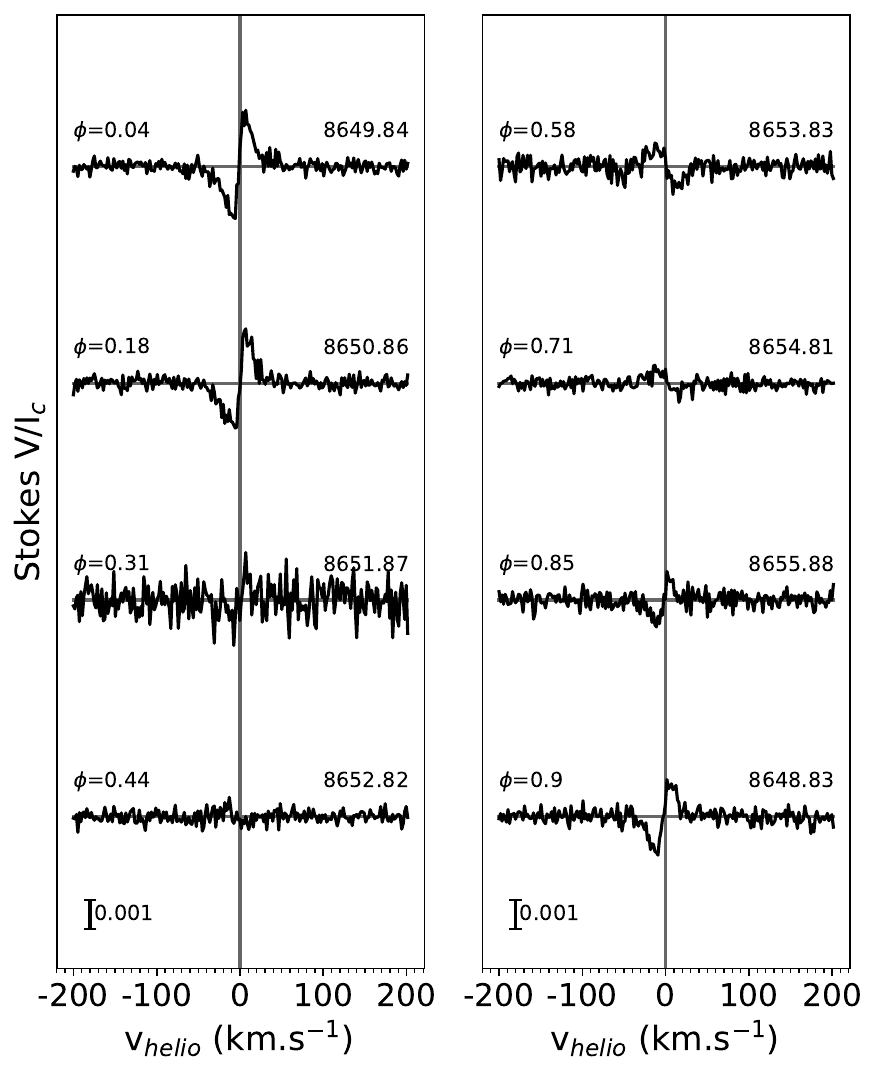}
                \caption{Same as Fig.~\ref{fig:LSD16} for SPIRou observations.}
                \label{fig:LSDIR}
            \end{figure*}

            In this section, we used the Least-Squares Deconvolution method \citep[LSD,][]{Donati97, Kochukhov10} to study the large-scale magnetic field.
            This method allows us to increase the S/N of the Stokes \textit{I} (unpolarised) and Stokes \textit{V} (circularly polarised) profiles, by using as many photospheric lines as possible.
            To compute the LSD profiles we used the \texttt{LSDpy}\footnote{\url{https://github.com/folsomcp/LSDpy}} Python implementation.
            We normalised our LSD weights using an intrinsic line depth, a mean Landé factor and a central wavelength of 0.2, 1.2, and 500 nm (respectively) for ESPaDOns, and 0.1, 1.2, and 1700 nm for SPIRou observations. 
            The photospheric lines were selected from a mask produced using the same \texttt{VALD} line list and \texttt{MARCS} atmospheric models as in Sect.~\ref{subsec:rv}, and removing the emission lines, the telluric and the heavily blended lines regions using the \texttt{SpecpolFlow}\footnote{\url{https://github.com/folsomcp/specpolFlow}} Python package. 
            About 12\,000 lines were used for ESPaDOnS, and 1600 for SPIRou observations.
            The LSD profiles of ESPaDOnS 2016, 2019, and SPIRou observations are presented in Figs.~\ref{fig:LSD16}, \ref{fig:LSD19}, and \ref{fig:LSDIR}, respectively, and the S/N of the profiles are available in Table~\ref{tab:logObs}. 
            Please note that the observation at HJD 2\,457\,551.87135 was remove from this analysis because of its low S/N.
            
            A clear Stokes \textit{V} signature is detected for all ESPaDOnS observations and most of the SPIRou observations.
            Furthermore, one can note that the SPIRou Stokes \textit{V} signatures are much weaker than ESPaDOnS and the signal at $\phi\sim$0.6--0.7 almost vanishes on SPIRou when maximal on ESPaDOnS.
            
\begin{figure}
                \centering
                \includegraphics[width=.45\textwidth]{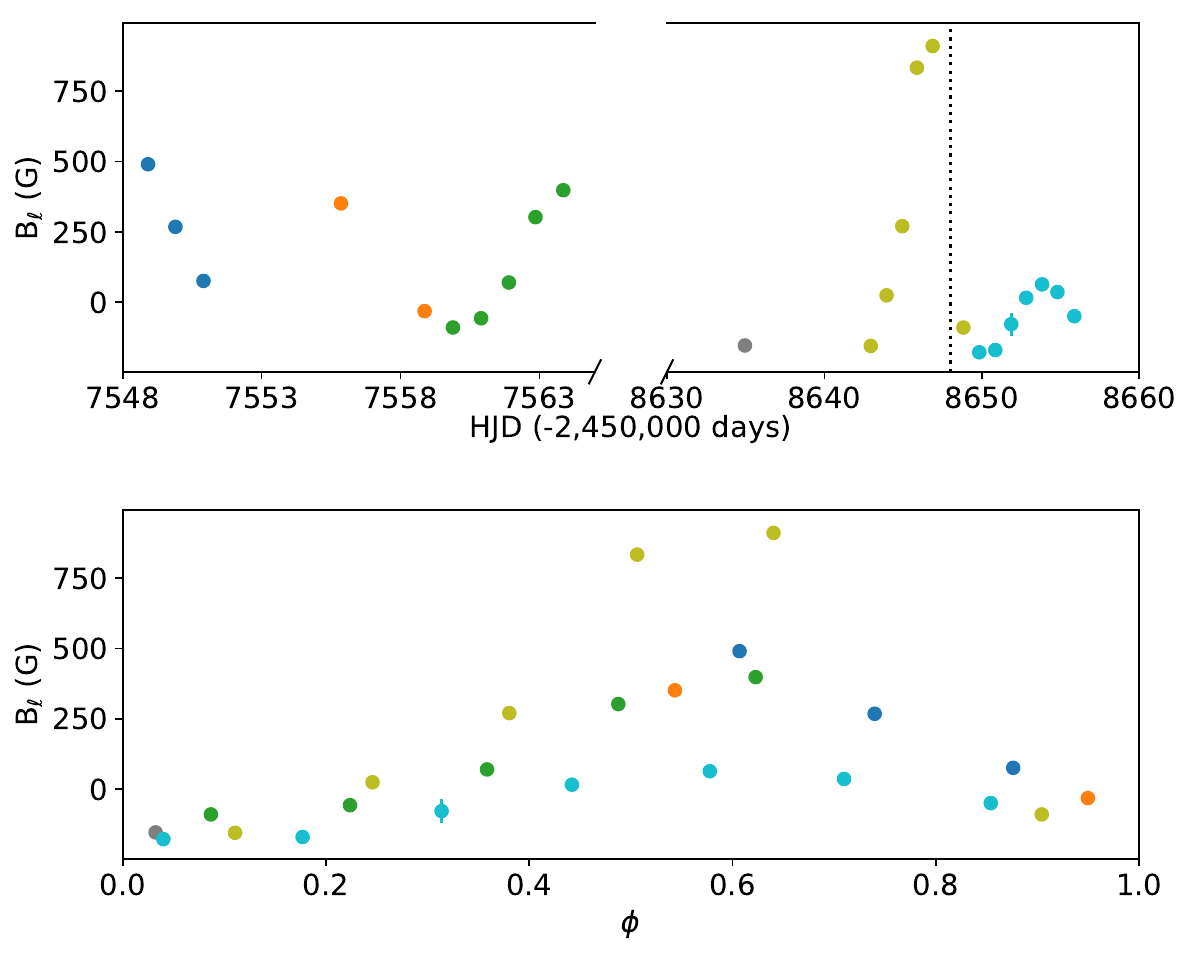}
                \includegraphics[width=.45\textwidth]{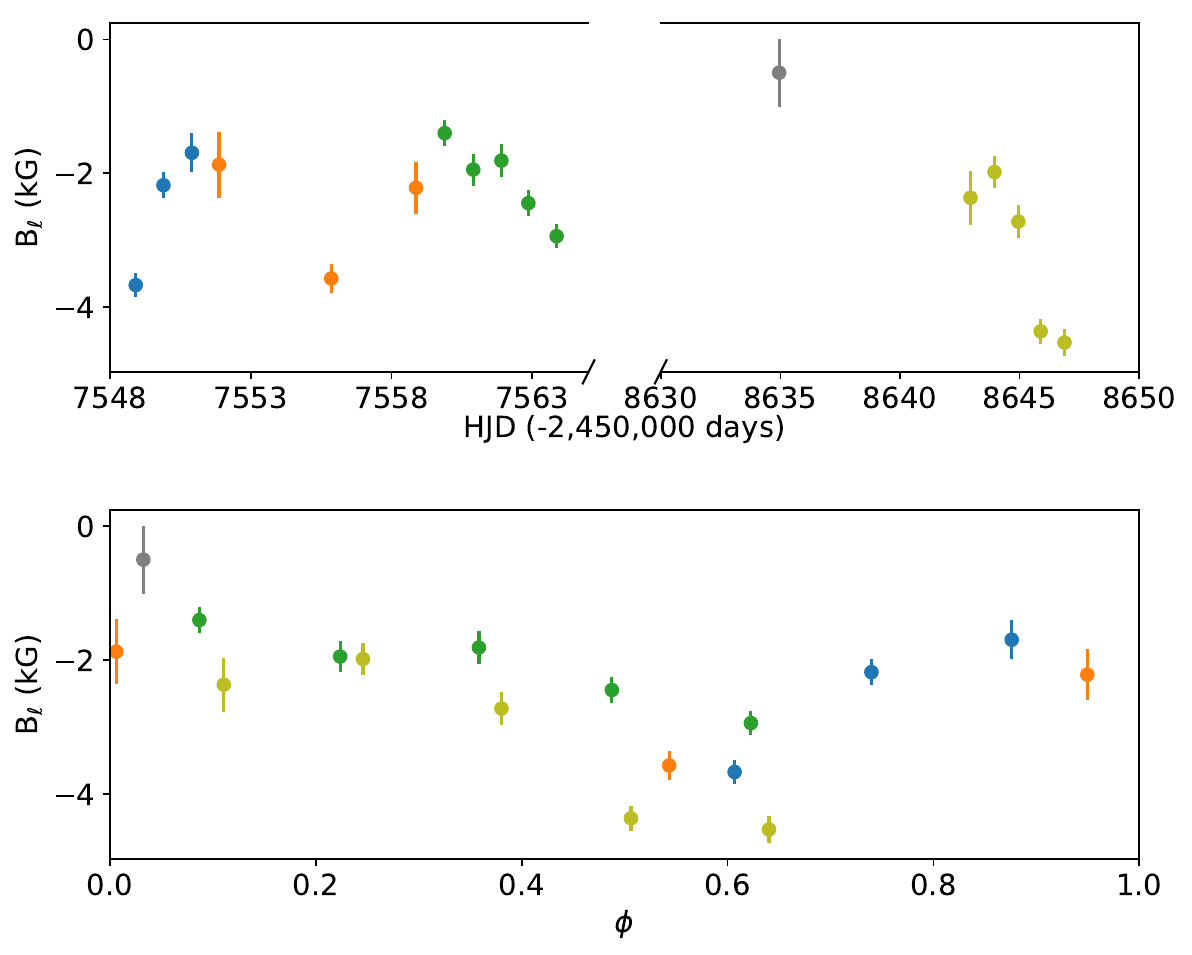}
                \caption{B$_\ell$ curves from LSD profile \textit{(top two panels)} and from \ion{He}{I} D3 line \textit{(bottom two panels)}.}
                \label{fig:Bl}
            \end{figure}

            Directly from the LSD profiles, one can estimate the surface averaged longitudinal magnetic field \citep[B$_\ell$,][]{Donati97, Wade01}:
            \begin{equation}
                B_\ell = -2.14 \times 10^{11} \times \frac{\int vV(v)dv}{\lambda g c \int (1-I(v))dv},
                \label{eq:bl}
            \end{equation}
            where B$_\ell$ is in Gauss, $v$ the velocity relative to the line centre, and $\lambda$ and $g$  the central wavelength and the mean Landé factor used for the LSD computation.
            The integration was performed on a $\pm$25~\kms\ ($\pm$35~\kms) velocity range around the stellar rest frame to minimise the uncertainties without losing any magnetic information on ESPaDOnS (SPIRou) observations.
            The B$_\ell$ curves are shown in Fig.~\ref{fig:Bl}.
            The three curves are modulated with the stellar rotation period and show a maximum around $\phi$=0.6.
            In the optical frame, the modulation's amplitude is slightly larger in 2019 than in 2016, which is reminiscent of the radial velocity behaviour (see Sect.~\ref{subsec:rv}).
            Finally, as expected from the Stokes \textit{V} signatures, the SPIRou measurements are far weakest, and mostly negatives.

            The same analysis can be performed on the NC of the \ion{He}{I} D3 line (on a $\pm$50~\kms\ velocity range), formed close to the accretion shock and thus gives access to the magnetic field strength at the foot of the accretion funnel flow.
            The B$_\ell$ obtained are shown in Fig.~\ref{fig:Bl} and range between $-$1.4 and $-$3.7 kG in 2016 and between $-$0.5 and $-$4.5 kG in 2019.
            As for the radial velocity (see Sect.~\ref{subsec:rv}), the amplitude of the modulation is larger in 2019, but both curves are in phase, with a minimum reached at $\phi$=0.6, which is also consistent with the emitting region's position obtained from the radial velocity modulation of the \ion{He}{I} D3 NC (see Sect.~\ref{subsubsec:heid3}).
            These measurements are also opposed in phase and sign with the LSD B$_\ell$. This is a common behaviour on cTTSs \citep[e.g., see the studies of S Cra N, TWA Hya, CI Tau by][respectively]{Nowacki23, Donati11, Donati20} reflecting that LSD and \ion{He}{I} D3 NC are two diagnostics probing different regions of different polarities.
            
                        
            Finally, we performed a complete ZDI analysis of the three data sets using \texttt{ZDIpy}\footnote{\url{https://github.com/folsomcp/ZDIpy}}\citep[described in][]{Folsom18} with a recent implementation of Unno-Rachkovsky’s solutions to polarised radiative transfer equations in a Milne-Eddington atmosphere \citep{Unno56, Rachkovsky67, Landi04} presented in \cite{Bellotti23} to used a more general description than the weak-field approximation used initially in \texttt{ZDIpy}. 
            The magnetic topology is reconstructed in two steps: (i) the building of a Doppler image (DI), starting from a uniformly bright stellar disk and iteratively adding dark and bright features to fit the observed LSD Stokes \textit{I} profiles, and (ii) fitting the LSD Stokes \textit{V} profiles to derive the magnetic topology by adjusting its spherical harmonic components \citep{Donati06}, here with a maximum degree of harmonic expansion $\ell_{\rm max}$=15.
            We used as input parameter P$_{\rm rot}$=7.417 d, $v\sin i$=4.4 \kms.
            Then we ran a grid of ZDI Stokes \textit{V} reconstruction over the inclination range derived in the literature (20-45$^\circ$) and used the minimal $\chi^2$ obtained to set our input inclination value (i=30$^\circ$).
            The resulting maps are presented in Fig.~\ref{fig:briMaps} and \ref{fig:magMaps} and the corresponding LSD profile fits are shown in Appendix~\ref{ap:ZDIfit}.

            The ESPaDOnS 2016's DI map shows a main dark spot around $\phi$=0.55 and extending between 70 and 20$^\circ$ latitude, which is consistent with \ion{He}{I} D3 emitting region, and a bright structure from $\phi$=0.95 to 0.20 around the equator which is certainly a plage as even the accretion shock is dark at the photospheric level.
            The magnetic topology, mostly toroidal (61\%) is dominated by the dipolar (60\%) and the quadrupolar (17\%) components, with a mean magnetic field strength of 1.00 kG (B$_{\rm max}$ = 3.12 kG), and a magnetic dipolar positive pole of 0.728 kG located at about 30$^\circ$ latitude and 166$^\circ$ longitude ($\phi$=0.46).

            The brightness map obtained from SPIRou revealed a main dark feature extended from $\phi$ = 0.0 to 0.25, and a bright plage around $\phi$ = 0.6, which is perfectly consistent with the 2016 map.
            However, the magnetic topology seems less complex than in 2016, almost fully poloidal (99\%) and more dominated by the dipole component (77\%).
            As expected, the recovered field strength is much smaller ($\langle$B$\rangle$ = 0.131 kG, B$_{\rm max}$ = 0.389 kG).
            The dipole negative pole is located at about 23$^\circ$ latitude, 329$^\circ$ longitude ($\phi$=0.91), with a $-$0.231 kG-strength.
            

            Given the poor rotational phase coverage of the ESPaDOnS 2019's data set ($\phi \in$[0.03,0.64]), we needed to guide the reconstruction instead of starting from a uniform map to avoid a too strong extrapolation on the missing phases.
            We do not expect a similar brightness contrast between ESPaDOnS and SPIRou, but given the very similar spectroscopic behaviour between the 2016 and 2019 ESPaDOnS data sets (see Sect.~\ref{subsec:rv} and \ref{subsec:emLines}), we could expect similar features, we thus used the ESPaDOnS 2016 brightness map as input to guide its reconstruction.
            To check this assumption, we reproduce the Stokes \textit{I} profiles resulting from the 2016 reconstruction on the 2019 phases, yielding a consistent behaviour (reduced $\chi^2$=1.1).
            The obtained final brightness reconstruction shows a main dark feature, slightly shifted in phase compared to 2016 ($\phi\approx$0.5) and less extended in latitude.
            A polar bright feature is also located around $\phi$=1, which is reminiscent of the SPIRou's maps.
            Concerning the magnetic reconstruction, we expect a similar topology of the magnetic field between SPIRou and ESPaDOnS 2019, with different magnetic strengths as pointed out by the B$_\ell$ analysis. 
            We thus used the SPIRou magnetic maps as input to guide the reconstruction.
            The resulting topology is less poloidal-dominated than SPIRou (91\%), and the dipolar component occupies a smaller fraction of this poloidal field (65\%) with a similar contribution of the quadrupolar and octupolar components (about 15\%).
            Surprisingly the mean magnetic field strength is similar to 2016 (1.08 kG) but the maximum field strength is much higher (4.79 kG), as well as the strength of the dipolar pole B$_{\rm dip}$=1.87 kG which is located at the same position ($\phi$ =0.43 and 30$^\circ$ latitude).
                

            \begin{figure*}
                \centering
                \includegraphics[width=.33\textwidth]{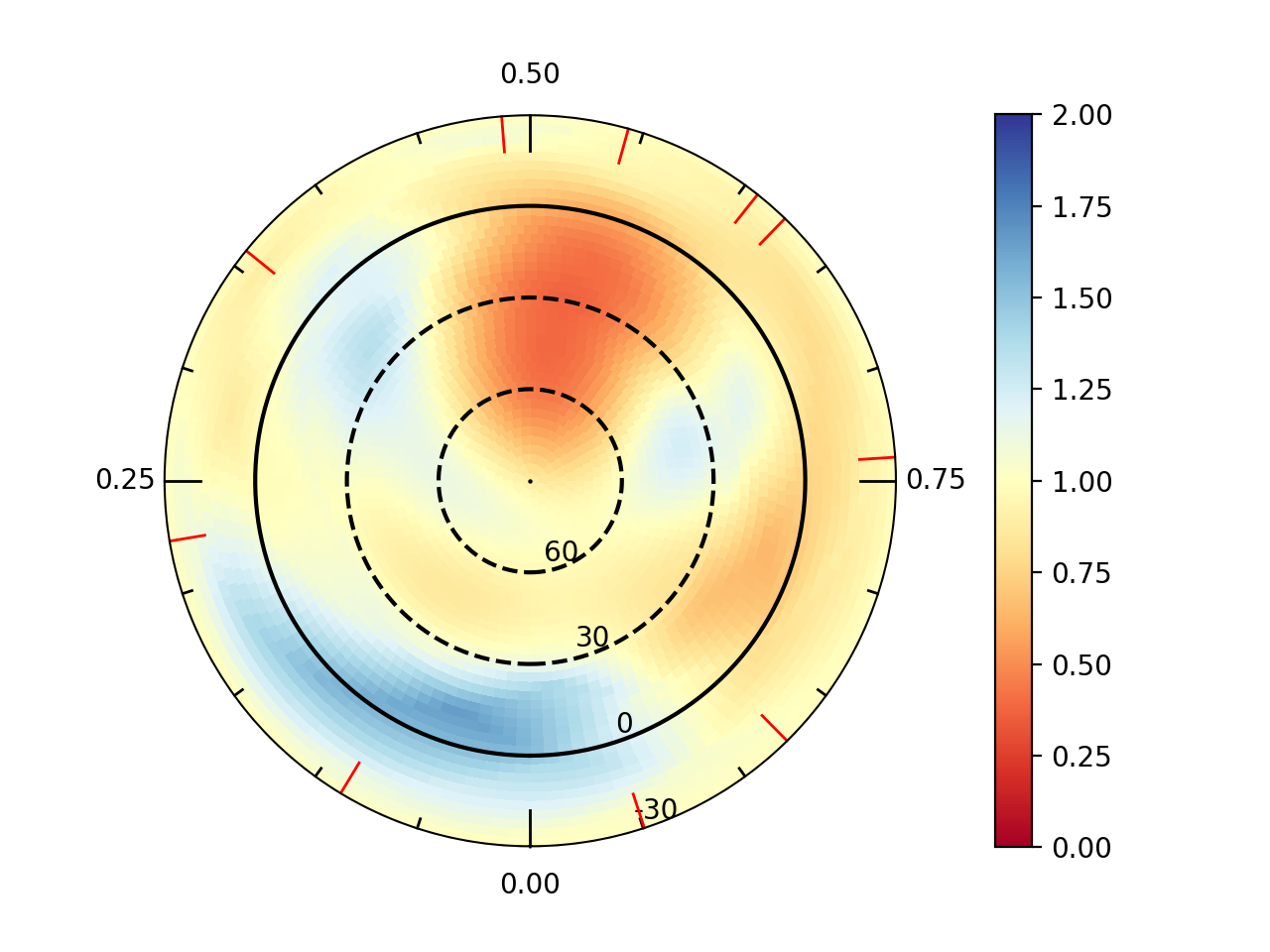}
                \includegraphics[width=.33\textwidth]{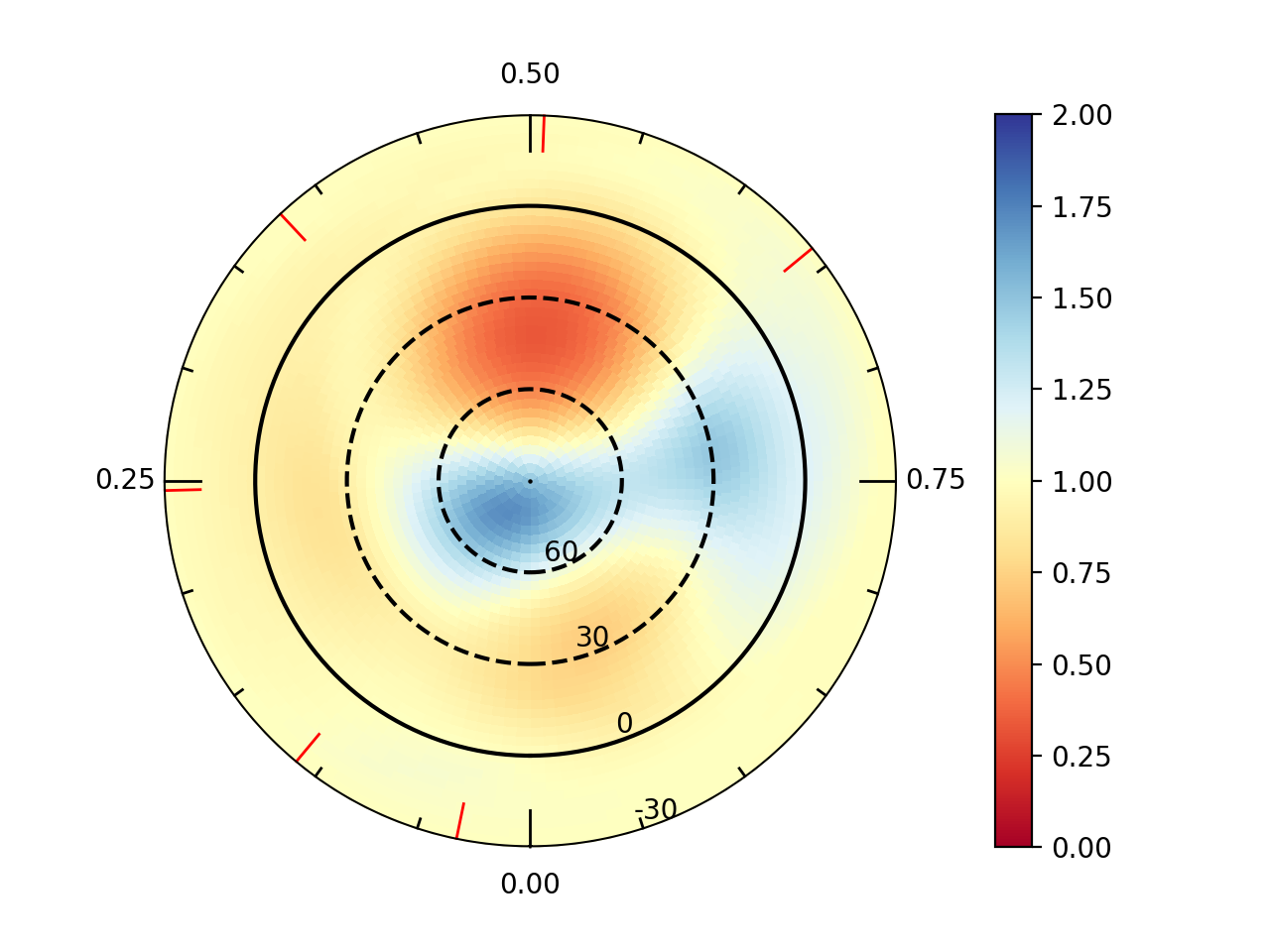}
                \includegraphics[width=.33\textwidth]{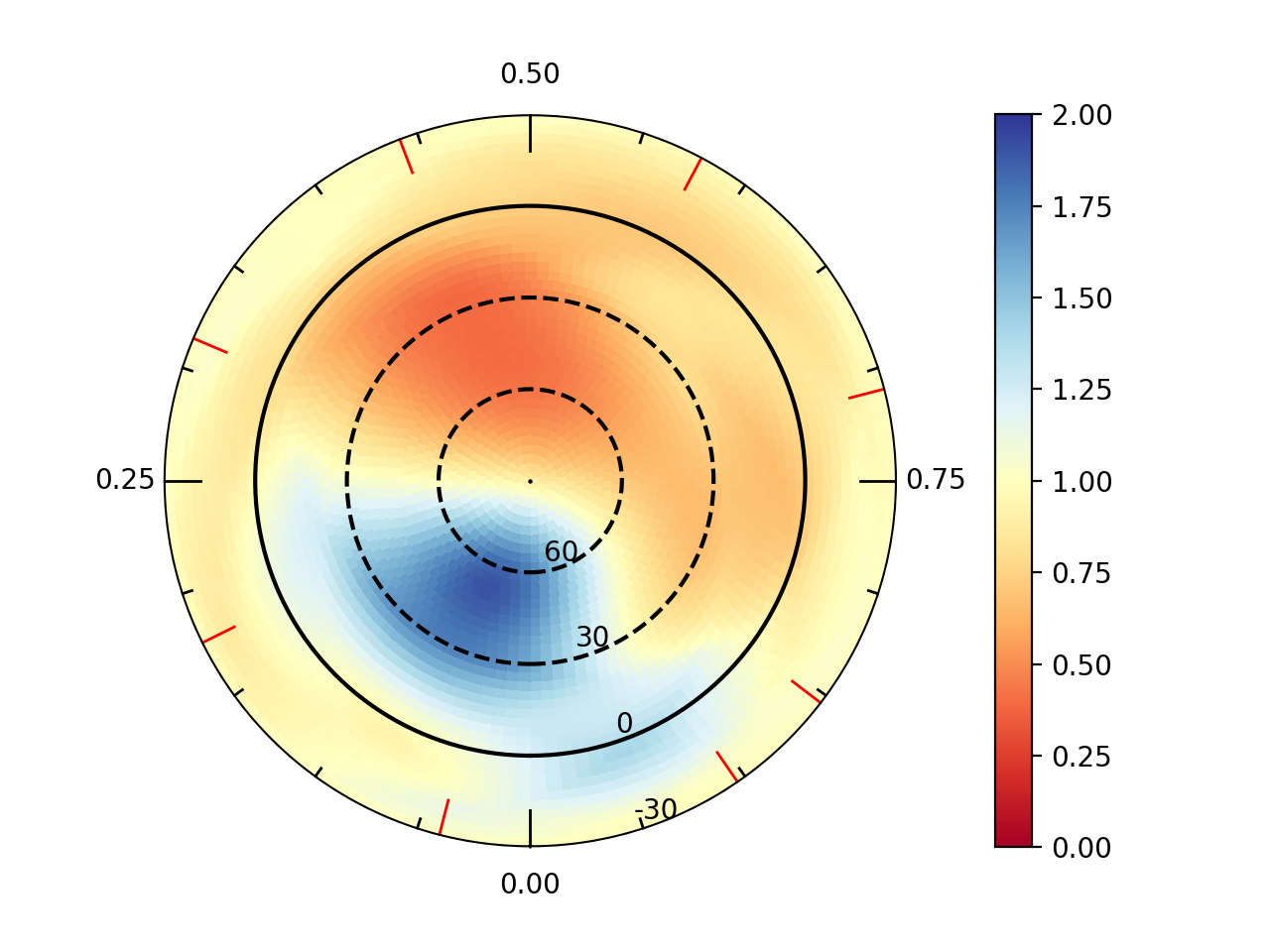}
                \caption{Brightness maps from ESPaDOnS 2016, 2019, and SPIRou data sets (\textit{left}, \textit{middle}, \textit{right}, respectively) on a flattened polar view. The central dot is thus the pole, the two dotted circles are latitude 60 and 30$^\circ$, and the solid circle represents the equator. The black ticks are the rotation phases going clockwise, and the red ticks represent the observed phases. The colour code indicates the brightness on a linear scale, where a value of 1.0 represents the quiet photosphere, values less than one are darker regions and values greater than one are bright.}
                \label{fig:briMaps}
            \end{figure*}  
            \begin{figure*}
                \centering
                \includegraphics[width=.33\textwidth]{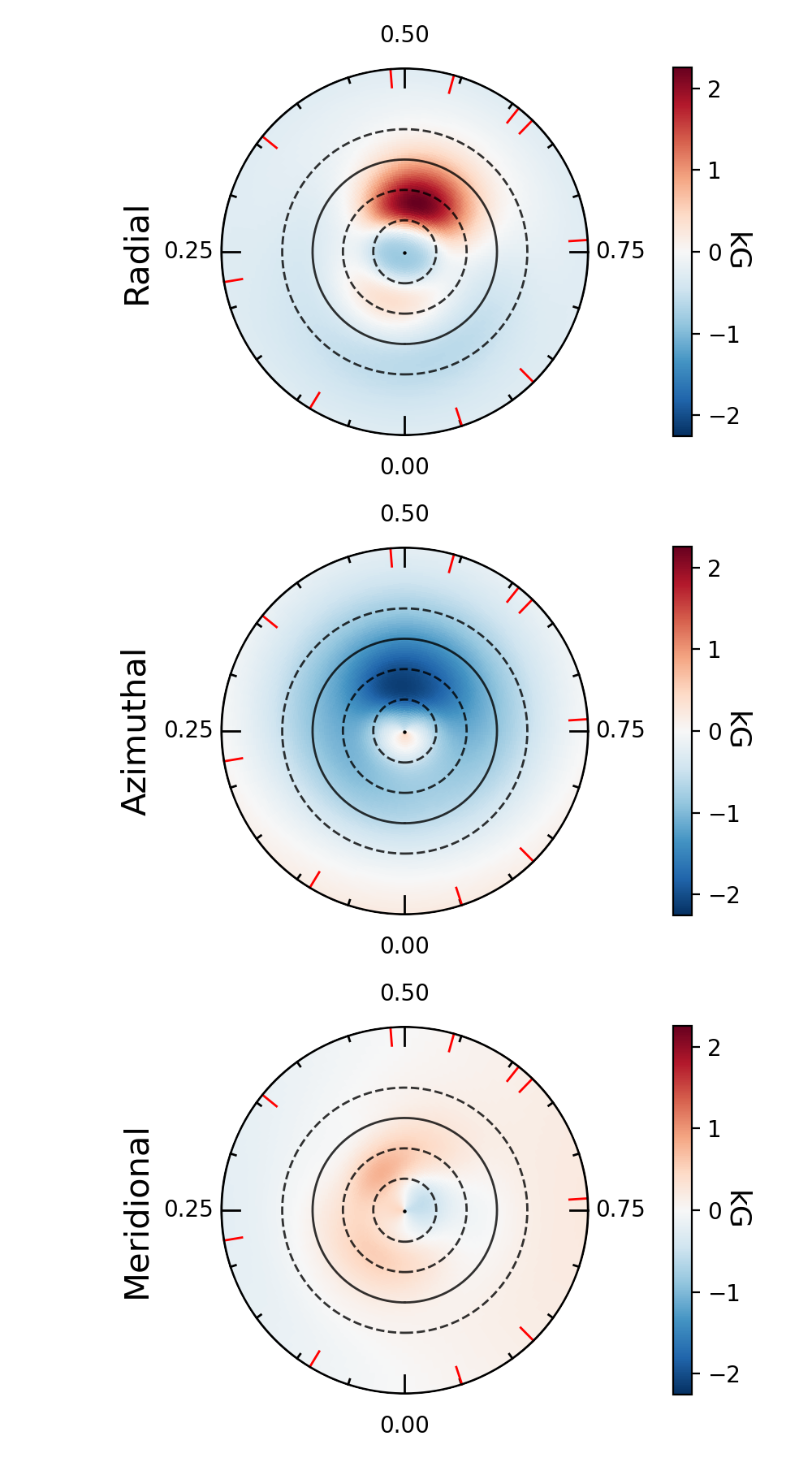}
                \includegraphics[width=.33\textwidth]{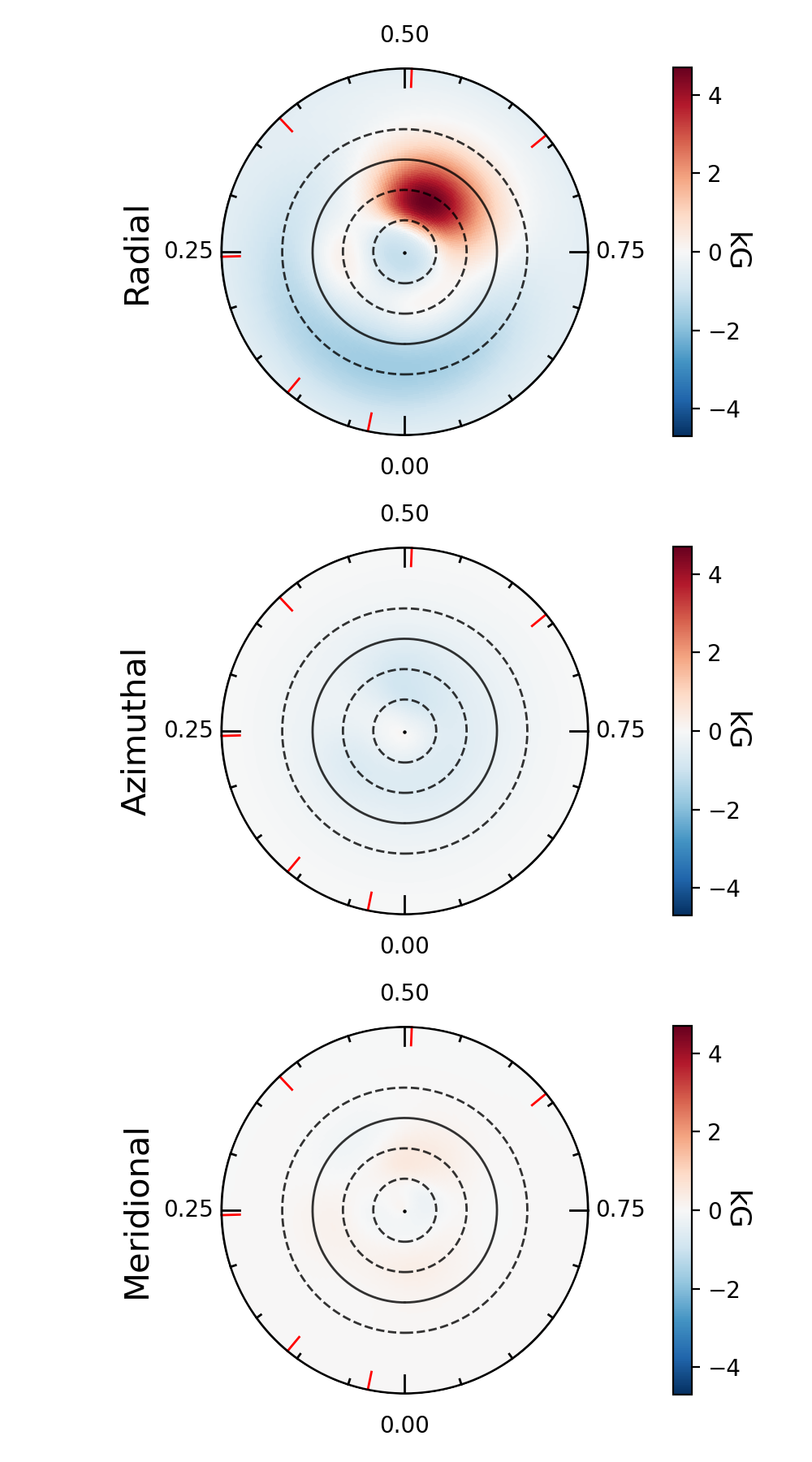}
                \includegraphics[width=.33\textwidth]{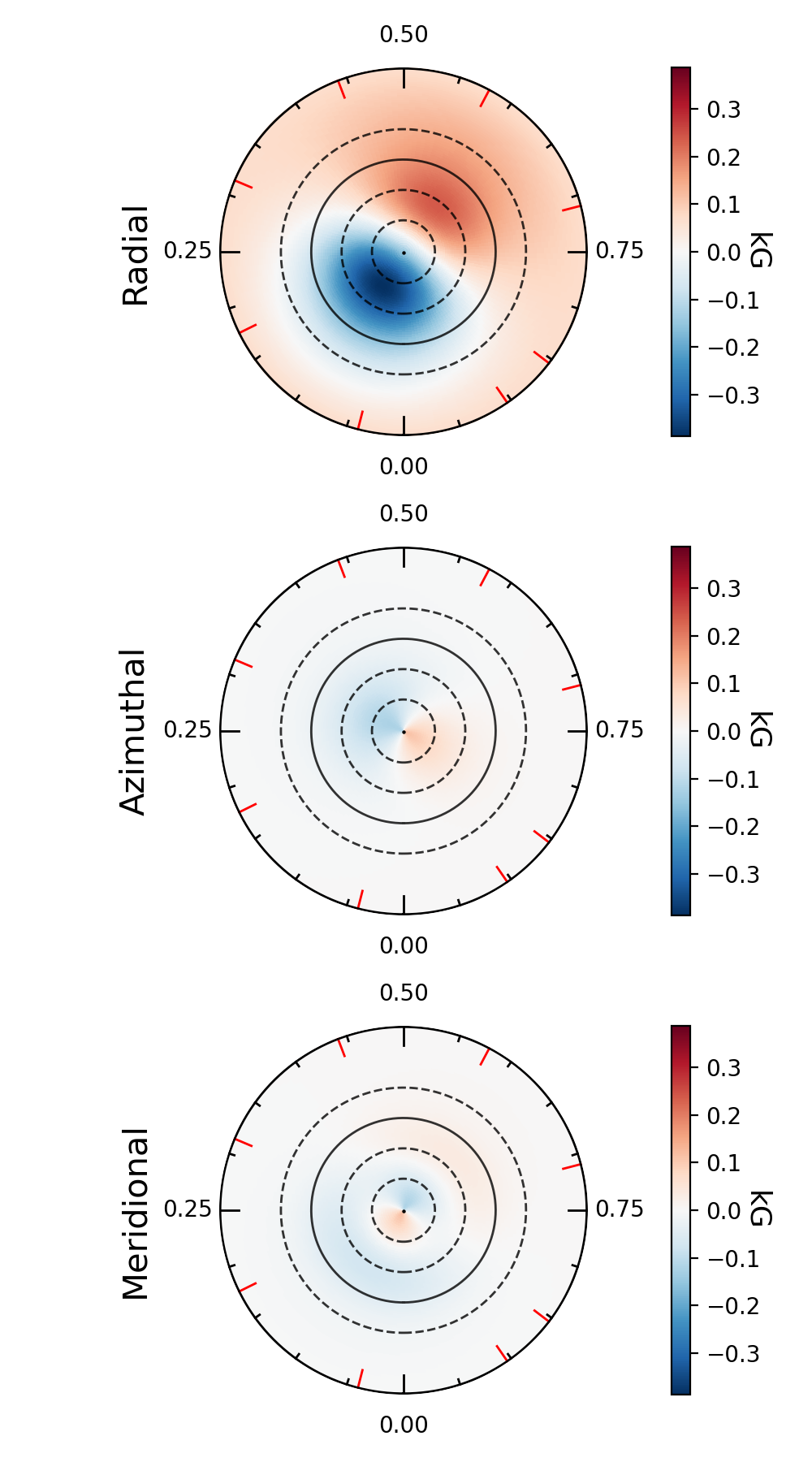}
                \caption{Radial \textit{(top row)}, azimuthal \textit{(middle row)}, and meridional \textit{(bottom row)} magnetic maps from ESPaDOnS 2016, 2019, and SPIRou data sets (\textit{left}, \textit{middle}, and \textit{right} columns, respectively) on the same flattened polar view as Fig.~\ref{fig:briMaps}. The colour code is scaling the magnetic field strength, going from dark blue for the strongest negative value to dark red for the strongest positive value.}
                \label{fig:magMaps}
            \end{figure*}

        \subsubsection{Small-scale}
            Even if the ZDI analysis gives access to the magnetic topology, it neglects the small-scale magnetic field, which contains a major part of cool stars’ magnetic energy.
            This is analysed by studying the change it induces in the shape of magnetically sensitive lines \citep{Kochukhov20}, a technique called Zeeman intensification.
            To perform this analysis, we used the algorithm of \cite{Hahlin21}, performing a Markov Chain Monte Carlo (MCMC) sampling, using the \texttt{SoBaT} library \citep{Anfinogentov21} on a grid of synthetic spectra produced by the \texttt{SYNMAST} code, a polarised radiative transfer code described by \cite{Kochukhov10}.
            This grid is computed from the \texttt{VALD} line lists used in Sect.~\ref{subsec:rv}, and \texttt{MARCS} atmospheric models \citep{Gustafsson08}.
            We parametrized a uniform radial magnetic field as a sum of magnetic field strength ranging from 0 to 6 kG, with a 2 kG step, weighted by filling factors representing the amount of stellar surface covered by this magnetic field.
            For ESPaDOnS observations, as in previous studies \citep{Hahlin22, Pouilly23, Pouilly24}, we used the 963.5--981.2 nm region.

                 

            \begin{table}
            \centering
            \caption{\ion{Ti}{I} lines used for the Zeeman intensification analysis.}
            \begin{tabular}{llll}
                 \hline
                 \hline
                 $\lambda$ & g$_{\rm eff}$ & $\lambda$ & g$_{\rm eff}$ \\
                 (nm) & & (nm) &\\
                 \hline
                 964.7370 & 1.53 & 978.3450 & 1.26  \\
                 967.5544 & 1.35 & 978.7686 & 1.50  \\
                 968.8870 & 1.50 & 2178.8866 &  1.29 \\
                 970.5665 & 1.26 & 2190.3353 &  1.16  \\
                 972.8405 & 1.00 & 2201.0501 &  1.00 \\
                 974.3606 & 0.00 & 2221.7280 &  2.08 \\
                 977.0298 & 1.55 & 2223.8911 &  1.66\\
                 
                 \hline
            \end{tabular}
            \tablefoot{The wavelengths in the optical frame are given in the air, the ones in the infrared frame are given in vacuum.}
            \label{tab:Tilines}
            \end{table}
            
            This region contains a group of \ion{Ti}{I} lines with different magnetic sensitivity (g$_{\rm eff}$ summarised in Table~\ref{tab:Tilines}), allowing us to disentangle the effect of the magnetic field on the equivalent widths from the effect of any other parameters, such as the \ion{Ti}{I} abundance. 
            However, this region also contains many telluric lines which are superimposed on the \ion{Ti}{I} lines from the stellar spectra and which need to be removed from the observed spectrum to perform the magnetic analysis. 
            To do so, we used the {\tt molecfit} package \citep{Smette15}, developed to model and remove telluric lines from spectra obtained with instruments at the European Southern Observatory, and which can be used on spectra from any instrument.

            Finally, as EX~Lup is an accreting star, showing a signature of an accretion shock, the veiling might disturb the inference results, in particular the abundance, the $v\sin i$ and the radial tangential macroturbulent velocity, v$_{\rm mac}$.
            We thus estimated the veiling using the magnetic null line of the \ion{Ti}{I} multiplet at 974.36 nm by performing a $\chi^2$ minimisation using \texttt{SYNMAST} synthetic spectra letting the abundance, the $v\sin i$, and the v$_{\rm mac}$ as free parameters and adding a fractional veiling defined as:
            \begin{equation}
                I_{veil} = \frac{I+r}{1+r},
            \end{equation}
            where I$_{\rm veil}$ is the veiled spectrum, I the spectrum without veiling, and r the fractional veiling.
            We stress to the reader that deriving a precise value of the veiling is outside the scope of the present study
            , our aim was only to get an estimation of the EX~Lup mean spectrum at a wavelength close to the \ion{Ti}{I} lines in order to minimise its effect on the other inferred parameters.
            For the 2016's ESPaDOnS mean spectrum, the minimal $\chi^2$ is reached for r=0.5, $v\sin i$=4.9 \kms, v$_{\rm mac}$=1.9 \kms, and a \ion{Ti}{I} abundance of $-$7.2.
            For the mean spectrum of the 2019's ESPaDOnS observations, we obtained r=0.48, $v\sin i$=5.0 \kms, v$_{\rm mac}$=2.5\kms, and a \ion{Ti}{I} abundance of $-$7.2.
            We will thus assume the values obtained for r and v$_{\rm mac}$, and use the others as initial guesses for the MCMC sampling.

            We assumed a multicomponent model given by:
            \begin{equation}
                S=\sum{f_i S_i},
            \end{equation}
            where $f_i$ are the filling factors, meaning the fraction of the stellar surface covered by a field strength $B_i$, and $S_i$ the synthetic spectra of the corresponding magnetic field strength.
            The averaged magnetic field is thus given by:
            \begin{equation}
                \langle B \rangle = \sum{f_i B_i}.
            \end{equation}
            To set the number of filling factor to use, we iteratively added filling factors, with a 2-kG step in the corresponding magnetic field strength, and used the Bayesian information criterion \citep[BIC,][]{Sharma17} to only include the filling factors that significantly improve the fit.
            The suitable solutions are components of 0, 2, 4 kG for ESPaDOnS 2016 and SPIRou observations, and 0, 2, 4, 6, 8 kG for ESPaDOnS 2019.
            The free parameters of the analysis are thus the following: $f_i$, $v\sin i$, v$_r$, and the \ion{Ti}{I} abundance, for which uniform prior were adopted.
            Finally, we used an effective sample size of 1000 \citep{Sharma17}.
            
\begin{figure*}
                \centering
                \includegraphics[width=.45\textwidth]{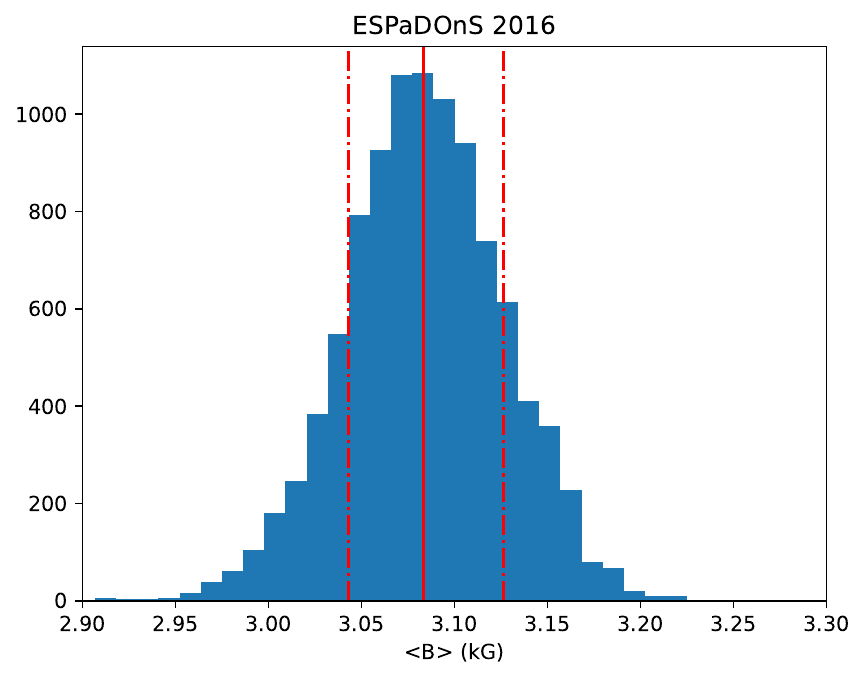}
                \includegraphics[width=.45\textwidth]{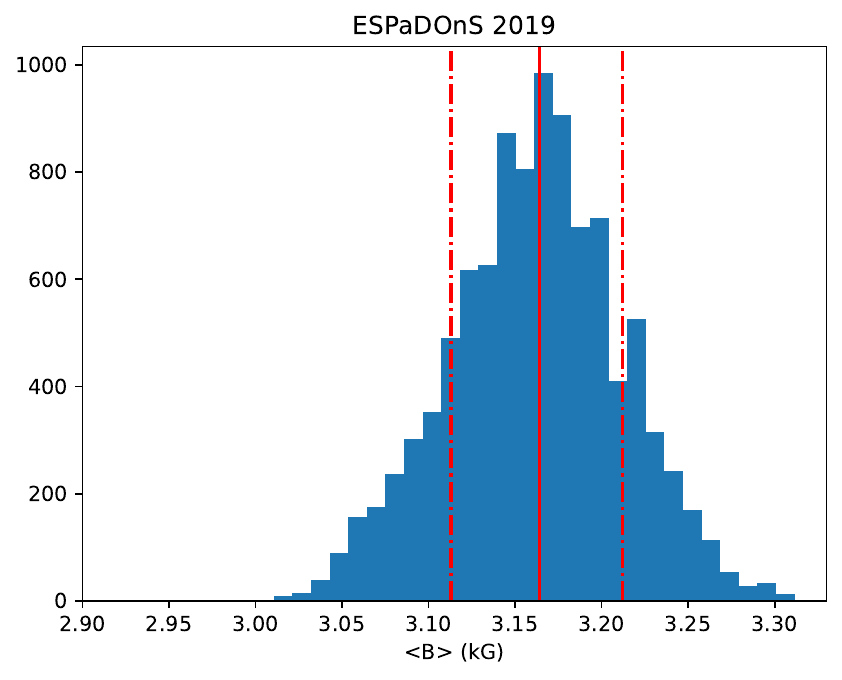}
                \includegraphics[width=.45\textwidth]{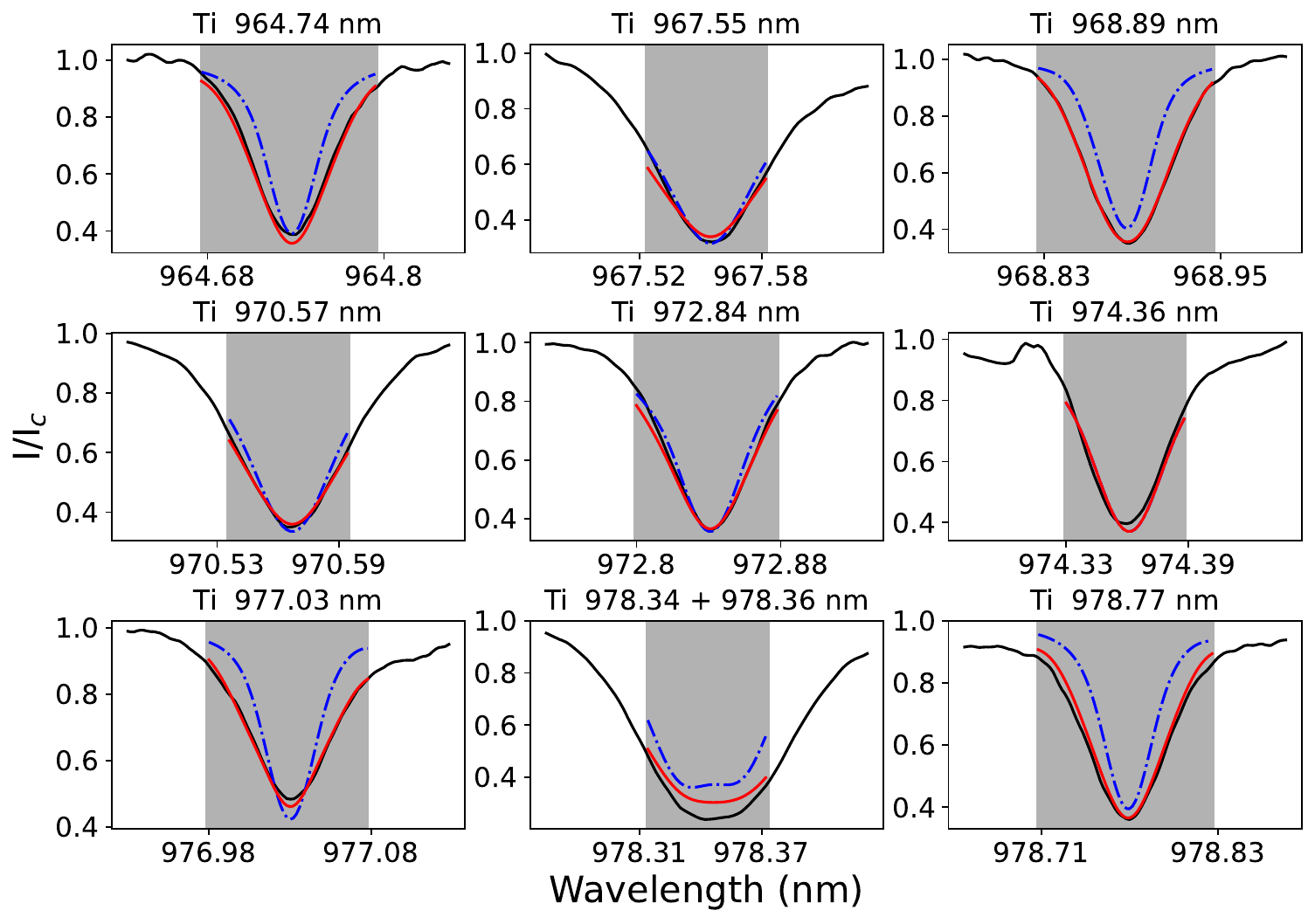}
                \includegraphics[width=.45\textwidth]{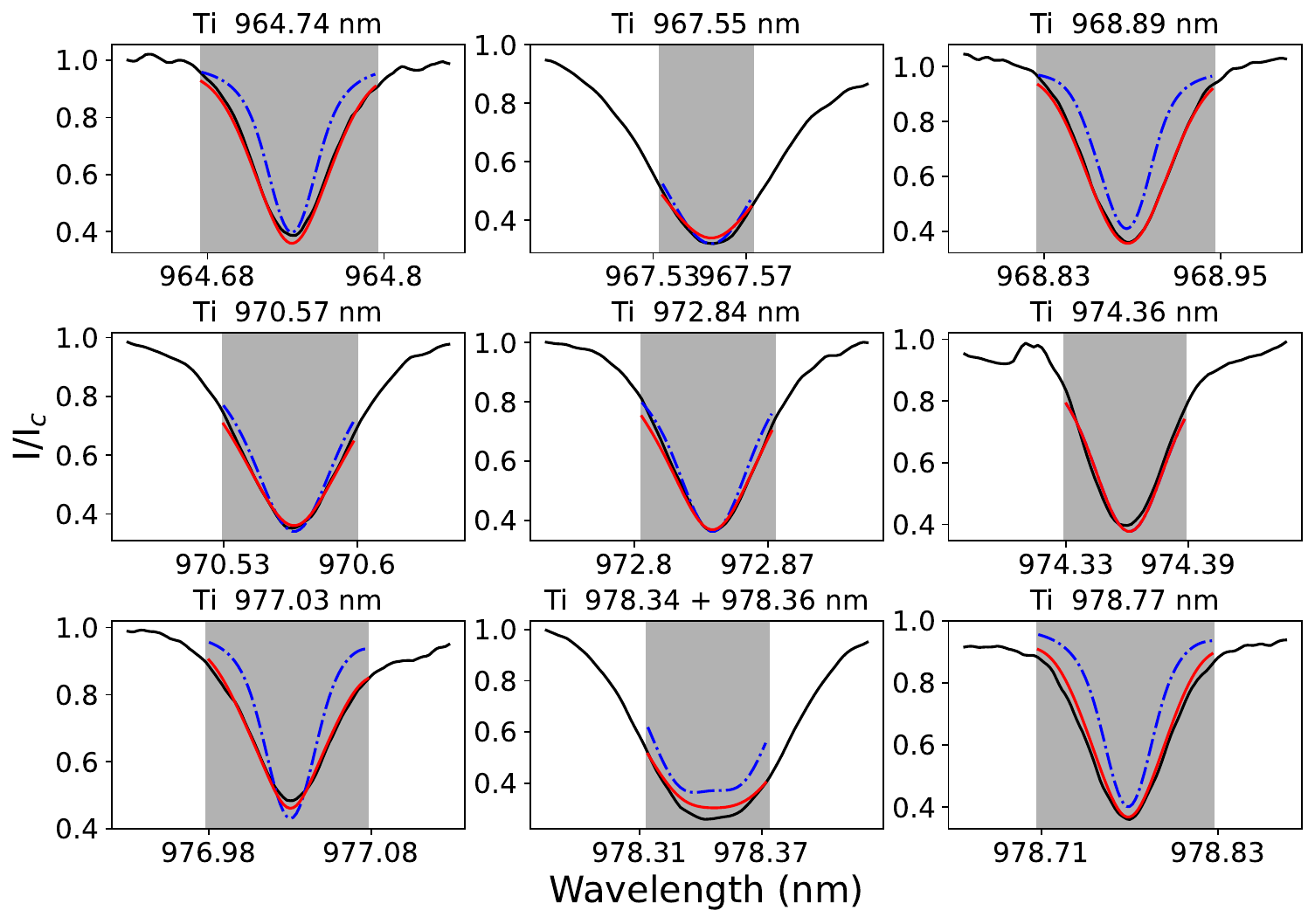} 
                \caption{Results of the ESPaDOnS Zeeman intensification analysis. \textit{Top:} Distribution of the average magnetic field of EX~Lup for 2016 \textit{(left)} and 2019 \textit{(right)} data sets. The solid red line represents the median and the dashed lines represent the 68 \% confidence regions. \textit{Bottom:} Fit to 2016 \textit{(left)} and 2019 \textit{(right)} optical \ion{Ti}{I} multiplet of EX~Lup. The solid red line shows the best fit to the observations in black, while the blue dashed line shows the non-magnetic spectra with otherwise identical stellar parameters. The shaded area marks the region used for the fit.}
                \label{fig:smallScale}
            \end{figure*}

            The resulting line fit and magnetic field strength posterior distributions are presented in Fig.~\ref{fig:smallScale}.
            The 2016's magnetic field strength (3.08 $\pm$ 0.04 kG) is consistent with 2019 within uncertainties (3.16 $\pm$ 0.05 kG). 
            The inferences of all parameters are summarised in Table~\ref{tab:magInf}.

\begin{figure}
                \centering
                \includegraphics[width=.45\textwidth]{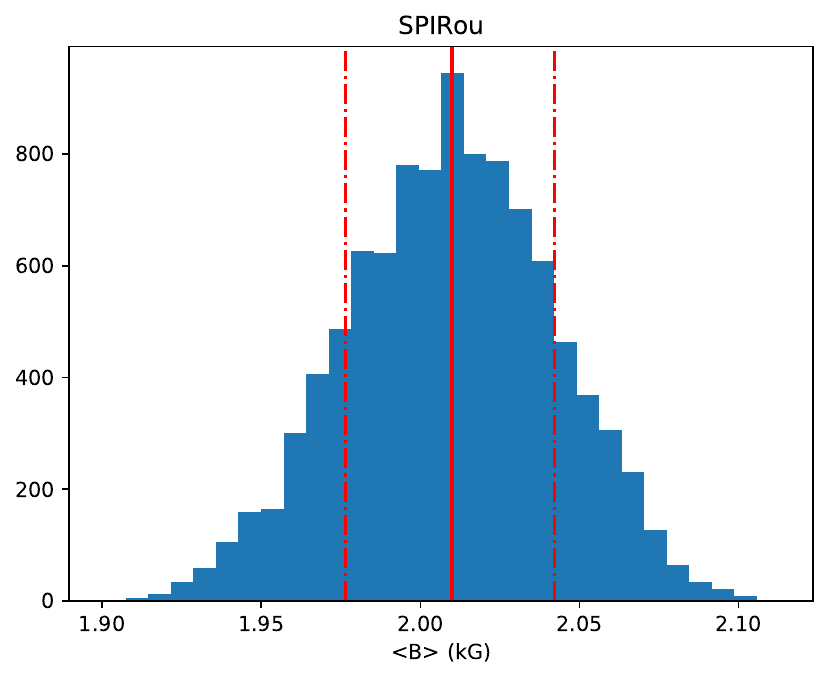}
                \includegraphics[width=.45\textwidth]{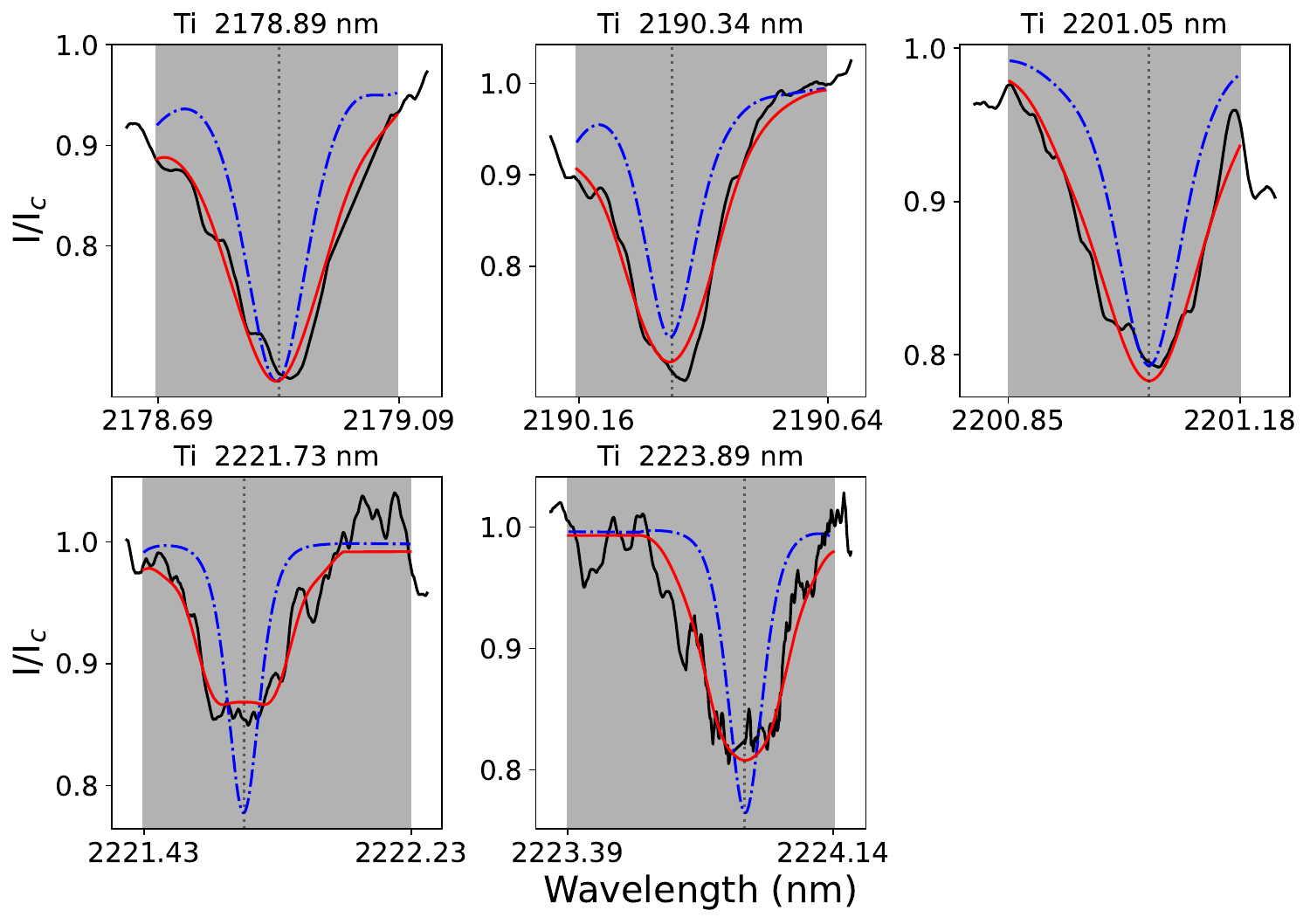}
                \caption{Same as Fig.~\ref{fig:smallScale} for SPIRou IR \ion{Ti}{I} multiplet.}
                \label{fig:smallScaleSPIR}
            \end{figure}

            For SPIRou observations, we used here again a set of \ion{Ti}{I} lines around 2200 nm (see Table~\ref{tab:Tilines}). 
            Unfortunately, this wavelength region does not contain any magnetically null line deep enough at our S/N to perform the veiling study we have done on ESPaDOnS observations.
            Only a visual inspection of the line at 974.4 nm is possible, indicating that at this wavelength the parameters obtained for ESPaDOnS are consistent with the SPIRou observations.
            We thus used the veiling values obtained for the 2019 ESPaDOnS data set, fixed the $v\sin i$ at the literature value and let the inference compensate for the eventual error with the non-magnetic parameters.
            The inferred parameters are summarised in Table~\ref{tab:magInf}, and the line fit and inferred magnetic field strength are shown in Fig.~\ref{fig:smallScaleSPIR}. 
            The magnetic field strength recovered (2.00$\pm$0.03 kG) is significantly lower than the values found for ESPaDOnS observations.
            As highlighted by \cite{Hahlin23}, the small-scale field might be overestimated when using optical wavelength. 
            In our case, a second explanation might come from the high v$_{\rm mac}$ obtained, probably needed to compensate for an underestimated veiling, which lowers the effect of the magnetic field.

            \begin{table*}
            \centering
            \caption{Inference results on the small-scale magnetic field on \ion{Ti}{I} lines.}
            \begin{tabular}{llll}
                 \hline
                 \hline
                 Parameter & ESPaDOnS 2016 & ESPaDOnS 2019 & SPIRou 2019  \\
                 \hline
                 $f_2$ & 0.45$\pm$0.02 & 0.51$\pm$0.02 & 0.57$\pm$0.02\\ 
                 $f_4$ & 0.55$\pm$0.02 & 0.42$\pm$0.03 & 0.21$\pm$0.01\\
                 $f_6$ & - & 0.01$^{+0.02}_{-0.01}$ & -\\
                 $f_8$ & - & 0.05$^{+0.01}_{-0.02}$ & -\\
                 Abundance & -7.15$\pm$0.01 & -7.14$\pm$0.01 & -7.27$\pm$0.01\\
                 $v\sin i$ (\kms) & 4.90$\pm$0.21 & 4.74$^{+0.22}_{-0.24}$ & (v$_{\rm mac}$) 11.82$^{+0.54}_{-0.49}$ \\
                 \hline
            \end{tabular}
            \tablefoot{The $f_i$ denote the filling factor for a $i$\,kG magnetic field strength.}
            \label{tab:magInf}
            \end{table*}
            
\section{Discussion}
\label{sec:discussion}
 
    This work aimed at characterising the accretion process of the prototypical EXor, EX Lup, in quiescence, together with its magnetic field at small and large scales.
    This study confirms that the typical magnetospheric accretion process of cTTSs is ongoing on this system, which seems stable between the two epochs studied (2016 and 2019), with a main accretion funnel flow connecting the disc to the stellar surface.
    This produces the IPC profile observed in the \ion{H}{} lines studied and their modulation with the stellar rotation period.
    The accretion shock at the stellar surface produces the emission of the NC of the \ion{He}{I} D3 and \ion{Ca}{II} IRT lines, which are modulated with the stellar period and anti-correlated with the IPC profiles as expected in the magnetospheric accretion scheme.
    This is consistent with the maximum IPC profile occurring at the same phase as the \ion{He}{I} D3 emitting region ($\phi \approx$ 0.6), indicating a funnel flow aligned with the accretion shock and inherently meaning that the truncation radius is located at the stellar corotation radius.
    As expected, this phase is also associated with an extremum of the B$_\ell$, for both LSD and \ion{He}{I} D3 measurements, and with the ESPaDOnS ZDI brightness and magnetic topology reconstructions,  showing the connection between accretion and magnetic field.
    To investigate the truncation radius $r_{\rm mag}$, we used the expression given by \cite{Bessolaz08}:
    \begin{equation} 
    {\frac{r_{\rm mag}}{R_{\star}}}  = 2 m_{\rm s}^{2/7} B_{\star}^{4/7} \dot{M}_{\rm acc}^{-2/7} M_{\star}^{-1/7} R_{\star}^{5/7}, 
    \end{equation}
    where the Mach number m$_s\approx1$, B$_{\star}$ is the equatorial magnetic field strength \citep[given from the dipole strength and the relation of ][]{Gregory11} in units of 140 G, \macc\ is the mass accretion rate in units of 10$^{-8}$~\msunyr, M$_\star$ the stellar mass in units of 0.8~M$_\odot$, and R$_\star$ the stellar radius in units of 2~R$_\odot$.
    As \macc\ in quiescence, we used the value before the 2022 outburst given by 
    \cite{Wang23}, 1.8$\times10^{-9}$~\msunyr.
    The stellar mass and radius are given by \cite{GrasVelazquez05} (0.6~M$_\sun$ and 1.6~R$_\sun$), and magnetic obliquity is obtained from ZDI analysis (30$^\circ$ for ESPaDOnS, 23$^\circ$ for SPIRou, see Sect.~\ref{subsubsec:largescale}).
    The dipole strengths needed to obtain $r_{\rm mag}$~=~$r_{\rm corot}$~=~8.5$\pm$0.5~R$_{\star}$ are thus 
    2.10$\pm$0.46~kG and 1.98$\pm$0.43~kG for ESPaDOnS and SPIRou, respectively.
    Only an estimate of the truncation radius can be obtained because we do not have a precise value of the dipole strength.
    Indeed, the B$_\ell$ measurements in the \ion{He}{I} D3 line and Zeeman intensification values also contain the higher-order components of the magnetic field, and the ZDI spherical harmonic decomposition from LSD might include flux cancellation lowering the large-scale strength obtained.
    Even if the ZDI results on the 2019 datasets indicate a topology largely dominated by the dipole component, using these values should yield an overestimation of $r_{\rm mag}$.
    The results using our magnetic field measurements are summarised in Table~\ref{tab:rmag}.
    \begin{table}
            \centering
            \caption{Magnetospheric truncation radius obtained using the various magnetic field measurements of this work.}
            \begin{tabular}{lll}
                 \hline
                 \hline
                 Measurement & B & $r_{\rm mag}$ \\
                 
                 & (kG) & (R$_\star$) \\
                 \hline
                 B$_\ell$ \ion{He}{I} D3 2016 & 3.7$\pm$0.2 & 11.7$\pm$1.2 \\ 
                 B$_\ell$ \ion{He}{I} D3 2019 & 4.2$\pm$0.2 & 12.6$\pm$1.2 \\
                 ZDI 2016 & 0.728 & 4.6$\pm$0.3 \\
                 ZDI 2019 & 1.87 & 7.9$\pm$0.5 \\
                 ZDI IR & 0.377 & 3.3$\pm$0.2 \\
                 Small-scale 2016 & 3.08$\pm$0.04 & 10.5$\pm$0.8 \\
                 Small-scale 2019 & 3.16$\pm$0.05 & 10.7$\pm$0.8  \\
                 Small-scale IR & 2.00$\pm$0.03 & 8.5$\pm$0.7 \\
                 \hline
            \end{tabular}
            \label{tab:rmag}
            \end{table}
    The values obtained from the B$_\ell$ and the small-scale field are not consistent with the corotation radius, as expected, except for the SPIRou measurement which can be explained by the smaller recovered field in the infrared domain. 
    However, the optical ZDI values yield truncation radii consistent with the corotation radius, due to the dipole-dominant magnetic topology of the system allowing us to recover a good estimate of the dipole pole strength using ZDI.
    Here again, the SPIRou value yields a lower truncation radius due to the lower magnetic field strength obtained.

    The difference in the results of the various magnetic analyses between the optical and infrared frames in 2019 needs to be discussed. 
    From the B$_\ell$ computed from the LSD profiles, the maximum values obtained are 911$\pm$26 and 64$\pm$20 G, for ESPaDOnS and SPIRou observations, respectively. 
    These maxima both occurred around $\phi$=0.6, where the Stokes \textit{V} signature is maximal with ESPaDOnS but almost vanishes with SPIRou, it is thus not surprising to have a large discrepancy here.
    Concerning the minimum B$_\ell$ values, they reach $-$155$\pm$11 and $-$178$\pm$13 G, for ESPaDOnS and SPIRou observations, respectively, both around $\phi$=0.1 and are thus consistent.
    This behaviour is also visible on the ZDI reconstruction, where the strong positive radial magnetic field region at phase 0.6 on ESPaDOnS map completely vanishes on SPIRou map.
    The two qualitative explanations we can provide are the following: (i) this maximum value is located in an optically dark region of the photosphere, which is consistent with simultaneous optical photometry (Kóspál et al., in prep.), and thus less contrasted in the SPIRou domain.
    The magnetic contribution of this region to the Stokes \textit{V} signal might be thus lowered in the infrared domain if the smaller-scale negative field is obscured in the optical domain.
    And (ii) the two wavelength domains allowing to trace different heights in the photosphere, that such a difference might be an indication of a vertical structure of the magnetic field.


    Concerning the two ZDI reconstructions obtained from the two ESPaDOnS epochs, they seem to point a similar brightness and magnetic radial maps even if the overall topology has evolved from a toroidal- to a poloidal-dominated state, as seen on previous objects \citep[e.g., DQ Tau A,][]{Pouilly23, Pouilly24}. 
    The strong positive radial field spot, associated with the dark spot of the stellar brightness and with the dipole pole, is also consistent with the modulation of \ion{He}{I} D3 NC radial velocity and B$_\ell$.
    However, the latter is pointing to an accretion shock associated with a region with a strong negative field.
    Given the large amplitude of the \ion{He}{I} D3 NC radial velocity variation compared to EX Lup $v\sin i$, the emitting region has probably a very small extent, meaning that its magnetic field information is lost at the photospheric level, explaining this disparity \citep{Yadav15}.  

    Despite this stable pattern between 2016 and 2019, we observed some disparities in some parameters.
    Even if the radial velocity and the B$_\ell$ (from LSD and \ion{He}{I} D3) modulations are well in phase, their amplitudes are slightly larger in 2019.
    If the stronger magnetic field obtained in the optical frame, at small- and large-scale, explains the larger amplitude of the B$_\ell$ modulation, the accompanying effect on the radial velocity points to a modulation by the hotspot.
    This is not a surprising behaviour, but the consistency between the ESPaDOnS and SPIRou radial velocity measurements is.
    Indeed, the stellar activity effect on the photospheric lines, producing the apparent radial velocity modulation, is a wavelength-dependent phenomenon, an amplitude consistency between the radial velocity measurements in the optical and infrared frames is thus not expected.
    
    To investigate in our data sets the relation between the stellar activity and the radial velocity modulation, we compared the latter to an activity indicator, the bisector inverse slope \citep[BIS, ][]{Queloz01}.
    The BIS was computed from the LSD profiles presented in Fig.~\ref{fig:LSD16}, \ref{fig:LSD19}, and \ref{fig:LSDIR}, and is defined as the difference between the mean velocity of the bisector at the top and at the bottom of the line.
    On ESPaDOnS profiles, the first 15\%, which contains the continuum and the wings, were ignored, as well as the last 15\%, where the noise or an activity signature splitting the profile into two parts can affect the computation.
    The top and bottom regions used to compute the BIS are the 25\% top and bottom parts of the remaining profile.
    For SPIRou profiles, the same conditions were used except that we had to ignore the first 25\% of the profiles as it was affected by stronger wings.
    In the case where the radial velocity modulation is only induced by the stellar activity, the line deformation, indicated by the BIS, is completely responsible for this modulation.
    This means that a strong linear correlation should appear between the BIS and the radial velocity with, in a perfect situation, a $-$1-slope and a BIS~=~0~\kms\ at the mean velocity.
    
    The BIS versus radial velocity plots are presented in Appendix~\ref{ap:bis}.
    For each data set, the Pearson correlation coefficient indicates a strong anti-correlation (2016: r=$-$0.81, p-value=0.005; ESPaDOnS 2019: r=$-$0.95, p-value=0.004; SPIRou: r=$-$0.72, p-value=0.05).
    However, the optical measurements show much shallower slopes ($-$0.51$\pm$0.13 and $-$0.43$\pm$0.07 in 2016 and 2019, respectively), and an intersect far from the mean velocity expected ($-$0.34$\pm$0.10 and $-$0.08$\pm$0.06). 
    Only the infrared measurements are consistent (slope:$-$0.68$\pm$0.27, intersect: $-$1.37$\pm$0.25), but only thanks to the larger uncertainties, probably due to the lower correlation.
    Furthermore, the much lower BIS obtained (exclusively negative) in the infrared compared to the optical frame, indicates chromatic effects that are not observed in the radial velocity modulation.
    Stellar activity is thus probably dominating the radial velocity modulation, but another effect, such as Doppler shift induced by a companion, can not be completely excluded and needs further investigations that are unfortunately beyond the scope of the present work.



    Finally, we would like to address the strong magnetic field recovered that drives the accretion process on EX~Lup.
    With the B$_\ell$ reaching 4.2 kG in the accretion shock, a dipole strength of 1.9 kG within a large-scale field of 1 kG reaching 4.8 kG locally, and a small-scale field exceeding 3 kG in the optical domain, EX~Lup has one of the strongest magnetic field among cTTSs, and the strongest among those with dipole-dominated topology.
    Such a configuration can set the suitable condition invoked by \cite{DAngelo10} for their hypothesis of an episodic accretion due to the magnetospheric accretion process itself.
    Indeed, with such a dipolar strength, the magnetospheric radius is outside but close to the corotation radius (see Table~\ref{tab:rmag}).
    In such a situation, instability arises due to the fact that angular momentum is transferred from the star to the disc (the so-called propeller regime), without being enough to drive an outflow.
    This magnetic interaction only prevents accretion, pilling up the gas in the inner disc and increasing its pressure, thus forcing the inner edge of the disc to move inward and cross the corotation radius allowing the accretion to occur.
    Once the gas reservoir has been accreted, the inner edge of the disc moves outward and another cycle starts \citep[see][]{DAngelo10, DAngelo11, DAngelo12}.
    This phenomenon is different from the magnetospheric inflation reported for other cTTSs \citep[e.g., see the studies of AA Tau or V807 Tau by][respectively]{Bouvier03, Pouilly21}, where a significant difference between the truncation and corotation radii induces a torsion of the magnetic field lines producing a toroidal field that inflates the whole magnetosphere. 
    This inflation goes up to an opening of the magnetic field lines that produces a magnetospheric ejection \citep{Zanni13, Pantolmos20} before reconnecting to the disc.
    Such a phenomenon is recurrent and produces an accretion variability as well.
    Still, the time scale of such a cycle is of the order of the stellar rotation period, far shorter than the gas pilling up invoked by \cite{DAngelo10}, and no signature of magnetospheric inflation, nor ejection, were detected in EX Lup.
    However, we would like to stress to the reader that no accumulation of matter at the magnetospheric radius was detected either, EX~Lup seems only to be in the same initial conditions as the latter authors' theory. 
    In addition, the BIS suggest another source for radial velocity variation, such as a companion. 
    Tidal interactions \citep{Bonnell92} or thermal instabilities in the disc \citep{Lodato04} induced by a companion are also existing hypotheses for EXor behaviour. 
    Moreover, recent works by \cite{Nayakshin24a, Nayakshin24b} favour the latter instability as the origin of the episodic accretion of FUor objects.
    Finally, the disc itself is not investigated in this work, instabilities in the disc are thus hypotheses that cannot be excluded \citep{Bell94, Armitage01}.


\section{Conclusions}
\label{sec:conclusion}
    
    EX~Lup is the prototypical EXor-type object whose recurrent bursts and outbursts were previously studied in detail using spectroscopy and photometry.
    However, until this work, no information about its magnetic field was derived, despite its key role in the accretion process of cTTS and its possible origin for episodic accretion.
    Here we provide the first spectropolarimetric time series, over two epochs (2016 and 2019) and two wavelength domains (optical and infrared), study of EX~Lup, the first EXor whose magnetic field is studied.

    We confirmed an ongoing magnetospheric accretion process as seen on many cTTSs.
    It is represented by an accretion funnel flow and an accretion shock corotating with the stellar surface and driven by a kG dipolar magnetic field.
    The funnel flow seems aligned with the accretion shock, itself located near the magnetic dipole pole, which is consistent with the stable pattern observed between the epochs studied.

    The magnetic field of EX~Lup has shown some disparities between wavelength domains, much weaker in the infrared.
    This can be understood as a wavelength dependency of the parameters studied, but can also point to a vertical structure of the magnetic field, as different wavelengths are tracing different heights in the photosphere.
    An expected small-to-large scale effect is also observed, by the different field strengths recovered using ZDI and Zeeman intensification, but also by the opposite polarity of the field associated with the accretion shock and the one recovered using LSD. 
    This indicates a very small accretion shock, as expected from the low mass accretion rate in quiescence and the large radial velocity variation of the emitting region.

    Finally, the multi-kG field recovered for EX~Lup is pointing to a magnetospheric radius being close but outside the corotation radius. 
    This configuration is suitable for disc instabilities induced by the magnetic field that yield accretion cycles.
    These cycles might explain the accretion bursts observed on EX~Lup, suggesting an inherently episodic magnetospheric accretion process.
    However, a definite identification of the origin of EXor behaviour is beyond the scope of this paper, and other theories implying the disc itself or a companion cannot be excluded.


\begin{acknowledgements}
    We would like to warmly thanks Oleg Kochukhov for useful discussions about the magnetic field of EX Lup, as well as Colin P. Folsom for his help in using the new version of his \texttt{ZDIpy} package.
    
    The \texttt{SpecpolFlow} package is available at \url{https://github.com/folsomcp/specpolFlow}.

    The \texttt{PySTEL(L)A} package is available at \url{https://github.com/pouillyk/PySTELLA}
    
    This research was funded in whole or in part by the Swiss National Science Foundation (SNSF), grant number 217195 (SIMBA). For the purpose of Open Access, a CC BY public copyright licence is applied to any Author Accepted Manuscript (AAM) version arising from this submission.
    
    Based on observations obtained at the Canada–France– Hawaii Telescope (CFHT) which is operated from the summit of Maunakea by the National Research Council of Canada, the institut National des Sciences de l’Univers of the Centre National de la Recherche Scientifique of France, and the University of Hawaii. The observations at the Canada–France–Hawaii Telescope were performed with care and respect from the summit of Maunakea which is a significant cultural and historic site.

    This work was also supported by the NKFIH excellence grant TKP2021-NKTA-64.
\end{acknowledgements}

\bibliographystyle{aa}
\bibliography{literature}

\begin{appendix}

    \section{Cross-correlation matrices of optical emission lines}
    \label{ap:CM}
        Here we present the cross-correlation matrices of the ESPaDOnS emission lines studied in this work.
        These matrices are discussed in Sect.~\ref{subsubsec:cmESP}.
        
        \begin{figure}[h!]
            \centering
            \includegraphics[width=0.34\textwidth]{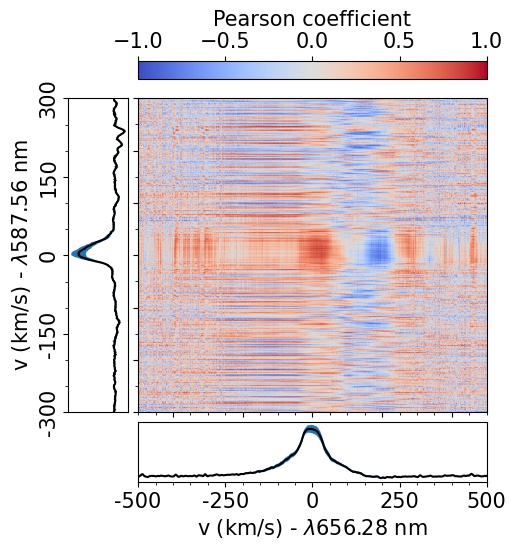}
            \includegraphics[width=0.34\textwidth]{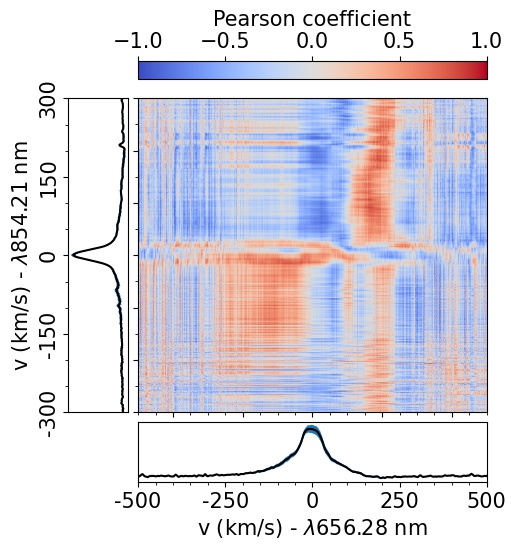}
            \includegraphics[width=0.34\textwidth]{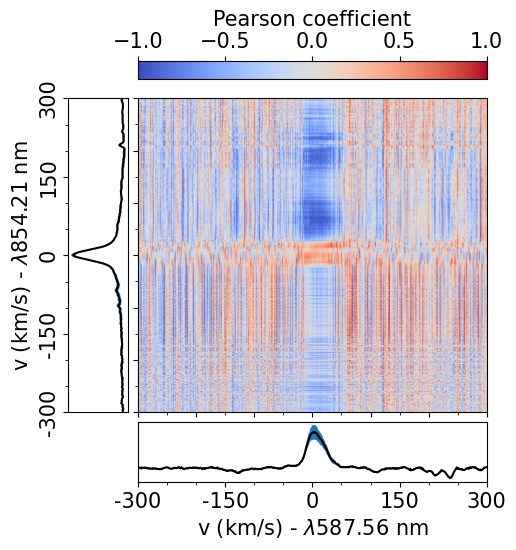}
            \caption{\ion{He}{I} D3 vs. H$\alpha$ \textit{(top)}, \ion{Ca}{II} IRT (854.2 nm) vs. H$\alpha$ \textit{(middle)}, and \ion{Ca}{II} IRT (854.2 nm) vs. \ion{He}{I} D3 \textit{(bottom)} correlation matrices for ESPaDOnS 2016 emission lines.}
            \label{fig:CM16}
        \end{figure}

        \begin{figure}
        \centering
            \includegraphics[width=0.34\textwidth]{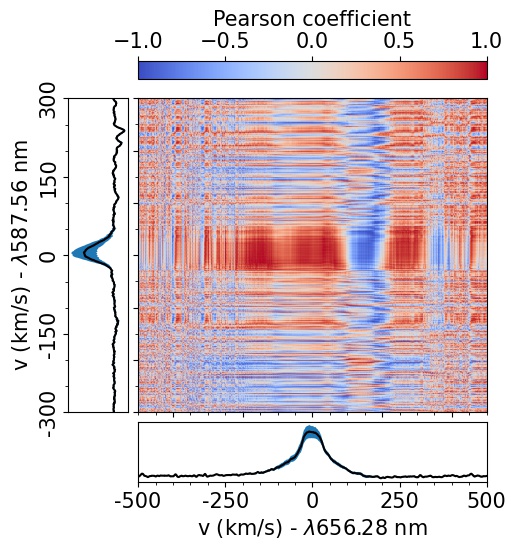}
            \includegraphics[width=.34\textwidth]{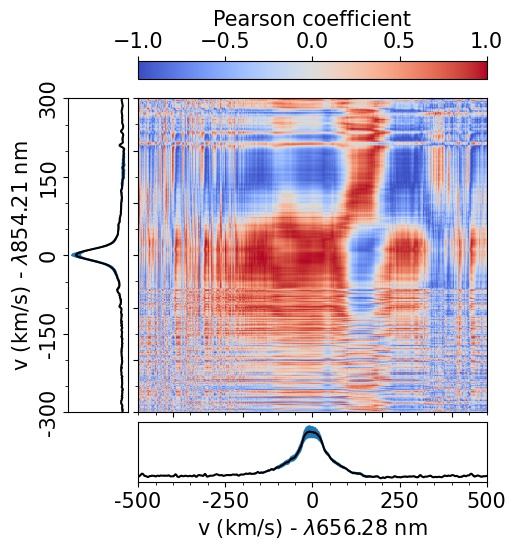}
            \includegraphics[width=.34\textwidth]{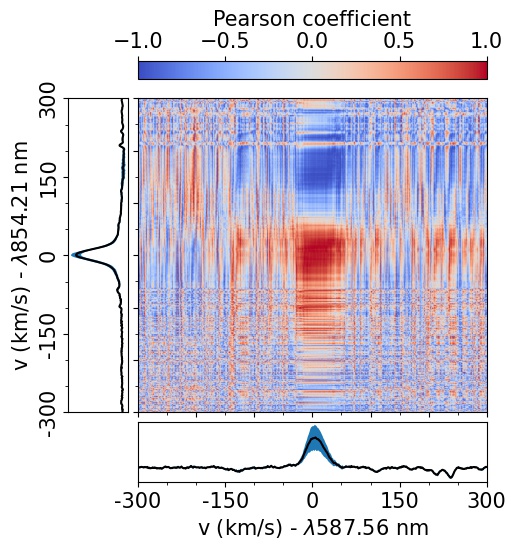}
            \caption{Same as Fig.~\ref{fig:CM16} for ESPaDOnS 2019 emission lines.}
            \label{fig:CM19}
        \end{figure}

    \section{Stokes \textit{I} and \textit{V} fit from ZDI reconstruction}
    \label{ap:ZDIfit}
\begin{figure*}
            \centering
            \includegraphics[width=.3\textwidth]{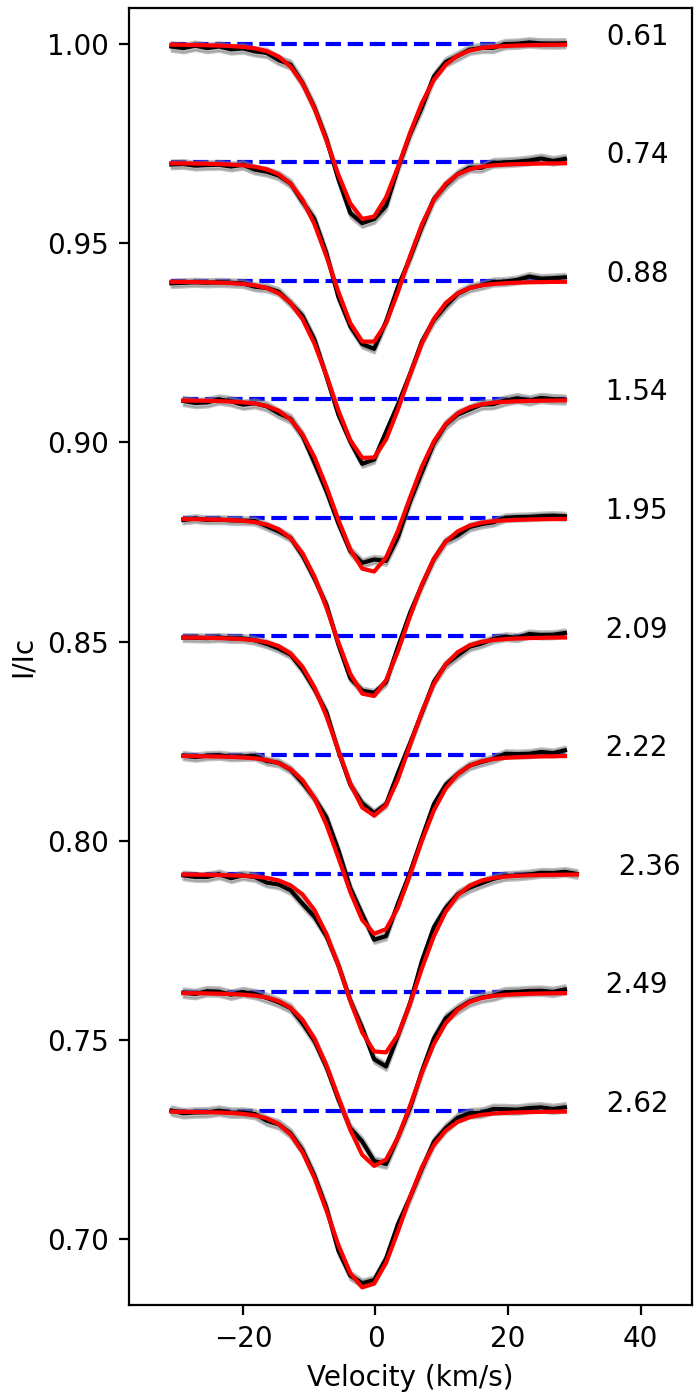}
            \includegraphics[width=.3\textwidth]{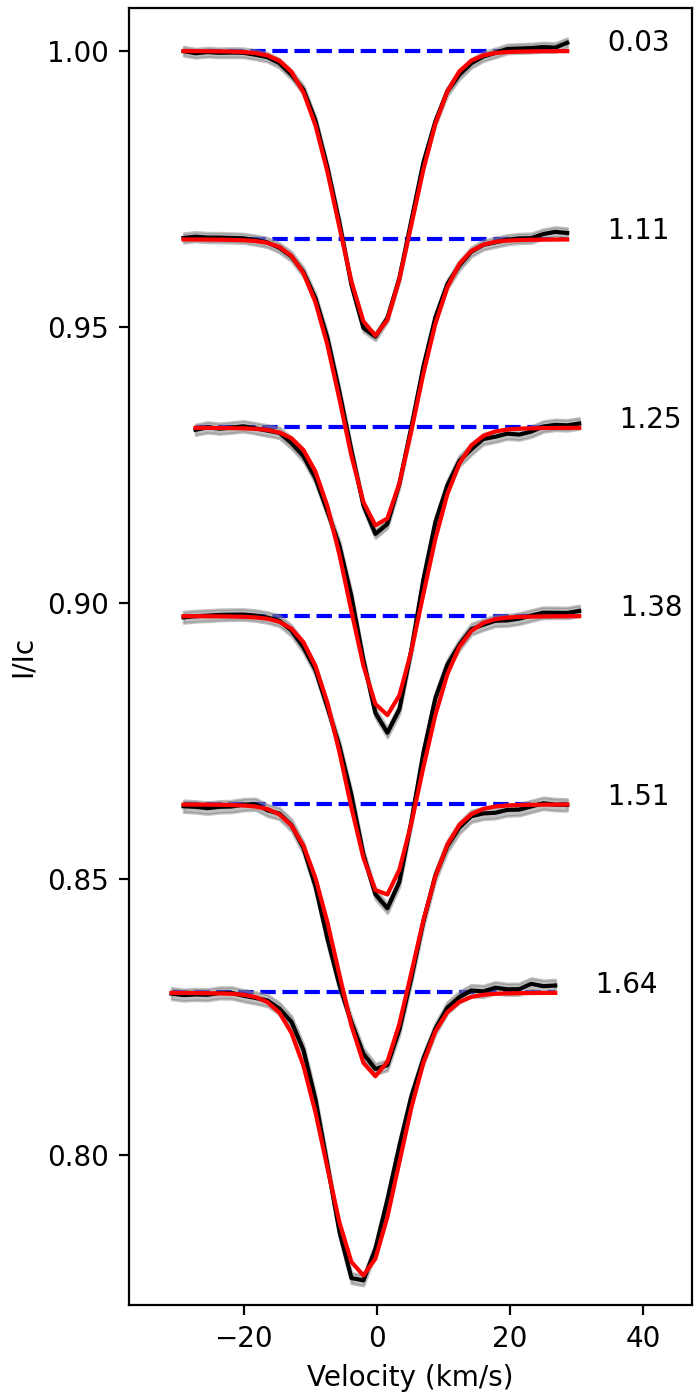}
            \includegraphics[width=.3\textwidth]{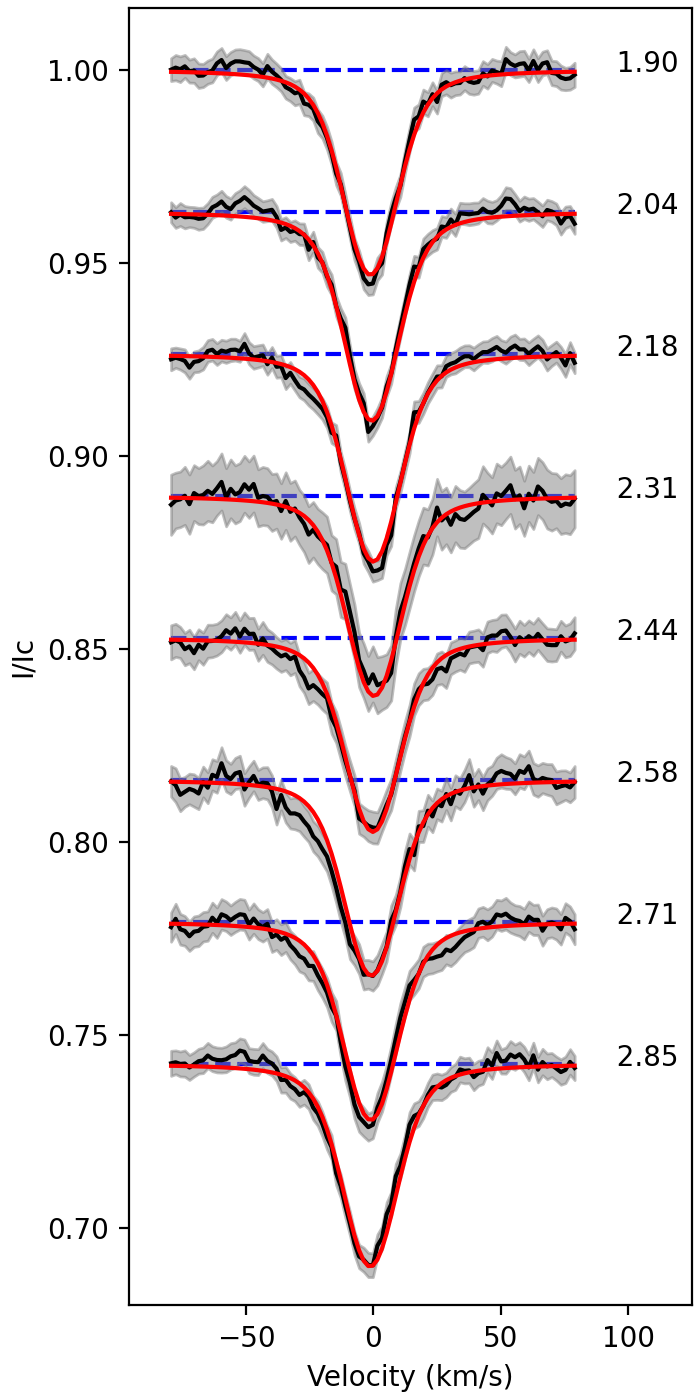}
            \caption{Fit \textit{(red)} of the observed \textit{(black)} Stokes \textit{I} profiles from the ZDI reconstruction for ESPaDOnS 2016 \textit{(left)}, 2019 \textit{(middle)} and SPIRou \textit{(right)} observations. The grey-shaded area corresponds to the uncertainties in the observation, and the number at the right of each profile indicates its rotation cycle.}
            \label{fig:zdiIfit}
        \end{figure*}

\begin{figure*}
            \centering
            \includegraphics[width=.3\textwidth]{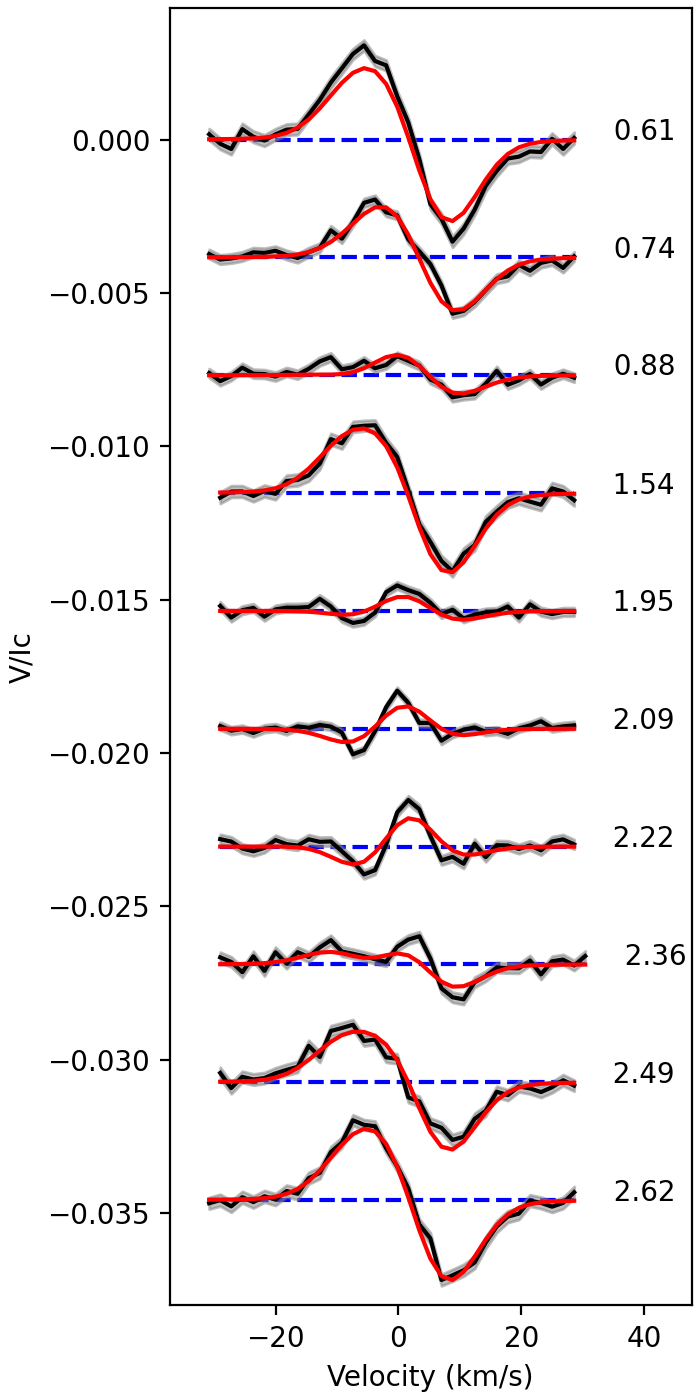}
            \includegraphics[width=.3\textwidth]{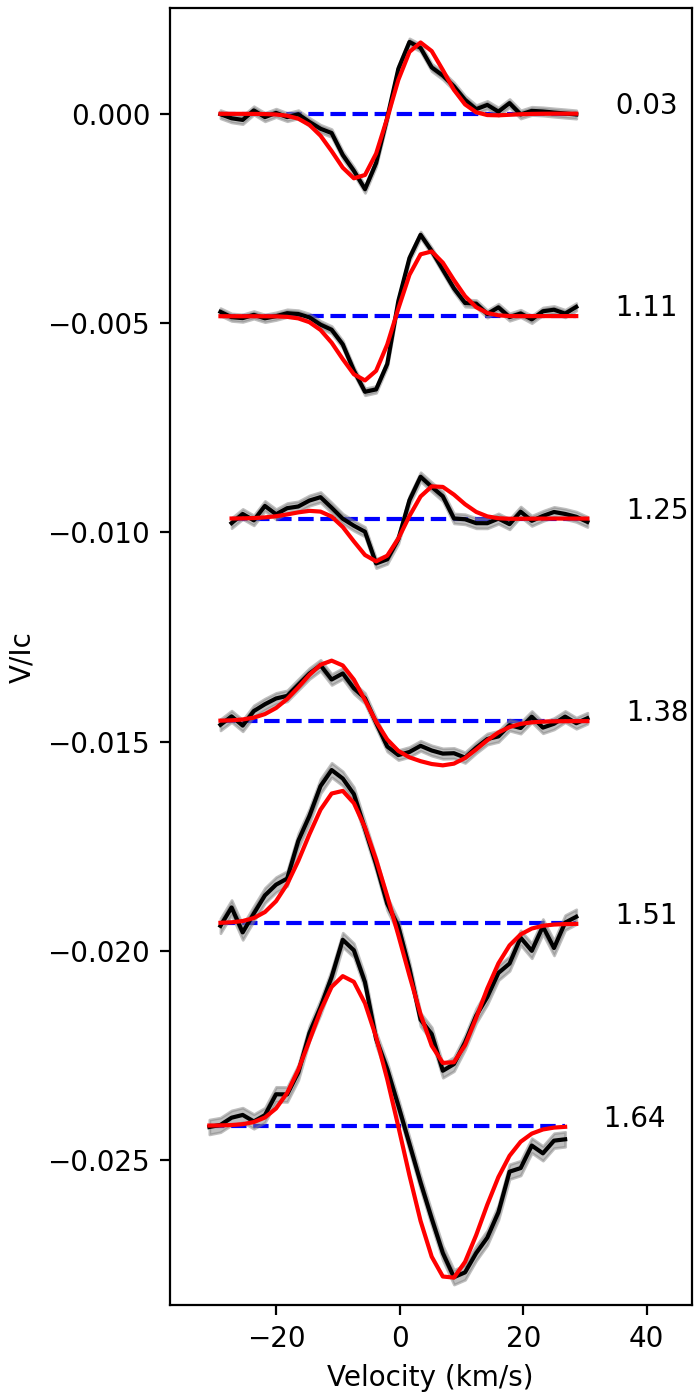}
            \includegraphics[width=.3\textwidth]{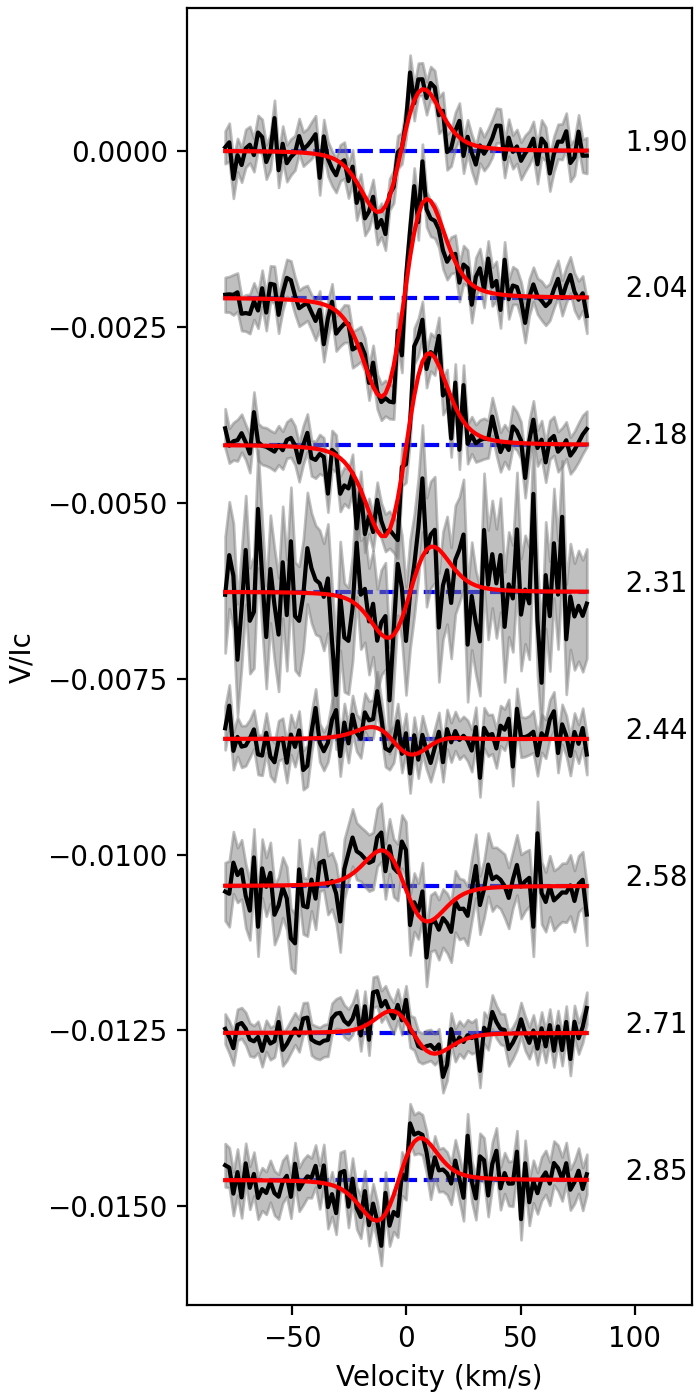}
            \caption{Same as Fig.~\ref{fig:zdiIfit} for Stokes \textit{V} profiles.}
            \label{fig:zdiVfit}
        \end{figure*}

        In this section, we show the fit of the LSD profiles by the ZDI reconstruction for the three data sets studied.
        The Stokes \textit{I} profiles are shown in Fig.~\ref{fig:zdiIfit} and the Stokes \textit{V} profiles in Fig.~\ref{fig:zdiVfit}.
    
    \section{BIS vs. radial velocity}
    \label{ap:bis}
        This appendix presents the analysis of the BIS and radial velocity correlation. 
        These results are discussed in Sect.~\ref{sec:discussion}.
        \begin{figure}[h!]
            \centering
            \includegraphics[width=.45\textwidth]{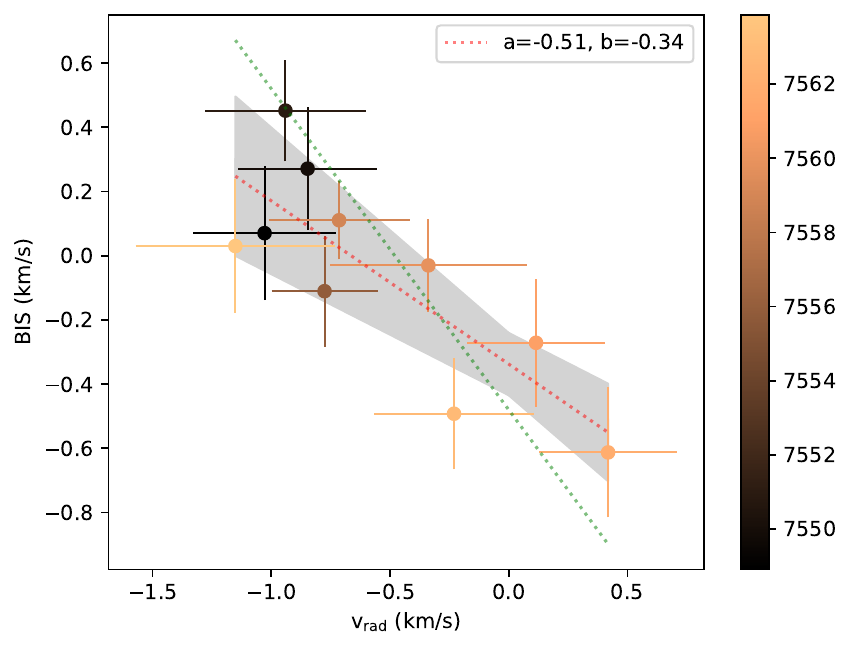}
            \includegraphics[width=.45\textwidth]{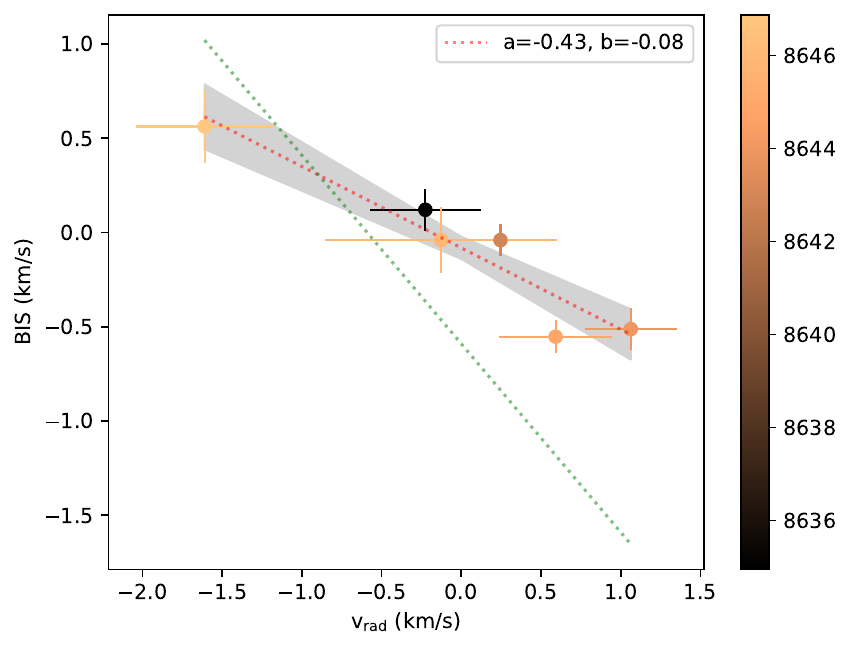}
            \includegraphics[width=.45\textwidth]{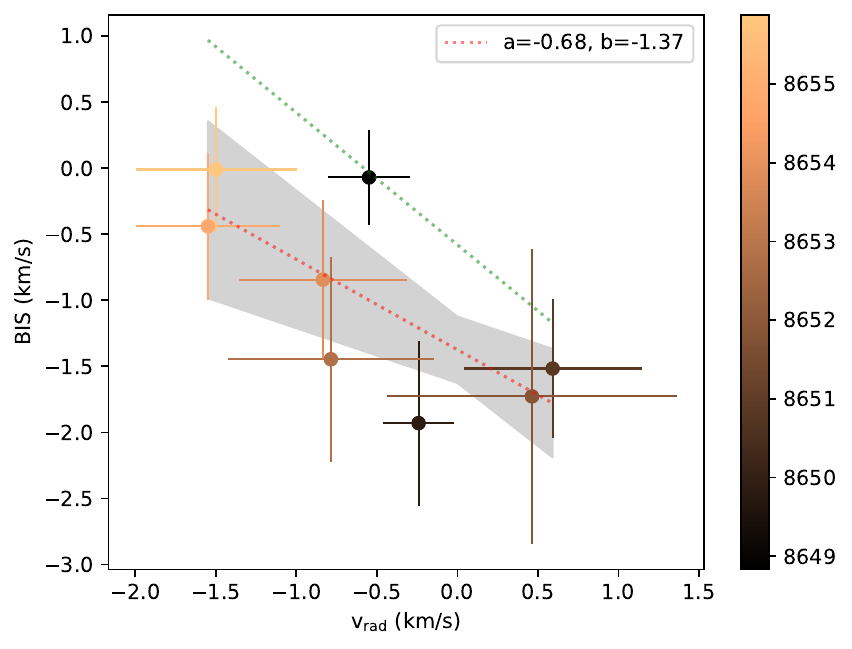}
            \caption{BIS vs. RV of ESPaDOnS 2016 \textit{(top)}, 2019 \textit{(middle)}, and SPIRou \textit{(bottom)} observations. The colour code scales the HJD of observations. The red dotted line shows the best linear regression with slope "a" and intercept "b" indicated in legend. The uncertainty on the regression is shown as the grey-shaded area The green dotted line has a slope of $-$1 and an intercept equal to the mean radial velocity.}
            \label{fig:BISxRV}
        \end{figure}

    
        
        
\end{appendix}

\end{document}